\documentclass[10pt,journal]{IEEEtran}
\usepackage[table]{xcolor}
\usepackage{amsmath, cite, float}
\usepackage{rotating, tabularx, booktabs}
\usepackage{tikz, url}
\usepackage{bbding}
\usepackage{pifont}
\usepackage{wasysym}
\usepackage{amssymb}
\usepackage{xcolor}
\usepackage{graphicx}
\usepackage{multirow}
\usepackage{comment}
\newcolumntype{Y}{>{\raggedright\arraybackslash}X} % 自动换行的左对齐列
\usepackage{caption}

\usepackage{pifont}

\newcommand{\cmark}{\ding{51}}
\newcommand{\xmark}{\ding{55}}

\usepackage{cuted}    % 提供 strip 环境，非浮动跨两栏
\usepackage{capt-of}  % 提供 \captionof{table}；不改变 IEEE 的标题样式

\usepackage{dblfloatfix} % 或 \usepackage{stfloats}

\usepackage[switch]{lineno}

\title{Explainable AI for Next-Generation Wireless Physical Layer: Basics, State-of-the-Art, and Open Challenges}
\author{Bingnan Xiao,
        Shuyan~Hu,~\IEEEmembership{Member,~IEEE}, 
        Xiaojing~Chen,~\IEEEmembership{Member,~IEEE},
        Zhiyuan~Zhai, Bingcong~Li, \\
        Wei~Ni,~\IEEEmembership{Fellow,~IEEE},
        Xin~Wang,~\IEEEmembership{Fellow,~IEEE},
        and Ekram Hossain,~\IEEEmembership{Fellow,~IEEE}
\thanks{B. Xiao, Z. Zhai, and X. Wang are with the Key Laboratory of EMW Information (MoE), College of Future Information Technology, Fudan University, Shanghai 200433, China (e-mail: \{22110720061, 22110720067\}@m.fudan.edu.cn, xwang11@fudan.edu.cn).

S. Hu is with the College of Electronics and Information Engineering, Tongji University, Shanghai 201804, China (e-mail: syhu@tongji.edu.cn).

X. Chen is with the Key Laboratory of Specialty Fiber Optics and Optical Access Networks, Shanghai University, Shanghai 200444, China 
(e-mail:~jodiechen@shu.edu.cn).

B. Li is with the Department of Computer Science at ETH Zurich, 8092 Zürich, Switzerland (email: bingcong.li@inf.ethz.ch).

W. Ni is with the School of Engineering, Edith Cowan University, Perth, WA 6027, Australia (e-mail: wei.ni@ieee.org).

E. Hossain is with the Department of Electrical and Computer Engineering, University of Manitoba, Winnipeg, MB R3T 2N2, Canada (e-mail:
ekram.hossain@umanitoba.ca).
}

}
\begin{document}

\maketitle

\begin{abstract}

Next-generation wireless systems are expected to be ``AI-native," with neural networks (NNs) embedded throughout the physical (PHY) layer protocol stack to improve spectral efficiency, latency, and network autonomy. 
However, the opacity of deep learning (DL) models raises increasing concerns about system reliability, safety, and privacy, especially under complex and time-varying network environments.
This survey studies explainable AI (XAI) in wireless PHY layers from the explainability perspective. 
We first formalize a series of responsibility-oriented goals for wireless XAI.
Then, we develop a systematic taxonomy of explainability approaches and distill practical criteria for deploying explanations in communication scenarios. 
% Building on a layer perspective, 
We provide a comprehensive review of where and how XAI can be applied throughout the PHY layer, connecting representative learning paradigms to appropriate explanation techniques, evaluation metrics, and deployment considerations. 
Open challenges and future directions are discussed, including explainability--performance tradeoffs, explainability-aware data processing, customized XAI for communication-specific structures, cross-layer explanation consistency, and emerging needs for explaining LLM- and Agentic-AI-driven PHY layers.

\end{abstract}

\begin{IEEEkeywords}
Explainable AI, wireless physical layer, deep learning, wireless communication.

\end{IEEEkeywords}

\begin{table*}[!t]
    \caption*{\textbf{List of abbreviations}}
    \label{table:abbr_fullwidth}
    \centering
    \fontsize{7.2}{8.64}\selectfont
    \setlength{\tabcolsep}{5pt}
    \begin{tabularx}{\textwidth}{@{} lY lY @{}}
    \toprule
    \textbf{Abbreviation} & \textbf{Full form} & \textbf{Abbreviation} & \textbf{Full form} \\ \midrule
    5G & The fifth generation & LDAMP & Learned Denoising-based Approximate Message Passing \\
    6G & The sixth generation & LIME & Local Interpretable Model-agnostic Explanations \\
    ACK & Acknowledgment & LLM & Large Language Model \\
    AE & Autoencoder & LRP & Layer-wise Relevance Propagation \\
    AI & Artificial Intelligence & MAML & Model-Agnostic Meta-Learning \\
    AI4Net & AI for Network & MCS & Modulation and Coding Scheme \\
    AoA & Angle of Arrival & MIMO & Multiple-Input Multiple-Output \\
    AMC & Automatic Modulation Classification & MISO & Multiple-Input Single-Output \\
    BER & Bit Error Rate & ML & Machine Learning \\
    BS & Base Station & MLP & Multilayer Perceptron \\
    CAM & Class Activation Mapping & mmWave & Millimeter Wave \\
    CEN & European Committee for Standardization & MSE & Mean Squared Error \\
    CNN & Convolutional Neural Network & MTL & Multi-Task Learning \\
    CS-DLMA & Carrier-Sense Deep Reinforcement Learning Multiple Access & NLDT & Nonlinear Decision Tree \\
    CSI & Channel State Information & O-RAN & Open Radio Access Network \\
    CSMA & Carrier Sense Multiple Access & OFDM & Orthogonal Frequency-Division Multiplexing \\
    DBN & Deep Belief Network & PIRL & Programmatically Interpretable Reinforcement Learning \\
    DDPM & Denoising Diffusion Probabilistic Model & PPO & Proximal Policy Optimization \\
    DDQN & Double Deep Q-Network & QoE & Quality of Experience \\
    DDT & Differentiable Decision Tree & QoS & Quality of Service \\
    DeepLIFT & Deep Learning Important FeaTures & RAN & Radio Access Network \\
    DeepRED & Deep Rule-Extraction & RIC & RAN Intelligent Controller \\
    DkNN & Deep k-Nearest Neighbors & RL & Reinforcement Learning \\
    DL & Deep Learning & RNN & Recurrent Neural Network \\
    DOA & Direction of Arrival & RSRP & Reference Signal Received Power \\
    DQN & Deep Q-Network & SBL & Sparse Bayesian Learning \\
    DRL & Deep Reinforcement Learning & SDT & Soft Decision Tree \\
    DRQN & Deep Recurrent Q-Network & SHAP & SHapley Additive exPlanations \\
    DSP & Deep Symbolic Policy & SINR & Signal-to-Interference-plus-Noise Ratio \\
    ETSI & European Telecommunications Standards Institute & SliceOps & Slice Operations framework \\
    EU & European Union & SNR & Signal-to-Noise Ratio \\
    FFDNet & Fast and Flexible Denoising Convolutional Neural Network & SSFM & Split-Step Fourier Method \\
    GAM & Generalized Additive Model & TDMA & Time Division Multiple Access \\
    GNN & Graph Neural Network & UAV & Unmanned Aerial Vehicle \\
    Grad-CAM & Gradient-weighted Class Activation Mapping & UE & User Equipment \\
    GRU & Gated Recurrent Unit & URLLC & Ultra-Reliable Low-Latency Communications \\
    IoT & Internet of Things & VIA & Validity Interval Analysis \\
    IRS & Intelligent Reflecting Surface & XAI & Explainable Artificial Intelligence \\
    ITU & International Telecommunication Union & XRL & Explainable Reinforcement Learning \\
    \bottomrule
    \end{tabularx}
\end{table*}

\section{Introduction}

% The advent of the sixth-generation (6G) mobile networks marks a transformative evolution in wireless communications.
% % Distinct from previous generational advances, 6G is
The sixth-generation (6G) mobile communication networks are envisioned to have artificial intelligence (AI) deeply integrated into its protocol stack, including the physical (PHY) layer, to unlock significant performance gains in spectral efficiency, latency, and network autonomy~\cite{
% khan2025artificial,
HongleiCOMST2026,XiaolinCOMST2026}.
% For instance, AI techniques at the physical layer can optimize transmitter/receiver operations and improve communication efficiency beyond what traditional models achieve~\cite{8663966}.
The high dimension and nonlinearity of deep neural networks (DNNs) often lead to opacity, raising concerns about reliability, accountability, and safety
% , especially in dynamic wireless environments~
\cite{10736556}.
It is critical to design understandable AI models
to facilitate transparent decision-making in 6G
for the following reasons.
\begin{itemize}
    \item \textit{Robustness: }Wireless channels are susceptible to a dynamic environment.
    It is necessary to identify key components influencing AI outputs
    and reveal the underlying features and reasoning patterns
    % for fast and accurate channel estimation and prediction,
    to improve the quality and resilience of communication systems against unpredictable radio conditions~\cite{10854503}.

    \item \textit{Personalization: }The real-time requirements, partial observability of radio environments, and multi-layered protocol architectures of 6G necessitate tailored, lightweight, and domain-specific explainability.
    To be meaningful to network operators, feature-importance explanations need to be mapped to domain-specific semantics, e.g., channel quality indicators, modulation patterns, or protocol logic, rather than to raw pixel- or token-level features~\cite{9322496}.

    \item \textit{Trustworthiness: }For quality- and safety-critical 6G applications, e.g., ultra-reliable low-latency communications (URLLC), autonomous driving, and remote surgery,
    stakeholders must trust AI-driven decisions~\cite{10.1145/3675392,LukasIOTJ2025}.
    Engineers, operators, and regulators can understand model behavior before deployment.
    % , for a broader acceptance in high-stakes environments.
    
\end{itemize}
% In response to this requirement, t
The concept of eXplainable AI (XAI) has emerged,
which refers to techniques and methodologies that make the decision-making processes of AI models understandable to humans.
XAI alleviates the unreliable deployment of complex DNNs by producing interpretable outputs or explanations for users to trace, validate, and understand model behavior~\cite{8466590},
and turns the AI ``closed box'' into a ``glass box'' by illuminating how inputs are transformed into outputs~\cite{rai2020explainable}.

\subsection{Explainable AI}

Explainability in AI can be categorized along two dimensions: the model structure and the learning paradigm.
From the perspective of model structure, XAI methods are typically divided into post-hoc explainability and inherently interpretable models~\cite{bhalla2023discriminative}.
Post-hoc methods interpret an already-trained, opaque model by analyzing its internal mechanisms or outputs (e.g., feature importance scores, surrogate models, or rule extraction), whereas inherently interpretable models are transparent by construction, enabling direct inspection of their decision processes~\cite{rudin2019stop}.
For example, decision trees and linear models are interpretable by design, while DNNs require post-hoc explanation techniques.

From the perspective of the learning paradigm, explainability techniques vary across supervised, unsupervised, and reinforcement learning (RL) settings.
In supervised learning, explainability is linked to identifying influential features, understanding decision boundaries, or assessing model confidence.
Model-agnostic methods (e.g., LIME~\cite{ribeiro2016should} and SHAP~\cite{NIPS2017_8a20a862}) have been developed to provide such explanations by approximating a complex model locally or attributing importance to input features~\cite{milani2024explainable}.\footnote{Model-agnostic methods do not consider the internal components of the model (i.e., weights and structure parameters)~\cite{10460345}, and can apply to any closed-box method. Model-specific methods are defined by the parameters of a model, e.g., explaining the weights of a linear regression or using the inference rules of a decision tree~\cite{speith2022review}. Using model-agnostic methods, developers can flexibly choose any ML model to generate explanations, which differs from generating decisions for the actual closed-box model~\cite{hassija2024interpreting}.}
In unsupervised learning, however, the lack of explicit ground truth makes it more challenging to validate explanations, shifting the focus toward interpreting latent structures (e.g., cluster formations or embedding spaces) and evaluating the consistency of learned representations.

RL introduces further complexity, as the explanatory targets may include not only state-action mappings but also long-term policy behaviors and weight assignment over time (i.e., which past actions or rewards most influenced the current decision). 
The XAI methods must be aligned with the learning task and its operational context, ensuring that the explanations are accurate and relevant for the domain at hand. These distinctions are pronounced in wireless communications due to the specific constraints and objectives of each learning task, e.g., signal detection and resource control.

In practice, XAI is measured by a set of attributes across different learning paradigms, instead of a universal score.
One main attribute is faithfulness to examine whether the generated explanation truly reflects the model's decision~\cite{yeh2019fidelity, agarwal2022openxai}.
It can be tested by perturbation-based evaluations. When features or regions highlighted by an explanation are masked or injected with noise, researchers then check whether the model output or task performance degrades accordingly.
Similarly, when a surrogate model is used as the explanation, its consistency with the black-box predictions provides a quantitative measurement.
Another attribute is stability/robustness, which examines whether explanations remain consistent under input perturbations, noise, or retraining randomness~\cite{alvarez2018robustness}.
It is quantified by the variance of attribution maps or the rank correlation of feature importance scores across nearby samples.

% Simplicity is also considered, where simpler rules and shallower trees are easier to inspect, but may sacrifice performance.
Cost is vital in communication settings, since producing explanations can introduce extra computation, memory, and latency~\cite{pavel2025onboard, choudhary2025edge}. This attribute is typically reported as additional runtime, energy, or feedback overhead under the same deployment constraints.
Meanwhile, these attributes may be weighted differently across learning paradigms.
For supervised learning, faithfulness and stability of feature attributions are frequently emphasized.
XAI attributes for unsupervised learning are often related to whether latent dimensions correspond to stable and meaningful generative factors.
For RL, explanation quality is often evaluated with the action trajectory, e.g., whether explanations remain consistent over multi-step interactions and whether counterfactual action changes lead to predictable return variations.

XAI presents distinctive challenges in wireless communications.
Time-varying channels, mobility-induced distribution shifts, and partial observability can perturb both model decisions and their explanations.
% Explanations should adapt to the dynamic environment and help identify invariant structure, for efficient adaptation and fine-tuning across cells.
Meanwhile, many conventional XAI methods are evaluated with generic proxies or produce raw pixel-/token-level saliency, which is insufficient for wireless tasks unless the explanation is mapped to radio semantics, e.g., channel statistics, modulation patterns, beam or resource states, and protocol logic.
% , for a better understanding of designers.
Moreover, computationally intensive explanation methods are difficult to deploy in communication scenarios that require real-time inference and low-latency feedback~\cite{DuXiaTDSC2026}.
% Explanations must be lightweight, embedded, and actionable for learning policies.
% to be returned within tight timing budgets.
Wireless XAI should be evaluated not only by human readability, but also by whether it supports communication-relevant diagnostics, e.g., causal reasoning, robustness assessment, and generalization under domain shift.
% {\color{red}
% 'The term XAI and the related concepts could be discussed in a bit more detail in Section 1.A. What are the measurable attributes of XAI? Are there different metrics/attributes for different AI/ML techniques? If yes, what are these attributes?'
% }

\subsection{Existing Literature Review}
Several recent surveys have delved into XAI and its applications across various domains, as summarized in Table~\ref{tab:related_surveys}. In~\cite{saeed2023explainable}, the authors systematically organized the prevalent challenges and future research directions, providing a consolidated guide for researchers. Another review~\cite{yang2023survey} categorized XAI techniques and examined their applications (e.g., healthcare and finance), offering an analysis of approaches, limitations, and prospects for XAI.

% In the wireless networking arena, 
XAI has emerged as a critical component for future communication systems. The authors of \cite{10854503} focused on XAI in 6G communication systems, with an emphasis on network slicing. Their survey categorized explainability methods (from model-agnostic to model-specific, and spanning pre- to post-model strategies) and discussed how these techniques improve transparency and reliability in complex, real-time network scenarios. In \cite{10772472}, the authors integrated XAI into the Open RAN (O-RAN) architecture. They reviewed how AI-driven RAN functions can be augmented with explainability (mapping learned models to XAI-enabled solutions) and highlighted O-RAN use cases (e.g., RAN automation and slicing) that benefit from XAI. The study \cite{10620685} investigated XAI for 5G/6G network security, noting that prior efforts on XAI were fragmented. It is one of the first holistic overviews of explainable techniques for security and trust in B5G networks.

The authors of \cite{10499970} focused on connecting explainability techniques with 6G operation across different vertical use cases.
In comparison, our paper focuses on explainability for wireless PHY learning tasks and PHY-supporting MAC/RAN resource-control functions, and examines how representative XAI techniques can be aligned with their corresponding interpretation targets and deployment constraints.
The authors of \cite{10158334} conducted a review of XAI within IoT systems.
A taxonomy of XAI techniques suitable for resource-constrained IoT environments was established, and the study proceeded to explore XAI applications in specific IoT domains, including smart healthcare and industrial automation.

The integration of explainability into wireless AI systems remains in its early stages.
Most existing efforts have focused on directly applying classical XAI techniques originally designed for vision or language tasks
to wireless scenarios.
Little attention has been paid to the radio semantics of wireless tasks, e.g., channel estimation, modulation classification, beam selection, and PHY-facing resource control, or to how explanations should be evaluated under communication constraints.
Moreover, the connection between explainability and responsible-AI diagnostics, e.g., causality tracing, robustness assessment, generalization under domain shift, and trust calibration, remains insufficiently organized for wireless PHY-oriented systems.
Challenges such as real-time explainability under latency constraints, lightweight explanations for edge/RAN deployment, and validation of neural models under resource limitations remain only partially addressed in the literature.

\begin{table*}[!t]
\caption{Summary and comparison of related surveys}
\label{tab:related_surveys}
\centering
\footnotesize
\setlength{\tabcolsep}{4pt}
\renewcommand{\arraystretch}{1.15}
\begin{tabularx}{\textwidth}{l|X|>{\centering\arraybackslash}p{1.95cm}|>{\centering\arraybackslash}p{1.95cm}|>{\centering\arraybackslash}p{1.7cm}|>{\centering\arraybackslash}p{2.5cm}}
\hline
\textbf{Ref.} & \textbf{Overview} & \textbf{General XAI} & \textbf{Wireless/6G XAI} & \textbf{PHY Focus} & \textbf{Communication Evaluation} \\
\hline
\cite{saeed2023explainable} &
Organize prevalent challenges and future research directions of XAI &
\color{green}\cmark & \color{red}\xmark & \color{red}\xmark & \color{red}\xmark \\
\hline
\cite{yang2023survey} &
Categorize XAI techniques and examine their applications in fields such as healthcare and finance &
\color{green}\cmark & \color{red}\xmark & \color{red}\xmark & \color{red}\xmark \\
\hline
\cite{10854503} &
Focus on XAI in 6G with an emphasis on network slicing, and discuss transparency in real-time scenarios &
\textit{Partially} & \color{green}\cmark & \color{red}\xmark & \textit{Partially} \\
\hline
\cite{10772472} &
Review explainability enhancement for AI-driven RAN functions, and highlight O-RAN use cases that benefit from XAI &
\textit{Partially} & \color{green}\cmark & \color{red}\xmark & \textit{Partially} \\
\hline
\cite{10620685} &
Provide an overview of XAI techniques for security and trust in B5G networks &
\textit{Partially} & \color{green}\cmark & \color{red}\xmark & \textit{Partially} \\
\hline
\cite{10499970} &
Connect XAI techniques with 6G operation across several vertical use cases &
\textit{Partially} & \color{green}\cmark & \textit{Partially} & \textit{Partially} \\
\hline
\cite{10158334} &
Conduct a review of XAI within IoT systems, and explore explainability enhancement in smart healthcare applications &
\textit{Partially} & \color{red}\xmark & \color{red}\xmark & \textit{Partially} \\
\hline
\textbf{\textit{Ours}} &
Provide a systematic survey on XAI for wireless physical-layer, covering learning models, radio-semantic explanation targets, PHY-facing resource control, and communication-aware evaluation &
\color{green}\cmark & \color{green}\cmark & \color{green}\cmark & \color{green}\cmark \\
\hline
\end{tabularx}
\end{table*}

\begin{figure}[tbp]
	\centering
	\includegraphics[width=0.82\linewidth]{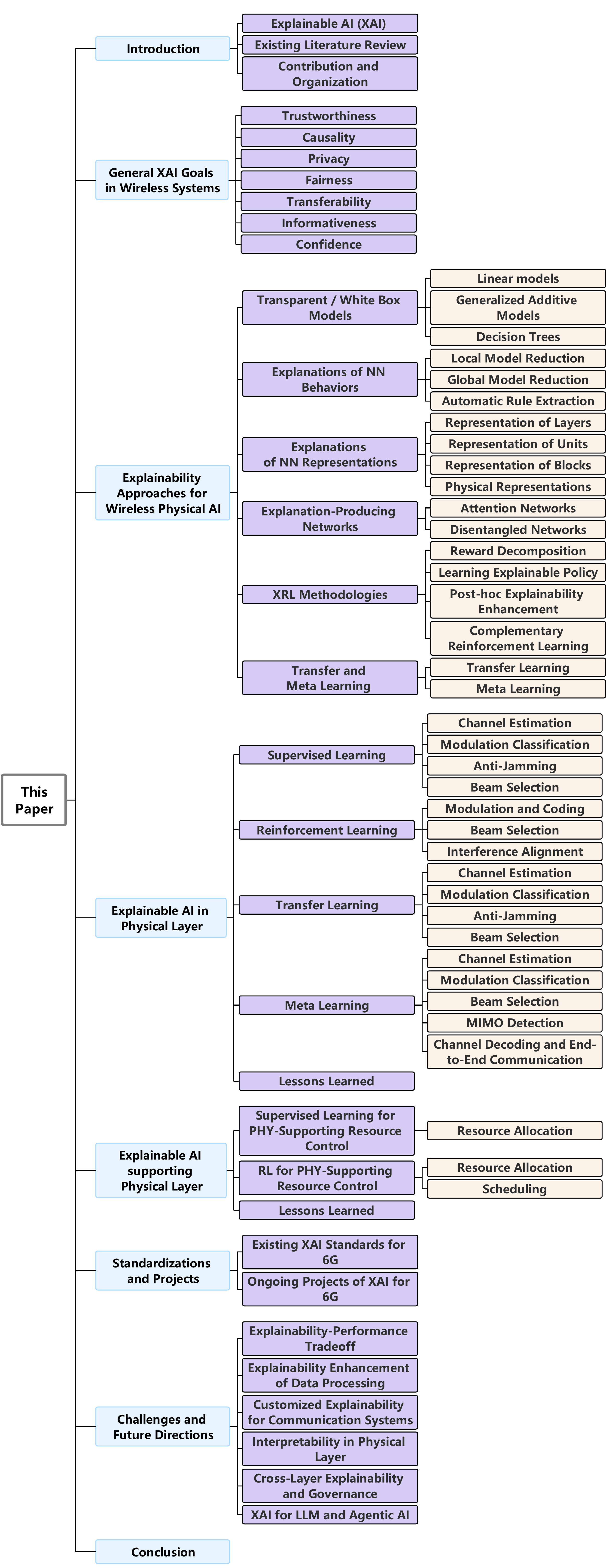}
	\caption{The organization of this survey.}
	\label{fig:organizational diagram}
\end{figure}

\subsection{Contribution and Organization}
This survey examines the interplay between explainability and wireless PHY layer to enable future, responsible AI-powered wireless communication systems.
We provide a comprehensive review of how explainability techniques can be applied in wireless PHY layer to enhance system-level reliability, fairness, and security in AI-enabled 6G networks.
We emphasize the need for explainability mechanisms that are aligned with radio semantics and practically viable under communication constraints, particularly in scenarios involving latency-sensitive decisions, time-varying channels, and partial observability.
We further examine how representative XAI techniques can be selected, interpreted, and constrained according to wireless task structures and communication-aware evaluation criteria.
% {\color{red}We emphasize the need for explainability mechanisms that are theoretically grounded and practically viable in communication scenarios, particularly those involving time-sensitive decisions, dynamic channel environments, and limited observability.
% We also dive deep into how XAI techniques can be adapted or extended to meet the unique demands of wireless tasks.}
The contributions of this survey are summarized as follows:
\begin{itemize}
    \item \textit{PHY-oriented interpretability perspective:} We develop a PHY-oriented view of explainability for wireless AI systems, focusing on how explanation methods can be aligned with PHY tasks and PHY-supporting MAC/RAN resource-control functions. This perspective avoids treating wireless XAI as a generic application of vision- or language-oriented explanation tools, and emphasizes communication-specific constraints such as latency, channel dynamics, and resource coupling.

    \item \textit{Responsibility-driven XAI analysis:} We establish a conceptual linkage between interpretability and key responsible-AI objectives, including trustworthiness, causality, privacy, fairness, transferability, informativeness, and confidence. We discuss how explanations can support model validation, debugging, risk assessment, and accountable deployment in wireless environments.

    \item \textit{Systematic taxonomy and future directions:} We provide a structured review of representative XAI techniques and their applicability to wireless learning paradigms, including supervised learning, reinforcement learning, transfer learning, and meta-learning. We further highlight open challenges such as real-time interpretability, domain-specific validation, communication-aware explanation design, cross-layer explanation consistency, and deployment-oriented standards and projects.
\end{itemize}

The rest of this survey is arranged as depicted in Fig.~\ref{fig:organizational diagram}.
Section II formalizes responsible-AI goals for wireless systems.
Section III reviews representative XAI methodologies, including transparent models, post-hoc explanations, representation analysis, explanation-producing networks, XRL, and transfer/meta-learning-based interpretation.
Section IV analyzes how XAI can be applied to PHY layer learning tasks, including channel estimation, modulation classification, anti-jamming, beam selection, interference alignment, MIMO detection, and end-to-end communication.
Section V discusses XAI for MAC/RAN resource-control functions that support PHY operation.
Section VI summarizes relevant XAI standards, regulatory frameworks, and 
% representative 6G/O-RAN 
projects.
% We highlight where and how XAI can be applied, and connect tasks to evaluation metrics and deployment considerations in each layer. 
Section VII synthesizes challenges and future directions of XAI in wireless systems, followed by conclusions in Section~VIII.
% Table~\ref{table:abbr_fullwidth} collates the abbreviations involved in the paper. 
% Table~\ref{tab:related_surveys} summarizes the existing surveys.
% Fig.~\ref{fig:organizational diagram} depicts the organization of this survey.

\section{General XAI Goals in Wireless Systems}

\begin{table*}[hbt] 
\centering
    \renewcommand\arraystretch{1.2}    
    \caption{Summary of XAI goals and criteria in communication systems 
    % {\color{red}put refs to support statements in the table}
    }
    \label{xai goals}
    \resizebox{\textwidth}{!}{
\begin{tabular}{|c|l|l|l|} 
\hline
\textbf{XAI Goals}       & \multicolumn{1}{c|}{\textbf{Description}}                                                                                                                              & \multicolumn{1}{c|}{\textbf{Target Party}}                                                                     & \multicolumn{1}{c|}{\textbf{Role in Wireless XAI}}                                                                                                                                                                 \\ 
\hline
\textbf{Trustworthiness} & \begin{tabular}[c]{@{}l@{}}The degree to which models' behavior \\justify confidence that it will act reliably \\and as intended across conditions.\end{tabular}       & \begin{tabular}[c]{@{}l@{}}Domain experts;\\Users affected by decisions\end{tabular}                           & \begin{tabular}[c]{@{}l@{}}
Serve as the foundation for network decisions under \\
tight latency and imperfect observations.\\
\textit{\textbf{Example (AI Beamforming):}} Verifying beam tracking \\
decisions in high-mobility V2X to prevent \\
misplaced confidence causing link failures.
\end{tabular}      \\ 
\hline
\textbf{Causality}       & \begin{tabular}[c]{@{}l@{}}Use explanations to distinguish \\cause–and–effect structure \\from correlations, and guide \\validation of causal hypotheses.\end{tabular} & \begin{tabular}[c]{@{}l@{}}Domain experts;\\Managers/executives;\\Regulatory entities/agencies\end{tabular}   & \begin{tabular}[c]{@{}l@{}}
Enable robustness by separating true channel \\
information from spurious environmental proxies.\\
\textit{\textbf{Example (Link Adaptation):}} Distinguishing packet loss \\
caused by deep fading vs. interference collision \\
to correct Modulation and Coding Scheme (MCS) selection.
\end{tabular}  \\ 
\hline
\textbf{Privacy}         & \begin{tabular}[c]{@{}l@{}}Use explainability to assess data \\and internal representations that may \\reveal about individuals or the network.\end{tabular}           & \begin{tabular}[c]{@{}l@{}}Users affected by\\model decisions; \\Regulatory entities/agencies\end{tabular}     & \begin{tabular}[c]{@{}l@{}}
Make privacy risks measurable and support \\
privacy-by-design deployments.\\
\textit{\textbf{Example (Wireless Sensing):}} Identifying and masking \\
sensitive features in WiFi CSI that reveal \\
user identity or specific activities.
\end{tabular}           \\ 
\hline
\textbf{Fairness}        & \begin{tabular}[c]{@{}l@{}}Ensure predictions do not introduce \\unjustified disparities across groups,\\and expose bias pathways via explanation.\end{tabular}        & \begin{tabular}[c]{@{}l@{}}Users affected by \\model decisions; \\Regulatory entities/agencies\end{tabular}    & \begin{tabular}[c]{@{}l@{}}
Reveal network proxies that induce inequity and \\
enable constraints in resource allocation.\\
\textit{\textbf{Example (User Scheduling):}} Exposing DRL policies \\
that systematically starve cell-edge users \\
to maximize aggregate spectral efficiency.
\end{tabular}                                        \\ 
\hline
\textbf{Transferability} & \begin{tabular}[c]{@{}l@{}}Clarify the applicability and reuse of \\learned representations across \\tasks, bands, hardware, and environments.\end{tabular}            & \begin{tabular}[c]{@{}l@{}}Domain experts;\\Data scientists\end{tabular}                                       & \begin{tabular}[c]{@{}l@{}}
Identify invariant structures to guide efficient \\
fine-tuning and safe deployment.\\
\textit{\textbf{Example (Channel Estimation):}} Isolating hardware- \\
agnostic channel features to enable model transfer \\
from synthetic datasets (Sim) to real-world deployment (Real).
\end{tabular}                             \\ 
\hline
\textbf{Informativeness} & \begin{tabular}[c]{@{}l@{}}Provide clear and task-relevant \\information that links model reasoning \\to human decisions.\end{tabular}                                 & \begin{tabular}[c]{@{}l@{}}Domain experts; \\Managers/executives; \\Regulatory entities/agencies\end{tabular}  & \begin{tabular}[c]{@{}l@{}}
Convert explanations into operational signals \\
to speed model debug and support decisions.\\
\textit{\textbf{Example (Fault Diagnosis):}} Mapping abnormal \\
signal degradation patterns to specific physical \\
causes like antenna misalignment or blockages.
\end{tabular}                                  \\ 
\hline
\textbf{Confidence}      & \begin{tabular}[c]{@{}l@{}}Quantify and clarify calibrated uncertainty\\about predictions to enable \\instance-level reliability judgments.\end{tabular}              & \begin{tabular}[c]{@{}l@{}}Domain experts; \\Developers; Managers;\\Regulatory entities/agencies\end{tabular} & \begin{tabular}[c]{@{}l@{}}
Clarify calibrated uncertainty to drive deferral \\
control and priority mechanisms.\\
\textit{\textbf{Example (Deep Receiver):}} Using uncertainty scores \\
in low-SNR regimes to trigger HARQ re-transmissions \\
or fallback to pilot-based estimation methods.
\end{tabular}           \\
\hline
\end{tabular}
}
\end{table*}

The goals of XAI typically include privacy, fairness, trustworthiness, transferability, and informativeness~\cite{2019Explainable}. 
Table~\ref{xai goals} summarizes these XAI goals along with their descriptions, target parties, and roles for Wireless XAI.

\subsubsection{Trustworthiness}
Trustworthiness denotes the degree to which a model’s behavior can instill justified confidence in its users~\cite{chinnaraju2025explainable}.
It reflects the assurance that the model acts reliably and as intended, and that its decision-making process is grounded in transparency, consistency, and robustness. 
Trustworthiness is a key motivation or even a goal of explainability, but they are not interchangeable. 
Explainability emphasizes interpretability.
Trustworthiness concerns the reliability and acceptability of those outputs, even in the absence of full interpretability.  
% ~\cite{zhu2025wireless}
% ~\cite{10579546}
% ~\cite{zhu2023pushing}
In communication systems, NN models often operate under tight latency~\cite{9318524}, imperfect observations~\cite{9540365}, and adversarial or rapidly changing environments~\cite{8449065,10403935}. 
An explainable yet untrustworthy model may foster misplaced confidence, which is particularly risky in communication systems where model outputs are translated into protocol actions in real time. 
When a fragile explanation is trusted, it can silently propagate to wrong adaptation and control decisions and eventually cause performance degradation.
The authors of \cite{10.1145/3704413.3764466} have demonstrated that models can appear interpretable through time-/frequency-domain activation maps, yet exhibit critical vulnerabilities such as over-reliance on power-related cues and sharp performance drops under practical impairments, underscoring that explanations alone do not guarantee trustworthy behavior. 
% {\color{red} 
% An explainable yet untrustworthy model may foster misplaced confidence, and lead to system failures~\cite{}. 
% 'how this is related to communications ?'}
A trustworthy model, whose decisions are not only stable and robust but also interpretable, supports timely and confident operational decisions~\cite{alzubaidi2023towards}.

\subsubsection{Causality}
Causality refers to a model’s ability to assist in uncovering cause-and-effect relationships (or causality). While deep learning (DL) models are adept at detecting statistical patterns, they often lack mechanisms to differentiate true causality from spurious associations
% , limiting their ability to generalize across changing environments~
\cite{rawal2025causality}. Machine learning (ML) alone cannot establish causality without prior knowledge or interventions; explainable models can identify potential causal links, validate causal inference outputs, or guide experts in interpreting data-driven relationships.
In wireless systems, non-stationarity, mobility, and hardware diversity are common. From PHY signal adaptation to MAC scheduling and routing decisions,
models that rely solely on correlation risk making unstable or incorrect decisions under domain shift~\cite{10499970}.
Integrating causal reasoning or causal information into AI models can enhance robustness and generalization.
% , and support resilient system behavior.
 
\subsubsection{Privacy}

If people cannot understand the meaning of models' stored data and internal representation, there is a potential for privacy leakage \cite{castelvecchi2016can,XinYuanTDSC2026}. 
Based on \cite{hleg2019high}, the privacy principle comprises: 
\begin{itemize}
    \item
\textit{Privacy and data protection:}
% AI systems must guarantee privacy and data protection~\cite{morley2021initial}. 
This includes the information initially provided by users, and the information generated about the users over the course of their interaction with the system (e.g., outputs that the AI system generated for specific users~\cite{wachter2019right}). To allow individuals to trust the data-gathering process, the data collected about them must not be used to unlawfully or unfairly discriminate against them~\cite{morley2021initial}. 
    \item 
\textit{Quality and integrity of data:}
% The quality of datasets used is paramount to the performance of AI systems. 
When data is gathered, it may contain socially constructed biases, inaccuracies, errors, and mistakes~\cite{hleg2019high,WeicaiTIFS2025}. This needs to be addressed prior to training. In addition, the integrity of data must be ensured. 
% Feeding malicious data into an AI system may change its behavior. 
Processes and datasets used must be tested and documented at each step, e.g., planning, training, testing, and deployment. 
    \item 
\textit{Access to data:}
In any organization that handles individuals’ data (whether someone is a user of the system or not), protocols governing data access should be in place~\cite{MengmengACM2025}. 
Only personnel whose roles and responsibilities specifically require such access are authorized to handle individual data.
% {\color{red}Only authorized personnel with competence and a need to access an individual’s data are allowed to do so.}
    
\end{itemize}

% \subsubsection{Basic concepts of privacy in XAI}
% \begin{itemize}
%     \item ...
% \end{itemize}
% \subsubsection{Privacy issues in data fusion}
% \begin{itemize}
%     \item ...
% \end{itemize}
% \subsubsection{The future of privacy issues}
% \begin{itemize}
%     \item ...
% \end{itemize}

\subsubsection{Fairness}

Models need to be designed and evaluated so their predictions avoid unjustified disparities among individuals or groups, particularly concerning sensitive attributes~\cite{0Principled,2017Conscientious,XiyuTMC2025}. Explainability is central to fairness; 
instead of treating the model as a closed box, it provides the necessary transparency to audit whether decisions are driven by legitimate qualifications or are influenced by sensitive attributes through proxy variables, thus ensuring the logic aligns with ethical non-discrimination standards.
% {\color{red} it exposes which features, interactions, or proxy variables drive decisions and whether these pathways correlate with protected attributes. 'how these are related to fairness?'}

Fairness interventions span three complementary families: 
\begin{itemize}
    \item 
% {\color{red}Pre-processing methods that modify data or learn fair representations before modeling \cite{kamiran2012data,zemel2013learning}, 'how these are related to fairness?' }
Pre-processing methods that mitigate intrinsic data biases by decorrelating sensitive attributes from decision-relevant features, thereby preventing the model from inheriting discriminatory patterns from the outset \cite{kamiran2012data,zemel2013learning}.
    \item 
In-processing that imposes fairness via constraints or adversarial objectives during training \cite{zhang2018mitigating,celis2019classification}, and
    \item 
Post-processing adjustments that alter decision thresholds or outputs to satisfy criteria, e.g., equalized odds \cite{hardt2016equality}.  
\end{itemize}
In wireless communication systems, schedulers, admission controllers, beam selectors, and handover policies can encode bias through radio-domain proxies (e.g., path loss and mobility class), systematically disadvantaging cell-edge users. Domain-aware explanations that surface these proxy pathways enable targeted fairness checks (e.g., subgroup error/throughput parity over time and geography).
% , counterfactual tests under radio-aware causal assumptions, and integration of fairness constraints into resource allocation and federated client selection. 
 
\subsubsection{Transferability}
Transferability explore the generalization ability of a model, i.e., whether the model can be applied to other (un)related tasks after learning one task.
By increasing the explainability of the model, the boundaries of the model's applicability to users become clear. Transferability can be facilitated using techniques, e.g., transfer learning. In wireless systems, AI models trained for one frequency band, antenna configuration, or propagation environment may underperform when transferred to other settings. For example, a channel estimator trained for urban may not generalize to rural or high-speed rail scenarios~\cite{10078433}. Explainability can rationalize such failures, and identify transferable structures~\cite{8715338,11125862}. 
% {\color{red}'ref is not related to channel estimation-fixed'}

\subsubsection{Informativeness}
An explainable AI system should be able to provide clear, unambiguous, and easy-to-understand information to users, stakeholders, or regulators to help them understand how the model works and handles tasks~\cite{2019Explainable}.
Almost all rule extraction techniques, e.g., \cite{2000Extracting,2002Medical}, substantiate their approach to the search for a simpler understanding of what the model internally does, stating that the knowledge (information) can be expressed in these simpler proxies that they consider explaining the antecedent
% . This is the most used argument found among the reviewed papers to back up what they expect from reaching explainable models
\cite{2019Explainable}.
In wireless systems, informativeness helps align AI models with operational needs. 
For PHY tasks such as channel estimation, decisions are made under latency, resource, and reliability constraints.
% Explanations lacking actionable insight are of limited  value.

\subsubsection{Confidence}
In XAI, confidence refers to a model’s estimated certainty about its predictions.
% , typically represented as a probability. 
% These scores help users gauge reliability at the instance level and reflect broader aspects of model robustness and stability.
When aligned with accuracy, high-confidence outputs can enhance user trust while transparent low-confidence alerts prevent blind reliance~\cite{le2023explaining}.
Explainable models should incorporate calibrated confidence assessments to support meaningful interpretation~\cite{TJOA2023126825}. Unstable models that produce erratic confidence values cannot yield trustworthy explanations.
% Wireless 
% confidence is crucial due to real-time and reliability demands. T
Tasks, e.g., modulation adaptation, channel estimation, or interference management, must contend with uncertainty from fading, mobility, and noise. Confidence-aware models can defer decisions under uncertainty, trigger fallbacks, or help prioritize actions~\cite{10.1145/3711896.3737010}. 
% More importantly, c
Confidence-integrated explanations inform not only what the model predicts but how certain it is, to support reliable deployment.

\begin{table*}[hbt] 
\centering
    \renewcommand\arraystretch{1.2}    
    \caption{Summary and comparison of the state-of-the-art XAI techniques }
    \label{xai tech}
    \resizebox{\textwidth}{!}{
\begin{tabular}{|c|c|c|c|c|c|} 
\hline
\textbf{Classification}                         & \textbf{XAI Techniques}                                                 & \textbf{Global/Local}                                          & \textbf{Model Specific/Agnostic}                                        & \textbf{Surrogate/Visual/Textual} & \textbf{Algorithms to Explain}  \\ 
\hline
\multirow{3}{*}{White Box Models}               & Linear Models                                                           & Global                                                         & Agnostic                                                                & N.A.                              & Transparent                     \\ 
\cline{2-6}
                                                & General Addictive Models                                                & Global                                                         & Agnostic                                                                & N.A.                              & Transparent                     \\ 
\cline{2-6}
                                                & Decision Trees                                                          & Global                                                         & Agnostic                                                                & N.A.                              & Transparent                     \\ 
\hline
\multirow{5}{*}{Local Model Reduction}          & LIME~\cite{ribeiro2016should}                                           & Local                                                          & Agnostic                                                                & Surrogate                         & Supervised                      \\ 
\cline{2-6}
                                                & SHAP/DeepSHAP~\cite{NIPS2017_8a20a862}                                  & \begin{tabular}[c]{@{}c@{}}Local \\ (also global)\end{tabular} & Agnostic                                                                & Surrogate                         & Supervised                      \\ 
\cline{2-6}
                                                & TreeSHAP~\cite{lundberg2018consistent}                                  & \begin{tabular}[c]{@{}c@{}}Local \\ (also global)\end{tabular} & \begin{tabular}[c]{@{}c@{}}Specific \\ (Tree-based models)\end{tabular} & Surrogate                         & Supervised                      \\ 
\cline{2-6}
                                                & DeepRED~\cite{2016DeepRED}                                              & Local                                                          & Agnostic                                                                & Surrogate and Textual             & Supervised                      \\ 
\cline{2-6}
                                                & Counterfactual~\cite{sharma2019certifai}                                & \begin{tabular}[c]{@{}c@{}}Local\\ (also global)\end{tabular}  & Both                                                                    & Surrogate                         & Supervised                      \\ 
\hline
\multirow{2}{*}{Global Model Reduction}         & SSL~\cite{wen2016learning}                                              & Global                                                         & Agnostic                                                                & N.A.                              & Supervised                      \\ 
\cline{2-6}
                                                & Distillation~\cite{2015Distilling,Haselhoff_2021_CVPR,LI2022108345}     & Global                                                         & Agnostic                                                                & Surrogate                         & Supervised                      \\ 
\hline
\multirow{2}{*}{Automatic-Rule Extraction}      & Decompositional Approaches~\cite{1991Rule, 2000Extracting}              & Local                                                          & Agnostic                                                                & Surrogate and Textual             & Supervised                      \\ 
\cline{2-6}
                                                & Pedagogical Approaches~\cite{2002Medical, 1995Extracting}               & Global                                                         & Agnostic                                                                & Textual                           & Supervised                      \\ 
\hline
\multirow{2}{*}{Representation of Layers}       & Layer Generalization Detection~\cite{yosinski2014transferable}          & Local                                                          & \begin{tabular}[c]{@{}c@{}}Specific\\ (CNN models)\end{tabular}         & N.A.                              & Supervised                      \\ 
\cline{2-6}
                                                & CAM/GradCAM/GradCAM++~\cite{Zhou_2016_CVPR,selvaraju2017grad,8354201} & Local                                                          & \begin{tabular}[c]{@{}c@{}}Specific\\ (CNN models)\end{tabular}         & Visual                            & Supervised                      \\ 
\hline
\multirow{3}{*}{Representation of Units}        & Linear Score Model~\cite{2013Deep_inside}                               & Local                                                          & Agnostic                                                                & Surrogate and Visual              & Supervised                      \\ 
\cline{2-6}
                                                & Deep Taylor Decomposition~\cite{2016Deep}                               & Local                                                          & Agnostic                                                                & Surrogate and Visual              & Supervised                      \\ 
\cline{2-6}
                                                & Network Dissection~\cite{bau2017network}                                & Local                                                          & Agnostic                                                                & Textual                           & Supervised                      \\ 
\hline
Representation of Blocks                        & DeepLIFT~\cite{shrikumar2017learning}                                   & Local                                                          & Agnostic                                                                & Surrogate                         & Supervised                      \\ 
\hline
\multirow{2}{*}{Explanation-Producing Networks} & LRP-based Relevance \cite{Chefer_2021_CVPR}                             & Local                                                          & Agnostic                                                                & Visual                            & Supervised                      \\ 
\cline{2-6}
                                                & CREATE~\cite{NEURIPS2023_1e118ba9,yang2024scaling,yu2023white}          & Global                                                         & Agnostic                                                                & Surrogate                         & Supervised                      \\ 
\hline
\multirow{4}{*}{Reward Decomposition}           & drQ~\cite{juozapaitis2019explainable}                                   & Global                                                         & \begin{tabular}[c]{@{}c@{}}Specific\\ (RL models)\end{tabular}          & N.A.                              & RL                              \\ 
\cline{2-6}
                                                & Dot-to-Dot~\cite{8968488}                                               & Global                                                         & \begin{tabular}[c]{@{}c@{}}Specific\\ (RL models)\end{tabular}          & Visual                            & RL                              \\ 
\cline{2-6}
                                                & “What-if" Explanations~\cite{bicalearning}                              & Local                                                          & Agnostic                                                                & Textual                           & RL                              \\ 
\cline{2-6}
                                                & Reward Processing~\cite{jenner2022preprocessing}                        & Global                                                         & \begin{tabular}[c]{@{}c@{}}Specific\\ (RL models)\end{tabular}          & Visual                            & RL                              \\ 
\hline
\multirow{4}{*}{Learning Explainable Policy}    & CUSTARD~\cite{topin2021iterative}                                       & Global                                                         & \begin{tabular}[c]{@{}c@{}}Specific\\ (RL models)\end{tabular}          & Surrogate                         & RL                              \\ 
\cline{2-6}
                                                & Deep Symbolic Policy~\cite{landajuela2021discovering}                   & Global                                                         & \begin{tabular}[c]{@{}c@{}}Specific\\ (RL models)\end{tabular}          & Surrogate and Textual             & RL                              \\ 
\cline{2-6}
                                                & Nonlinear Decision Tree~\cite{9805655}                                  & Global                                                         & \begin{tabular}[c]{@{}c@{}}Specific\\ (RL models)\end{tabular}          & Surrogate and Visual              & RL                              \\ 
\cline{2-6}
                                                & Differentiable Decision Tree~\cite{silva2019optimization}               & Global                                                         & \begin{tabular}[c]{@{}c@{}}Specific\\ (RL models)\end{tabular}          & Surrogate                         & RL                              \\ 
\hline
\multirow{2}{*}{Post-hoc Enhancement}           & PIRL~\cite{verma2018programmatically}                                   & Global                                                         & \begin{tabular}[c]{@{}c@{}}Specific\\ (RL models)\end{tabular}          & Surrogate                         & RL                              \\ 
\cline{2-6}
                                                & SDT~\cite{coppens2019distilling}                                        & Global                                                         & \begin{tabular}[c]{@{}c@{}}Specific\\ (RL models)\end{tabular}          & Surrogate                         & RL                              \\ 
\hline
Complementary RL                                & Complementary QS agent~\cite{junghoonlee}                               & Global                                                         & Agnostic                                                                & Surrogate                         & RL                              \\
\hline
\end{tabular}
}
\end{table*}

\section{Explainability for Wireless PHY AI}
In line with \cite{gilpin2018explaining}, we categorize XAI approaches into explanations of NN actions, explanations of NN representations, and explanation-producing networks, as collated in
Table~\ref{xai tech}.
% summarizes and compares the state-of-the-art XAI techniques.

% {\color{red}
% 'I feel that this Section III is not much related to communications. The contents are rather generic to XAI, which can be already found in other surveys easily. Some reviewer may comment on this
% Thus, it is good to make Section III more communication, we can add context and/or example related to communications.'
% }

% , which facilitates understanding of the networks for human users.
% % \subsection{Some Concepts w.r.t XAI}
% % Model-agnostic vs. model-specific: 
% Model-agnostic methods do not consider the internal components of the model (i.e., model weights and structure parameters)~\cite{10460345}. They can be applied to any closed-box method. In contrast, model-specific methods are defined by the parameters of a single model, e.g., explaining the weights of a linear regression or using the inference rules of a decision tree specifically trained~\cite{speith2022review}. Advantages of model-agnostic methods are prominent, as developers can flexibly choose any ML model to generate explanations, which differs from generating decisions for the actual closed-box model~\cite{hassija2024interpreting}.

\subsection{Transparent and White Box Models}
Some AI models are simple and self-explanatory. 
They provide direct access to their internal reasoning, enabling users to trace how input features contribute to predictions.
\subsubsection{Linear Models}
Explainability of linear models involves a linear combination of feature values, adjusted by the coefficients of the models~\cite{10.1145/3561048}. Logistic regression is one of the most interpretable linear ML models for a certain class of events. The seaborn~\cite{Waskom2021}, matplotlib~\cite{6997585}, sklearn~\cite{pedregosa2011scikit}, and accumulated local effects (ALE)~\cite{apley2020visualizing} libraries can unfold and visualize a logistic regression model. 
% However, its performance in characterizing complex models is often unsatisfactory.

\subsubsection{Generalized Additive Model (GAM)}
GAMs extend generalized linear models, with smooth (often spline-based) functions applied to individual predictors; the model outputs the sum of these component functions. GAMs offer a middle ground between transparent linear models (e.g., logistic regression) and closed-box models (e.g., DNNs): They capture nonlinearities better than linear models and are more interpretable than closed-box models. Each univariate smooth term of GAMs reflects the marginal contribution of its features to the prediction, facilitating visualization and audit.

\subsubsection{Decision Trees}
Decision trees are another example of a model that can easily fulfill every constraint for \textit{transparency}. In a simple decision tree, nodes represent values of specific attributes, edges are rules, and leaf nodes represent classes. For simple decision trees, it is easier to understand the  decision by following the rules along the edges and nodes, and eventually the final decision of the leaf nodes. It provides an if-else rule summarizing a decision, making it interpretable.

\subsection{Explanations of NN Behaviors}
It is difficult to explain the decisions generated by DNNs, due to their sheer sizes. One way to tackle this is to find simpler functions to approximate the behaviors of DNNs to provide understandable explanations.

\subsubsection{Local Model Reduction}
\begin{figure}[t] 
\centering
\includegraphics[width=3.3in]{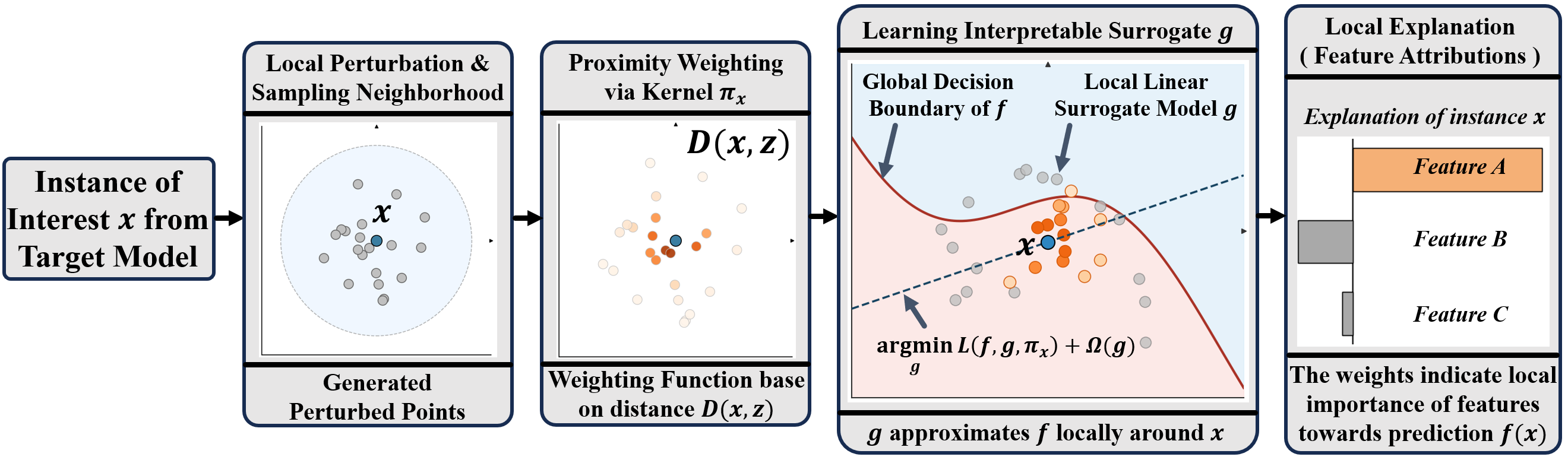}
\caption{\small Illustration of LIME~\cite{ribeiro2016should}, which constructs interpretable local surrogate models to explain the predictions of DNNs. By employing a perturbation-based sampling strategy and kernel-weighted regression, it approximates the model's global non-linear decision boundary with a linear function specifically within the vicinity of the instance of interest. LIME provides intuitive local interpretation, visualized as a simplified linear separator (dashed blue line) that is locally faithful to the complex global boundary (solid red curve), and generates explicit feature attributions, summarized as a bar chart where the magnitude of weights reveals the contribution of individual input features to the specific prediction. }
\label{lime illustation}
\end{figure}

Sparse linear models are inherently more interpretable than models with dense non-linear layers \cite{1996Regression}. One can create a proxy model that provides a simpler explanation of the global model. 
Local Interpretable Model-Agnostic Explanations (LIME)~\cite{ribeiro2016should} 
% implements this idea by explaining 
explains the predictions of closed-box classifiers and assessing their trustworthiness; see Fig.~\ref{lime illustation}. The process collects an instance sample for the closed-box model and generates new sample points by perturbing samples. Anticipated values from the closed-box model are obtained for these new data points to train an interpretable model, e.g., linear regression or decision trees, providing a local approximation of the closed-box model. 
% With such local approximations, LIME can enhance explainability for different supervised models and is compatible with different data types (i.e., text, tabular data, images, graphs).

\begin{figure}[b] 
\centering 
\includegraphics[width=3.3in]{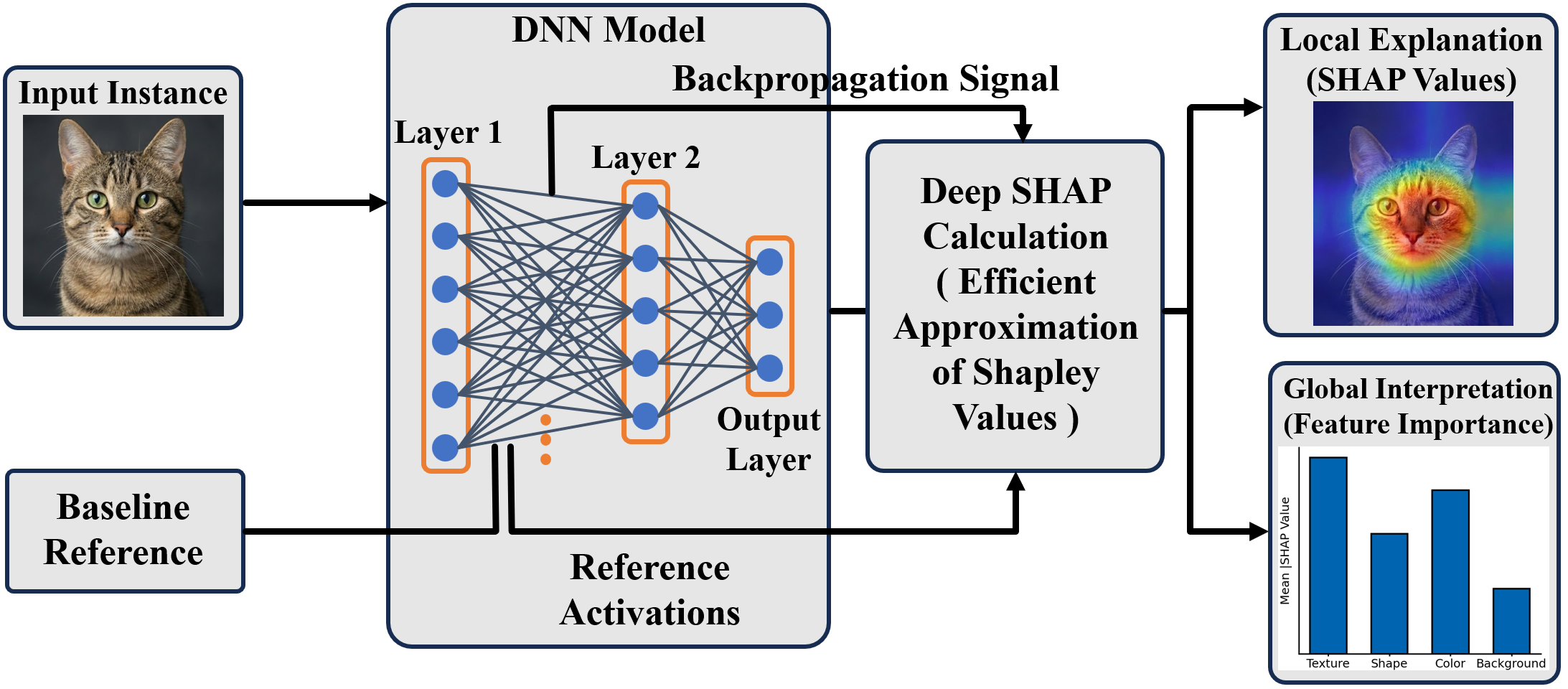}
\caption{ \small
Illustration of Deep SHAP \cite{NIPS2017_8a20a862}, which approximates Shapley values to interpret the predictions of DNNs. 
By employing a backpropagation-based technique, it attributes the model's output to individual input features, such as pixels in an image. Deep SHAP enables both local interpretation, visualized as a heatmap where red areas contribute positively and blue areas negatively to the prediction for a specific instance, and global interpretation to some extent, summarized as a bar chart showing the overall importance of different features across the dataset.
}
\label{deepshap illustration}
\end{figure}

As illustrated in Fig.~\ref{deepshap illustration}, SHapley Additive exPlanations (SHAP) \cite{NIPS2017_8a20a862} generate feature importance using Shapley values derived from cooperative game theory \cite{shapley1953value}. Unlike LIME, SHAP provides explanations at both instance and global levels, making it model-agnostic and outperforming LIME. Specialized versions of SHAP, e.g., TreeSHAP \cite{lundberg2018consistent} and Deep SHAP~\cite{NIPS2017_8a20a862}, are tailored for decision trees and DNNs, respectively. Another approach, DeepRED~\cite{2016DeepRED}, employs a decompositional technique to extract rules from DNNs, improving their interpretability.
% of their decision processes.

The Counterfactual algorithm \cite{sharma2019certifai} enhances the interpretability of predictor algorithms by identifying minimal adjustments in input feature values responsible for alterations in the initial prediction. Operating in both model-agnostic and model-specific variants, this algorithm elucidates predictions and pinpoints subtle modifications in input feature values that induce shifts in the original prediction.

\subsubsection{Global Model Reduction}
Previous research indicates that NNs often contain redundant parameters. Achieving a similar function approximation is feasible by removing some parameters \cite{wen2016learning}. The Structured Sparsity Learning (SSL) method~\cite{wen2016learning} learns a compact structure from a large DNN, reducing computational costs, and promoting hardware-friendly structural sparsity to efficiently accelerate DNN evaluations. 

Distillation \cite{2015Distilling} is another technique that effectively transfers knowledge from an ensemble or large, highly regularized model into a smaller, distilled model. This simplifies the network and enhances its interpretability. For example, the authors of \cite{Haselhoff_2021_CVPR} propose a post-hoc explainability technique based on a knowledge distillation process to generate an explainable surrogate model. In \cite{LI2022108345}, a shallow graph neural network is trained with knowledge distillation to display explicit ``contribution" between nodes. 

\subsubsection{Automatic Rule Extraction}
This technique
% , known as rule extraction, 
distills the decision-making process of an NN into interpretable rules represented by Boolean functions \cite{ANDREWS1995373}. Rule extraction methods are typically classified into two categories.

The first category, decompositional approaches, approximates each neuron's behavior using Boolean functions of the inputs. For instance, in \cite{1991Rule}, a threshold-based if-then rule is applied to approximate each neuron's behavior. However, this approach suffers from exponential time complexity, and is prohibitive for DNNs. The authors of \cite{2000Extracting} approximate Boolean functions with reduced complexity, achieving polynomial-time complexity while maintaining accuracy.

The second category, pedagogical approaches, treats the NN as a closed box and directly maps inputs to outputs. For example, in \cite{2002Medical}, the authors utilize changes in input and output unit levels to extract rules in medical diagnostics. Thrun et al. \cite{1995Extracting} introduce validity interval analysis (VIA) to extract symbolic knowledge from NNs trained with backpropagation. VIA partitions the activation ranges of units into intervals and iteratively refines the intervals to determine if network activations fall within specified intervals. Unlike decompositional approaches, pedagogical approaches disregard the network's internal structure and focus on the input-output relationship.

\subsection{Explanations of NN Representations}
\begin{figure}[b]
\centering
\includegraphics[width=2.8in]{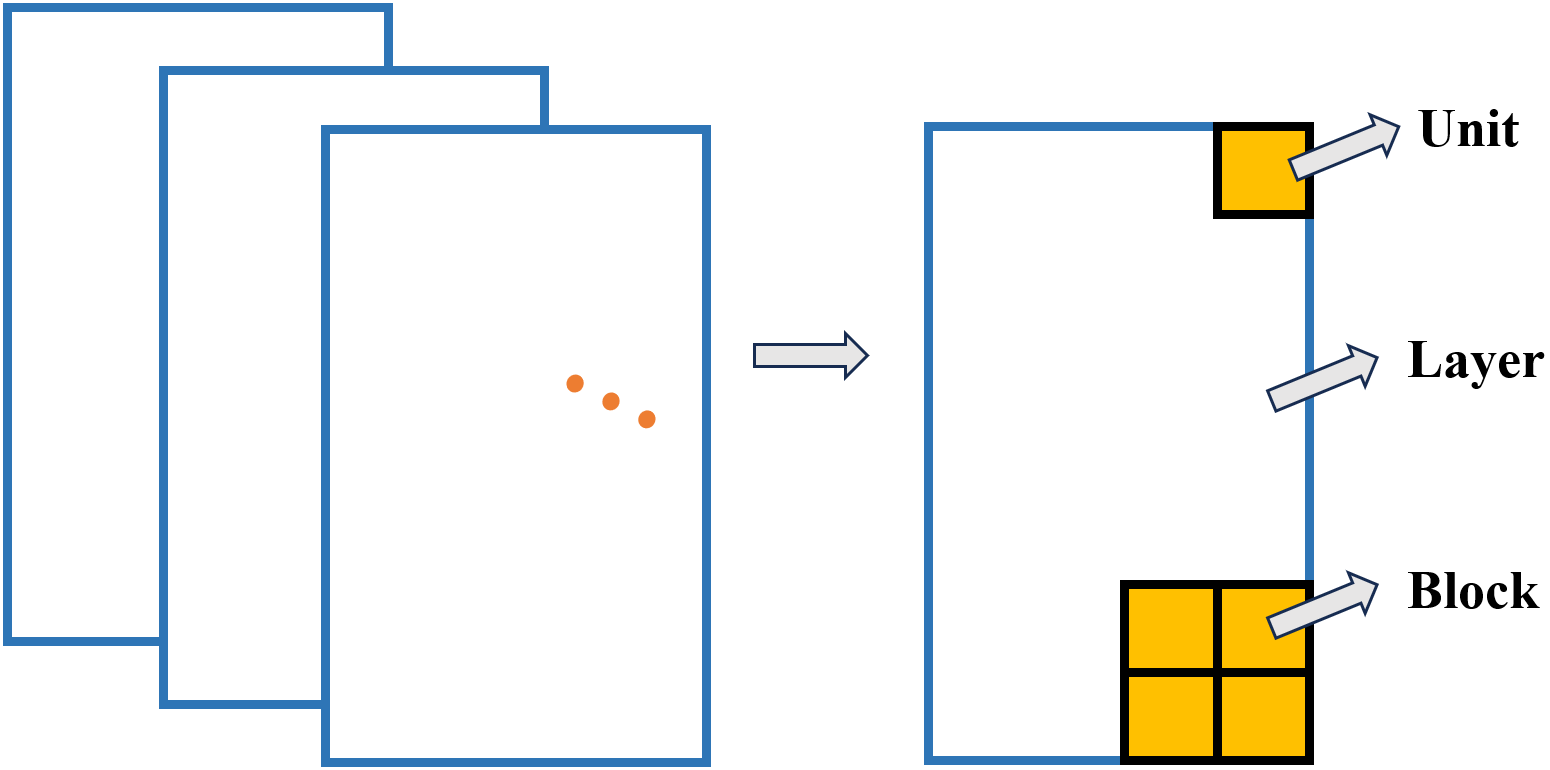}
\caption{\small The concepts of layers, units, and blocks. The concept of layers is defined as the processing modules of a network; the concept of units is defined as the neurons of a network; blocks are an aggregation of a bunch of units.}
\label{layer unit block}
\end{figure}

Unlike explanations of NN behaviors, which approximate the NN's functions~\cite{gilpin2018explaining}, explanations of representations enhance understanding of the NN's architecture and components. It delves into the meaning and significance of layers, units, and blocks within the NN, as depicted in Fig.~\ref{layer unit block}.

\subsubsection{Representation of Layers}
\begin{figure}[t] 
\centering
\includegraphics[width=3.3in]{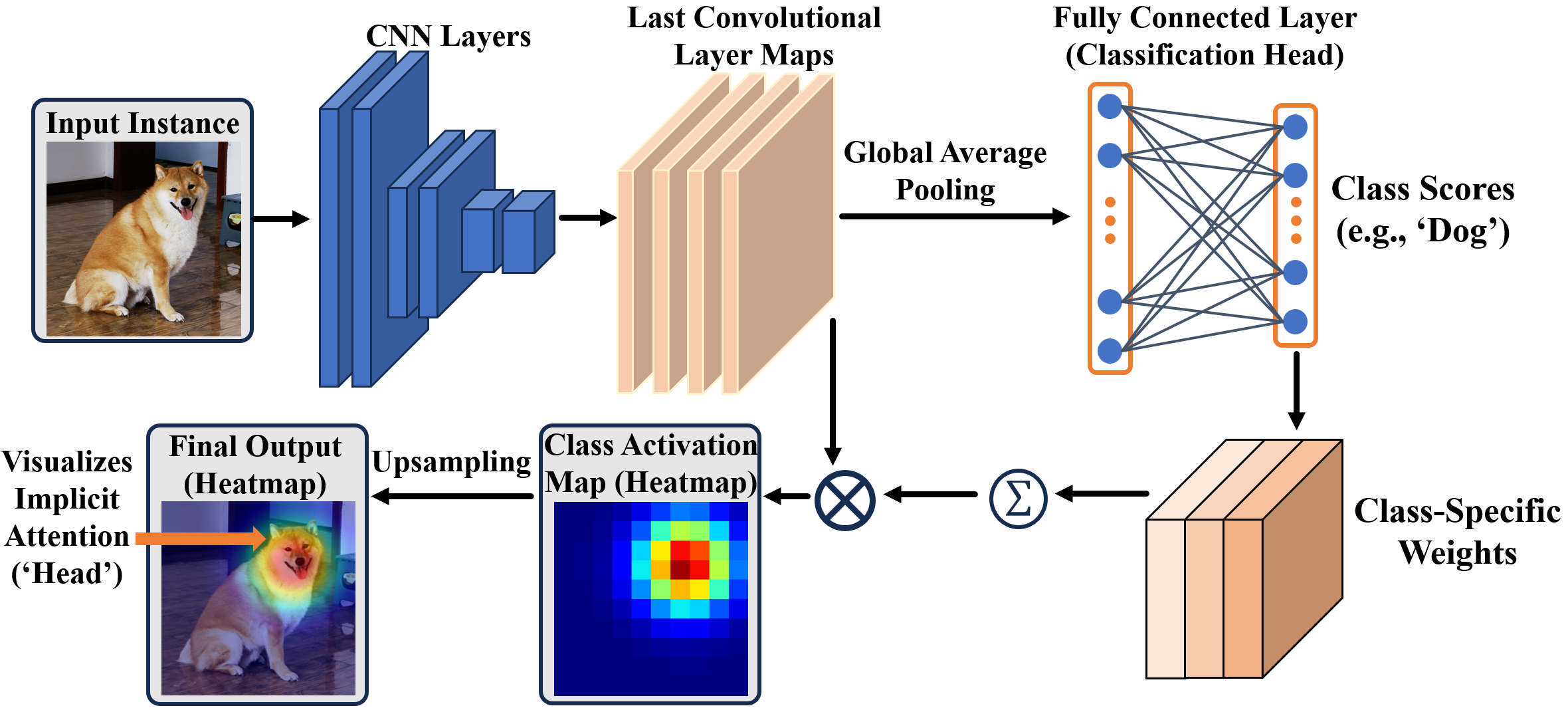}
\caption{\small Illustration of CAM \cite{Zhou_2016_CVPR}, which enhances explainability by visualizing a CNN's implicit attention for specific predictions. By computing a weighted sum of the last convolutional layer's feature maps using class-specific weights from the final fully connected layer, it generates a localization heatmap. CAM enables local interpretation for a specific instance, visualized in the final output where red, high-activation areas are superimposed on the image, indicating the regions (e.g., the dog's head) that contributed most positively to the classification decision. }
\label{cam illustration}
\end{figure}

Understanding the decisions made by an NN can be facilitated by examining its individual layers. This can be achieved through transfer learning, wherein the functionality of each layer is assessed by applying it to a different task.
For example, the authors of \cite{sharif2014cnn} conduct experiments using the network OverFeat, originally trained for object classification with the ILSVRC13 dataset, on various recognition tasks. 
% It is demonstrated that utilizing the extracted features from different layers of OverFeat leads to significant improvements in recognition tasks, without the need to train a new network. 
In \cite{yosinski2014transferable}, it is observed that many NNs trained on natural images exhibit similar features in their first layers, termed ``general" features, while their last layers possess distinct features specific to particular task classes, referred to as ``specific" features. 
% A method is developed to quantify the generality versus specificity, or transferability, of a given layer within an NN. 
CAM \cite{Zhou_2016_CVPR} and its variants \cite{selvaraju2017grad,8354201} enhance the interpretability of CNNs by visualizing the response of a particular layer to input data; see Fig.~\ref{cam illustration}.

\subsubsection{Representation of Units}
Each layer of an NN can be dissected into units or neurons.
% , allowing for a more granular understanding of network behavior. 
The terms ``neuron" and ``unit" are used interchangeably in this context.
The role of individual units can be elucidated by visualizing input patterns that maximize the response of a single unit. 
The authors of \cite{2013Deep_inside} propose a linear score model that computes the importance of image pixels by taking derivatives of the score function with respect to each pixel in the image. The authors of \cite{2016Deep} introduce a novel approach, called deep Taylor decomposition, which assesses the importance of pixels in image classification. 
% By generating heatmaps, one can gain insights into the significance of input pixels in classifying unseen data points. 
% Theoretical connections between Taylor decomposition and rule-based relevance propagation techniques are established to capture the relationship between the two approaches for certain types of NNs. 
On the other hand, the quantitative role of units can be evaluated by assessing their performance on alternative tasks. In~\cite{bau2017network}, the authors quantify the ability of individual hidden units within intermediate convolutional layers to discern implicit concepts present in the original training dataset. 
% This method enables the characterization of the information encoded by each unit in the network.

\subsubsection{Representation of Blocks}

Apart from the role of individual units, we can comprehend the network on the block level.  One technique to explore the representation of blocks is salience mapping, which, first introduced in \cite{MD2014Visualizing}, maps the feature activities back to the input pixel space to figure out what input pattern causes a given activation in the feature maps. The authors of  \cite{MD2014Visualizing} occlude different portions of an input image and monitor the changes in the output to explore the relationship between blocks of input and feature maps.
Important features can be visualized after training, according to the weights or gradients of local nodes in the NNs.  
Deep Learning Important FeaTures (DeepLIFT) \cite{shrikumar2017learning} is a method for decomposing an NN's output prediction on a specific input by backpropagating the contributions of all neurons in the network to each feature of the input to evaluate the contribution of different neurons.

\subsubsection{Physical Representations}
The premise of this technique lies in the notion that physics-based/informed approaches offer realistic explanations~\cite{2019Explainable}. The split-step Fourier method (SSFM) serves as a traditional technique for mitigating nonlinear channel loss~\cite{Ezra2008Compensation}. The study in \cite{hager2018nonlinear} proposes a DNN by unrolling the iterations of SSFM and approximating each span inversion in the form of digital backpropagation.
In \cite{9521808}, a physics-informed Gaussian process regression approach integrates a physical model into ML. This method leverages the physical model to establish targets and optimize hyperparameters in conjunction with experimental data through multi-task learning.
% The explainability of this approach is substantiated by its reliance on prior knowledge, as the kernel function employed is transparently understood.
Notably, the physical representation technique may not be universally applicable, as many real-world problems across various domains lack precise mathematical definitions and cannot be accurately modeled using DL  \cite{d2022underspecification}. 

\subsection{Explanation-Producing Networks}
Explanation-producing networks can offer ease of human understanding using attention and disentangled networks. 

\subsubsection{Attention Networks}
Attention networks, named for their focus on capturing human attention, elucidate the decision-making process by highlighting specific parts of input data~\cite{de2022attention}. They assign weights to input pixels or patterns, spotlighting information crucial for other parts of the network. 
% In fine-grained image classification, attention networks discern both the category and subordinate-level category of an image by leveraging distinctive features. 
In \cite{2014The}, object-level attention models use selective search to generate candidate patches. The selected patches are fed into a convolutional neural network (CNN) trained to output the subordinate category. Part-level attention models employ spectral clustering; each partitioned cluster acts as a part detector. 
% These detectors generate detection scores for raw patches, enabling the identification of image parts critical for classification. 
In \cite{2017Attention}, the authors propose the Transformer architecture, by replacing recurrence and convolution with stacked self-attention layers to capture sequence dependencies. 
% To enhance Transformer explainability, 
The authors of \cite{Chefer_2021_CVPR} introduce a Deep Taylor Decomposition-based method that propagates local relevance scores through attention layers and skip connections, enabling class-specific visualizations of input influence.
Recent works \cite{NEURIPS2023_1e118ba9,yang2024scaling,yu2023white} propose CREATE, a glass-box Transformer model optimized via sparse rate reduction, revealing interpretable internal structures.
% by alternating optimization.

Beyond attention-based discriminative models, interpretability in generative settings shifts from feature attribution to \emph{trajectory} attribution.
Denoising Diffusion Probabilistic Models (DDPMs)~\cite{ho2020denoising} are a canonical case, as the model explains samples via a stepwise denoising path. They learn to reverse a gradual noising procedure, so explanations should track how evidence accumulates along the denoising trajectory.
% The forward process is defined as
% \begin{align}
%     q(x_t \mid x_{t-1}) = \mathcal{N}(x_t; \sqrt{\alpha_t} x_{t-1}, (1-\alpha_t)\mathbf{I}),
% \end{align}
% and the reverse denoising is learned as
% \begin{align}
%     p_\theta(x_{t-1} \mid x_t) = \mathcal{N}(x_{t-1}; \mu_\theta(x_t, t), \sigma_\theta(x_t, t)\mathbf{I}).
% \end{align}
Combining DDPMs with Transformers, as in DALL$\cdot$E~\cite{ramesh2021zero}, has enabled powerful text-to-image generation.
% While inherently opaque, emerging research aims to improve interpretability in such diffusion-based models.

\subsubsection{Disentangled Networks}

Traditional NNs represent high-layer filters with a mixture of object patterns. Disentangled networks decouple these patterns, describing filters with independent features. In \cite{2017Interpretable}, a loss is applied to the feature map of each filter after ReLU. During forward propagation, the CNN selects a template as a mask to filter out noisy activations, pushing the filter to represent a specific object part. During backward propagation, the filter is off for images of other categories by minimizing the mutual information between the feature maps and template masks. 
% The resulting model yields single-part filters with high semantic purity and stable localization across images.
Similarly, the authors of \cite{2014Auto} propose a stochastic variational inference and learning algorithm by fitting an approximate inference model to the intractable posterior using information-theoretic measures. 

\subsection{XRL Methodologies}

RL enables systems to autonomously interact with their environment and learn. Deep RL (DRL) integrates RL with DNNs. Akin to other ML algorithms, RL often lacks explainability. The domain of RL poses intricacies that require a deeper understanding \cite{exDRL}. In what follows, we delineate explainable RL (XRL) methods employed in RL/DRL.

\subsubsection{Reward Decomposition}
The reward function often comprises naturally decomposable components. Traditional RL overlooks this and aggregates individual rewards into a single scalar value. To glean deeper insights, reward decomposition techniques offer a promising avenue for agents. This method, pioneered by \cite{van2017hybrid}, expedites training or adapts to multi-agent scenarios. Recent advancements, e.g.,~\cite{juozapaitis2019explainable}, have delved into decomposed reward Q-learning (drQ) to provide explanations. 
By delineating a set of reward components $C$, a Markov Decision Process (MDP) integrates reward decomposition, defining a vector-valued reward function $\overrightarrow{R}: S\times A\rightarrow \mathbb{R}^{|C|}$, alongside a vector-valued Q-function $\overrightarrow{Q^{\pi}}$. The goal is maximizing the overall reward function $R(s,a)=\sum_{c\in C}^{}R_{c}(s,a)$. 

Notably, the overall Q-function can be decomposed as $Q^{\pi}(s,a)=\sum_{c\in C}^{}Q_{c}^{\pi}(s,a)$. The convergence of each $Q_{c}^{t}(s,a)$ to $Q_{c}^{\pi^*}(s,a)$ is established, ensuring accurate explanations of $\pi^*$. Two metrics for pairwise action explanations are introduced: Reward Difference Explanations and Minimal Sufficient Explanations, leveraging the learned decomposed Q-function components. This approach can be extended to the Deep Q-Network (DQN) setting, where function approximators represent the decomposed Q-functions.

The work \cite{8968488} employs Q-value decomposition to visualize per-subtask Q-values as heatmaps over the environment. While this technique provides explanations, it is challenging to interpret and lacks evaluation with end users. Bica et al. \cite{bicalearning} address ``what-if" questions by modeling an expert's reward function based on preferences over alternative outcomes. 
% They aim to learn the expert's unknown reward function. 
The feature map uses counterfactual reasoning to explain the trade-offs associated with different actions.
% $R^*(h_t, a_t) = w^* \cdot \phi(h_t, a_t)$, where $h_t$ represents the history of observations and actions, and $\phi$ is the feature map. The feature map uses counterfactual reasoning to explain the trade-offs associated with different actions: $\phi(h_t, a_t) = \mathbb{E}[Y_{t+1}[a_t] \mid h_t]$, where $\mathbb{E}[Y_{t+1}[a_t] \mid h_t]$ is the expected outcome of action $a_t$.
Another work \cite{jenner2022preprocessing} focuses on learning reward functions from human feedback by simplifying them into equivalent but more interpretable forms.

% \begin{figure}[b]
% \centering
% \includegraphics[width=3.2in]{Illustration/illustration_whatif.png}
% \caption{ What-if }
% \end{figure}

\subsubsection{Learning Explainable Policy}
Some techniques derive policies (models) that are inherently interpretable, with no need for additional transformation steps.

CUSTARD \cite{topin2021iterative} directly generates an interpretable policy in the form of a decision tree. It does this by augmenting the original MDP into an IBMDP, which includes both the actions from the original MDP ($\mathcal{A}_M$) and additional actions ($\mathcal{A}_I$) for constructing the decision tree. This makes the resulting policy adopt a decision tree structure while supporting NN training.

\begin{figure}[t] 
\centering
\includegraphics[width=3.2in]{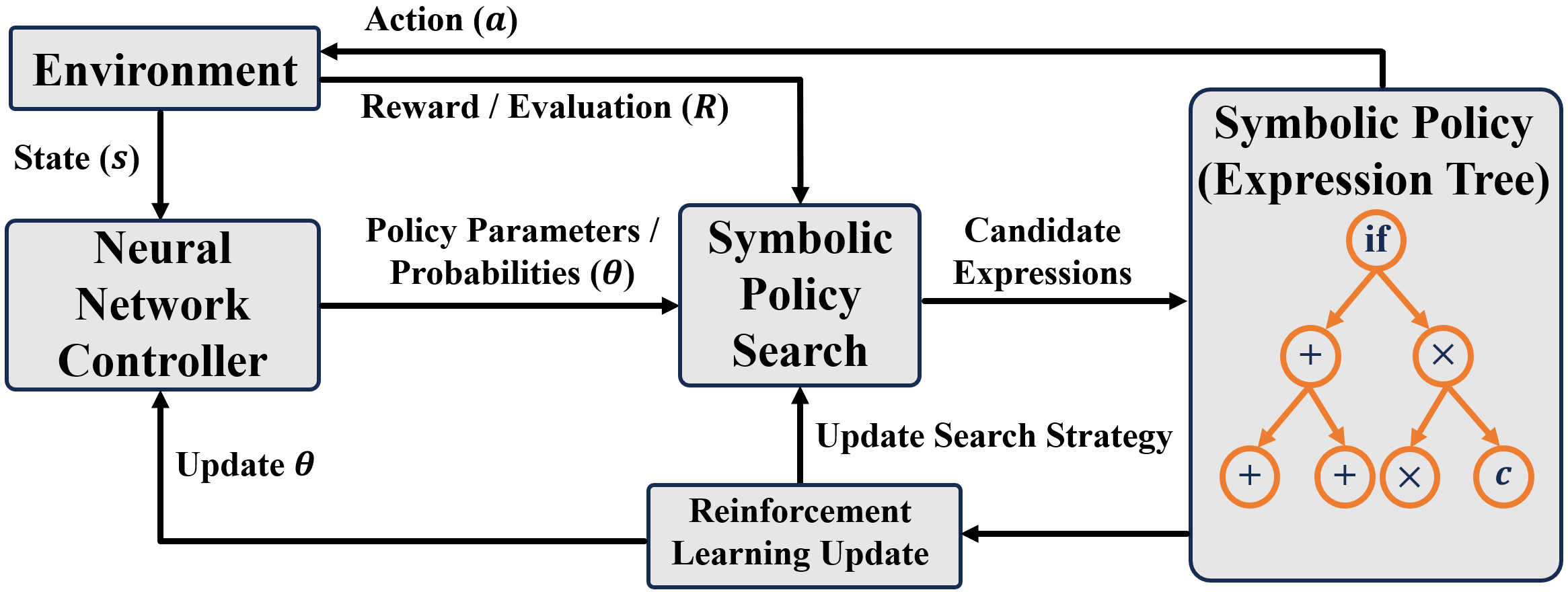}
\caption{\small Illustration of DSP~\cite{landajuela2021discovering}.
The ``Neural Network Controller" handles the complex mapping from ``State $S$" to search probabilities $\theta$. 
% However, this closed box is used solely to guide a "Symbolic Policy Search." 
The crucial mechanism for explainability is that the final resulting agent is the ``Symbolic Policy (Expression Tree)," instead of the neural network. Since ``Action ($a$)" is derived from a structured, human-readable tree composed of explicit mathematical and logical operators (e.g., if, +, x), experts can directly inspect and validate the precise rules governing the agent's behavior. }
\label{dsp illustration}
\end{figure}

Several structures based on decision trees have been explored to improve applicability, while maintaining adequate interpretability of the resulting strategies. For instance, DSP~\cite{landajuela2021discovering} directly searches for symbolic control policies using an autoregressive RNN and a risk-seeking policy gradient, while an ``anchoring" algorithm scales it to multi-dimensional action spaces by distilling pre-trained NN policies into symbolic forms, as illustrated in Fig.~\ref{dsp illustration}.
In \cite{9805655}, a nonlinear decision-tree (NLDT) is combined with evolutionary computation and bilevel optimization to derive interpretable hierarchical control rules from a pre-trained closed-box DRL agent. 
% Techniques can characterize more complex functions since they are more expressive than the standard Boolean comparisons.
To enable gradient-based training in decision trees, a differentiable decision tree \cite{silva2019optimization} is proposed, where sigmoid functions replace the Boolean splits. 
% This differentiable tree can later be discretized to approximate a standard decision tree, though with some performance trade-off.  

\subsubsection{Post-hoc Explainability Enhancement}

To make uninterpretable RL policies explainable, one approach is imitation learning \cite{abbeel2004apprenticeship}. By constructing a surrogate model $\hat{w}$ as the learner, the task is to correctly classify which action (label) to take from a state with the given examples from the RL policy $w^*$ as the expert.
Inspired by the classic DAGGER algorithm, PIRL~\cite{verma2018programmatically} conducts a local search for interpretable programmatic policies by minimizing the discrepancy between the outputs of the expert DRL policy $w^*$ and the programmatic policy $\hat{w}$ on a set of heuristically selected states. PIRL identifies a program $\hat{w}$ that replicates the expert policy.

Soft Decision Tree (SDT) is a hybrid model that combines a predetermined NN with binary decision trees \cite{frosst2017distilling}. This amalgamation allows for the creation of a surrogate model capable of mimicking the output of DNNs, unveiling hierarchical information and facilitating interpretability for decision-making tasks. 
Coppens et al.~\cite{coppens2019distilling} distill policies derived from DRL algorithms into SDTs, to gain deeper insights into the decision-making processes of complex policies. Initially, a DRL agent is trained using the Proximal Policy Optimization (PPO), followed by the distillation of state-policy samples from the deep policy network into an SDT.

One distinguishing characteristic of SDTs lies in their approach to data classification, which relies on hierarchical decisions rather than hierarchical features. 
Each node within the decision tree processes the entire input sample, enabling decisions to be made at a level of abstraction that is immediately comprehensible. 
% % Throughout the tree structure, b
% branching nodes apply different weights to filter each feature, hence altering the focus on input samples at each level.
By examining the learned filters from root to leaf, one can discern the features of SDTs when assigning an action distribution to a specific state. However, SDTs lack human-readable abstractions, and they fall short of providing explanations in a symbolic form~\cite{blockeel2023decision}.

\subsubsection{Complementary Reinforcement Learning}
% To better understand the mechanisms of RL, the
The authors of \cite{junghoonlee} propose a complementary quasi-symbolic (QS) agent that makes decisions based on simple rules derived from the RL agent. The QS agent comprises a matching network that records state transitions $\Delta S$ and a value network that stores the rewards obtained from these transitions. During training, the QS agent identifies the most valuable transitions leading to ``hub states" and searches for a sequence of actions to reach them. The RL and QS agents complement, as the matching and value networks are updated based on the RL agent's behavior. 
% The QS agent can evaluate the actions proposed by the RL agent and plan future actions accordingly. 
% This approach allows for decision-making based on simple rules, facilitating an easier understanding of the underlying mechanisms. 
Notably, the method aims for generality, and QS agents may not perform optimally in all RL problems~\cite{10238788}. 

\subsection{Transfer and Meta Learning }
\subsubsection{Transfer Learning}
Transfer learning utilizes a feature extractor \(g\) trained on a source domain \(D_S\), together with a classifier \(h\), to enhance performance on a target domain \(D_T\), particularly when labeled target data is limited \cite{WuTrustworthyTL}. To improve interpretability in transfer pipelines, one can consider a three‑stage framework:
\begin{itemize}
    \item \textit{Interpretable modules embedded before transfer:} During source-domain training, integrate attention-guided prototype layers (e.g.\ ProtoAttend) or sparse decomposition mechanisms. So, prototypical features are discoverable and attention-weighted, exposing decision paths~\cite{ArikProtoAttend}.
    \item \textit{Post-hoc interpretability after fine-tuning:} Apply attribution methods, e.g., Grad‑CAM, Saliency Maps, or SHAP/LIME, to the adapted model to analyze how feature relevance shifts after transfer, exposing domain-specific bias or spurious correlations.
    \item \textit{Trust‑aware evaluation using quantitative metrics:} Adopt the trustworthiness framework from the trustworthy transfer learning survey, i.e., measuring transferability via distribution discrepancy or task-level estimators, and trustworthiness via fairness, privacy, and adversarial robustness, to audit a transfer process  \cite{WuTrustworthyTL}.
\end{itemize}
Combining transparent architectural design with post-hoc analysis and trust metrics yields transfer learning systems that are not only effective but also auditable, robust, and aligned with responsible AI principles.

\subsubsection{Meta Learning}

Understanding how a meta-learning model acquires and applies knowledge across tasks can provide insight into its generalization behavior. This is particularly important in few-shot learning scenarios, where decisions must be made with limited data. A common approach to meta-learning is to train a model across a range of tasks so that it can adapt quickly to new ones. For example, the authors of \cite{Finn2017MAML} propose Model-Agnostic Meta-Learning (MAML), which aims to find an initialization that enables fast adaptation with only a few gradient steps. While effective, the model’s decision-making process remains opaque.
% , and it is often unclear how adaptation differs from task to task.
In~\cite{Shao2022FIND},
% To address the lack of transparency, 
a framework named FIND is introduced, which enhances interpretability by linking dataset meta-features to algorithm recommendations. 
% Their results show that such correlations allow practitioners to better understand why specific models are chosen under certain data conditions. 
In a similar direction, the authors of~\cite{Woznica2020Towards} develop a post-hoc analysis method that applies LIME and SHAP to meta-learned surrogate models. 

% Instead of analyzing models after training, s
Some researchers have focused on building interpretability directly into the meta-learning process. The authors of~\cite{Spinelli2021MATE} propose MATE, a meta-training framework designed to learn internal representations that are compatible with downstream explanation tools. In~\cite{Zhang2022MetaDT}, the authors approach the problem from a symbolic learning perspective. They propose MetaDT, a meta-learning model that generates interpretable decision trees for few-shot classification. The structure of the tree is adapted per task and incorporates visual attention to highlight which features contribute most to each decision path. 

% In general, interpretability in meta-learning can be achieved through post-hoc analysis of learned models or by designing learning processes and structures that favor transparency. Both approaches contribute to meta-learning models more accountable and trustworthy, for wireless systems where rapid adaptation must be accompanied by clear justifications.

\begin{table*}[t]
\caption{Summary of XAI applications in the PHY Layer of wireless systems.}
\centering
\label{xai phy layer}
\setlength{\tabcolsep}{3pt}
\renewcommand{\arraystretch}{1.05}
\scriptsize
\begin{tabularx}{\textwidth}{|p{3.6cm}|Y|p{6.75cm}|}
\hline
\textbf{Use Case} & \textbf{AI Algorithm and Ref.} & \textbf{Recommended XAI Techniques} \\ \hline

Channel Estimation (XAI-native; perturbation mask)
& STA-FNN / TRFI-FNN post-processing with learned noise-mask probing~\cite{10368353}
& Perturbation-based relevance map via noise-mask learning: per-subcarrier sensitivity visualization~\cite{10368353} \\ \hline

Channel Estimation (SBL unfolding; angular-delay sparsity)
& SBL-unfolded DNN for hybrid mmWave massive MIMO~\cite{10075639}
& Unfolded-step interpretability; exemplar-based inspection via DkNN-style support comparison~\cite{10075639} \\ \hline

Channel Estimation (Radar-assisted unfolding; meta-learning DOA)
& Deep unfolding-based radar-assisted CE with MAML + CNN for DOA~\cite{11165341}
& DkNN exemplar-based latent similarity explanations across hidden layers~\cite{11165341} \\ \hline

Block-structured PHY receiver / CE pipeline
& MBDL RSMA receiver~\cite{10091798}; CS frontend + DnNet/DnLSTM denoisers~\cite{9779569}
& SHAP~\cite{NIPS2017_8a20a862} with pilot/subcarrier occlusion~\cite{MD2014Visualizing} to separate linear CS and nonlinear denoising contributions \\ \hline

Channel Estimation (LDAMP; SE analysis)
& LDAMP network with state-evolution analysis for MIMO CE~\cite{he2018deep}
& DeepLIFT~\cite{shrikumar2017learning} for iteration-wise relevance decomposition aligned with SE analysis~\cite{he2018deep} \\ \hline

Channel Estimation (Attention / Beamspace learning)
& Attention-aided massive MIMO CE~\cite{gao2021attention}; beamspace amplitude estimation and selection~\cite{2019CS}
& SHAP~\cite{NIPS2017_8a20a862} on pilots/measurements to attribute beamspace coefficients and dominant-entry selection \\ \hline

Channel Estimation (Switch-based beamspace selection)
& Deep-learning beamspace CE with switch-based selection network~\cite{10570856}
& Support-consistency checks and saliency over candidate switches or beamspace entries; Grad-CAM~\cite{selvaraju2017grad} only when convolutional feature maps are available \\ \hline

Image-view CSI SR / Denoising
& FFDNet-based CE~\cite{8815888}; progressive-resolution CSI recovery~\cite{10319696}
& Grad-CAM~\cite{selvaraju2017grad} for time--frequency patch saliency in CSI image domain \\ \hline

Image SR+IR cascaded CE
& SR network cascaded with denoising IR network~\cite{soltani2019deep}
& Occlusion-based sensitivity analysis~\cite{MD2014Visualizing} on pilot ``images'' to quantify SR vs. IR roles \\ \hline

RIS-assisted CSI SR + denoising
& SR CNN (coarse features) + denoising CNN (coefficient recovery)~\cite{10025776}
& Grad-CAM~\cite{selvaraju2017grad} on SR and denoising branches for regional attribution \\ \hline

Denoise-then-LS / LS-refined CE
& LS-refined DNN~\cite{20215G}; train-free denoise--LS~\cite{balevi2019deep}; LEO CE denoising CNN~\cite{10200015}
& SHAP~\cite{NIPS2017_8a20a862} on LS/time--frequency grids; DeepLIFT~\cite{shrikumar2017learning} for layer-wise contribution tracing \\ \hline

End-to-End Autoencoder Communications
& Joint CE/detection~\cite{2017Power}; AE with SNR/energy feedback~\cite{kang2020deep,kang2018deep};
attention-based denoising/residual AEs~\cite{10120965,10042061}
& Global model reduction and surrogate distillation~\cite{wen2016learning,2016DeepRED};
attention-weight inspection for feature importance~\cite{10120965,10042061} \\ \hline

Modulation Classification (Concept / RNN / Transformer / Fusion)
& CB-AMC~\cite{9376108,pmlr-v119-koh20a};
LSTM/CNN-LSTM AMC~\cite{9822385,hong2017automatic,10298111,10380547,yu2019review};
Transformer AMC~\cite{10038598,liu2022convnet,yang2022lite,10130733,10478085};
glass-box Transformer~\cite{NEURIPS2023_1e118ba9};
fusion AMC~\cite{zheng2019fusion}
& SHAP~\cite{NIPS2017_8a20a862} for per-symbol/per-timestep attribution;
glass-box Transformer token-to-logit tracing~\cite{NEURIPS2023_1e118ba9};
CKA~\cite{ni2023learning} to verify complementary fused representations \\ \hline

Modulation Classification (Feature image / Cascaded CNN / NAS / DBN)
& Feature image-based CNN~\cite{8669002};
cascaded CNN~\cite{wang2019data};
NAS-designed AMC~\cite{10224342};
DBN AMC~\cite{ma2016dbn,mendis2019deep}
& Grad-CAM~\cite{selvaraju2017grad} for feature visualization;
DeepLIFT~\cite{shrikumar2017learning} for cascaded/NAS attribution;
layer probing for DBN transfer \\ \hline

Anti-Jamming (CNN-based classification)
& Bilinear CNN with spatial attention~\cite{xiao2021active};
Siamese CNN~\cite{9079547};
sliding-window CNN~\cite{liu2019pattern}
& Grad-CAM~\cite{selvaraju2017grad} saliency over time--frequency or sample-time domains \\ \hline

Anti-Jamming (Sequential / State-driven)
& GRU/RNN jammer-channel prediction~\cite{10004699};
movable-antenna repositioning via MLP~\cite{10955213}
& SHAP~\cite{NIPS2017_8a20a862} on temporal histories and predicted slots \\ \hline

Beam Selection (CNN-based / Multi-modal)
& Dual-variable CNN~\cite{9027103};
RSRP predictors~\cite{10118716};
LIDAR-aided CNN~\cite{8642397};
Omni-CNN~\cite{10388426}
& Grad-CAM++~\cite{8354201};
occlusion-based saliency~\cite{MD2014Visualizing};
SHAP~\cite{NIPS2017_8a20a862} for modality-level attribution \\ \hline

Beam Selection (LSTM-based tracking)
& Group-based LSTM with missing SSBs~\cite{9736600};
selective beam tracking~\cite{10149176,10511063}
& SHAP~\cite{NIPS2017_8a20a862} on sequential SSB/CSI inputs for per-timestep attribution \\ \hline

Beam Prediction / Classification (MLP)
& Two-step MLP for AoA mapping~\cite{9013188};
sub-6\,GHz$\rightarrow$mmWave MLP with dropout~\cite{9121328}
& Theory-based interpretability via UAT~\cite{yarotsky2017error} \\ \hline

MCS Selection / Link Adaptation (RL)
& Q-learning / bandit / DRL MCS selection~\cite{10826945,8703432,8721074,9490648,10026820,10024770,10841838}
& Reward decomposition (drQ)~\cite{juozapaitis2019explainable};
DSP~\cite{landajuela2021discovering};
PIRL~\cite{verma2018programmatically} \\ \hline

Beam Selection / Tracking (RL and bandit)
& DQN/DDQN/PG beam selection~\cite{10032173,9766078,9925080,9042821};
joint beam selection + precoding~\cite{9448095};
bandit beam--rate tracking~\cite{10041945,10643601}
& Integrated Gradients~\cite{pmlr-v70-sundararajan17a};
PIRL~\cite{verma2018programmatically};
DSP~\cite{landajuela2021discovering} \\ \hline

Interference Alignment (RL)
& DRL beamforming and power control~\cite{10052069};
DQN for IA~\cite{he2017optimization}
& Symbolic policy distillation via PIRL/DSP-style rule extraction \\ \hline

Transfer Learning (Channel Estimation)
& UL$\rightarrow$DL CE transfer~\cite{9175003};
low-resolution MIMO CE transfer~\cite{9388873}
& LRP~\cite{Chefer_2021_CVPR} to visualize informative subbands \\ \hline

Transfer Learning (Modulation Classification)
& CNN-based AMC transfer~\cite{meng2018automatic};
2D-CNN+GRU transfer AMC~\cite{10102600}
& Grad-CAM~\cite{selvaraju2017grad};
LRP;
integrated gradients;
SHAP~\cite{NIPS2017_8a20a862} \\ \hline

Transfer Learning (Anti-Jamming)
& CNN/Transformer/RNN transfer~\cite{9250656,10873840,10419906}
& Prototype layers / attention~\cite{Chefer_2021_CVPR,NEURIPS2019_adf7ee2d};
CKA~\cite{ni2023learning} for representation shift \\ \hline

Transfer Learning (Beam / IRS)
& IRS phase optimization~\cite{9367008};
mmWave beam-count adaptation~\cite{10279177}
& SHAP~\cite{NIPS2017_8a20a862} relevance comparison pre/post adaptation \\ \hline

Meta-Learning (Channel Estimation)
& Meta CE and CSI feedback~\cite{10308721,9878313,8761319,10623434}
& t-SNE~\cite{t-SNE};
pilot/gradient saliency;
SHAP~\cite{NIPS2017_8a20a862} on gradient entries \\ \hline

Meta-Learning (Modulation Classification)
& Few-shot and contrastive meta AMC~\cite{10049409,10681537,10423669}
& t-SNE~\cite{t-SNE};
CKA~\cite{ni2023learning} \\ \hline

Meta-Learning (Beam Selection)
& Meta beam alignment/prediction~\cite{10847789,9257198,9954418}
& CKA~\cite{ni2023learning} for pre/post adaptation comparison \\ \hline

Meta-Learning (MIMO Detection)
& Unfolded detector + LSTM optimizer~\cite{9234100}
& LRP~\cite{Chefer_2021_CVPR} through unrolled computation \\ \hline

Meta-Learning (Encoder--Decoder)
& Meta-learned encoder--decoder systems~\cite{9053252,9290055,10978447}
& SHAP~\cite{NIPS2017_8a20a862} on decoder outputs before/after meta-updates \\ \hline

\end{tabularx}
\end{table*}

\section{Explainable AI in Wireless PHY}
% The use of AI introduces opportunities to optimize, e.g., channel estimation, modulation classification, beam selection, and interference alignment. 
% % The gain is often fuzzy since many NNs cannot provide an analytical expression. 
Explanations at this layer should reveal which physical elements (e.g., pilots, subcarriers, or time–frequency patches) drive the decisions of a channel estimator, detector, or beam selector, and whether the learned behavior is consistent with basic propagation structures, e.g., sparsity, reciprocity, and Doppler patterns. 
% Since physical-layer errors propagate upward to MAC and network decisions, an explainable model at this layer helps diagnose when an estimation or classification failure stems from lack of information (e.g., insufficient pilots or missing SSBs), from adverse conditions (e.g., strong jamming or blockages), or from a mismatch between the training distribution and the current channel regime. 
A suitable choice of XAI tools is perturbation- and attribution-based methods that operate directly in signals or CSI domains, e.g., SHAP~\cite{NIPS2017_8a20a862} and occlusion-based sensitivity analysis~\cite{MD2014Visualizing}, which quantify the contribution of pilots, subcarriers, or time–frequency tiles to estimation and classification outputs. Moreover, gradient-based saliency and heatmap techniques, e.g., Grad-CAM~\cite{selvaraju2017grad}, can provide spatial or beamspace visualizations that align with time–frequency grids or angle–delay maps. Layer-wise relevance propagation and DeepLIFT~\cite{shrikumar2017learning} decompose channel estimates into contributions from intermediate residuals and denoiser outputs. 
Representation-probing tools, e.g., t-SNE~\cite{t-SNE} and CKA~\cite{ni2023learning}, can examine whether latent embeddings cluster according to physically meaningful factors (e.g., path geometry, Doppler, or modulation) and how these embeddings evolve under transfer and meta-learning.
In Table~\ref{xai phy layer}, we present a summary of the existing studies on AI in wireless PHY layer, and provide practical XAI techniques to enhance model explainability.

% \begin{figure}[tbp] 
% \centering
% \includegraphics[width=2.2in]{Illustration/ref_table1.png}
% \caption{ {\color{red}Change table 4-6 like this.} }
% \end{figure}

% \begin{figure}[tbp] 
% \centering
% \includegraphics[width=2.2in]{Illustration/ref_table2.png}
% \caption{ {\color{red}Change table 4-6 like this.} }
% \end{figure}

\subsection{Supervised Learning}
% We first discuss the application of XAI methods for 
Supervised learning has been applied to the PHY of wireless systems: channel estimation, modulation classification, and anti-jamming.

\subsubsection{Channel Estimation}
Data-driven channel estimation methods are increasingly studied. They do not rely on \textit{a-priori} knowledge of channel models.
% , and can adapt to diverse scenarios.

\paragraph{XAI Models for Channel Estimation}

\begin{figure}[tbp] 
\centering
\includegraphics[width=3.2in]{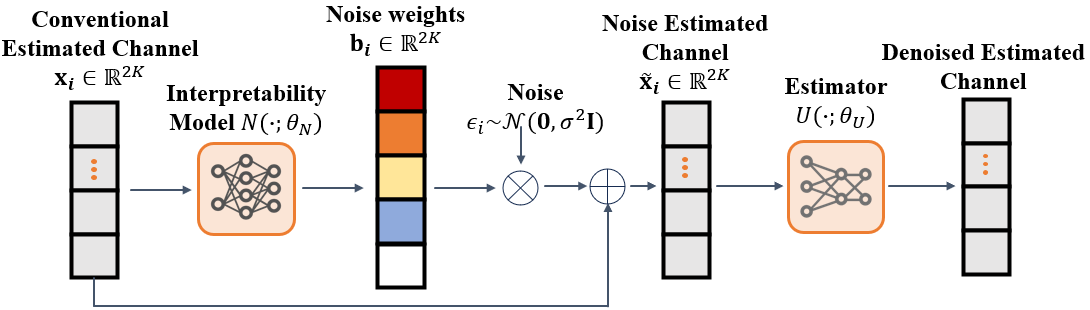}
\caption{\small The block diagram of XAI-CHEST in \cite{10368353}.
Based on the perturbation approach, the interpretability Model learns to generate a noise weight mask for the input features. With the premise that injecting noise into ``irrelevant" features does not degrade the Estimator's performance, the model identifies "relevant" subcarriers by maximizing noise weights while maintaining estimation accuracy, visualizing the decision-making logic of the closed-box model.}
\label{xaichest}
\end{figure}

Some studies have attempted to design DNNs that provide detailed explainability.
In \cite{10368353}, the authors investigate the explainability of two-channel estimation frameworks based on feed-forward NNs (FNNs) as the post-processing modules: spectral temporal averaging (STA)-FNN and time-domain reliable test frequency domain interpolation (TRFI)-FNN channel estimators. 
The authors uncover the interpretability by externally analyzing the input-output relationship by adding perturbation noises to determine the model inputs as relevant and irrelevant, as depicted in Fig.~\ref{xaichest}. 
\textit{The intuition is that if a subcarrier is correlated with the decision of a trained closed-box model, adding a high weight of noise to this subcarrier negatively affects the accuracy of the model}.

% {\color{red}
% The review of each related work should explain
% - motivations
% - contributions
% - briefly about the approaches
% - insights and findings
% - benefits and limitations

% Please check some examples of related work reviews in
% https://arxiv.org/abs/2508.18725
% https://arxiv.org/abs/2508.09561
% https://arxiv.org/abs/2507.14633
% }

\textit{Building on this setup, explainability is cast as learning a noise mask that reveals which subcarriers the frozen estimator truly depends on. For sample $i$ with real-stacked input $\mathbf{x}_i \in \mathbb{R}^{2K}$ (from $K$ active subcarriers) and reference output $\mathbf{y}_i \in \mathbb{R}^{2K}$, the interpretability network $N(\cdot;\theta_N)$ produces elementwise weights $\mathbf{b}_i \in [0,1]^{2K}$:
% \begin{align}
$\mathbf{b}_i  =  N(\mathbf{x}_i;,\theta_N)$,
% \end{align}
where a large $b_{i,k}$ indicates that subcarrier $k$ can tolerate stronger perturbations (hence is less critical), while small $b_{i,k}$ signals high sensitivity. The probe input is formed by gating i.i.d. Gaussian noise through this mask:
% \begin{align}
$\tilde{\mathbf{x}}_i = \mathbf{x}_i + \mathbf{b}_i \odot \boldsymbol{\epsilon}_i,
\quad \boldsymbol{\epsilon}_i \sim \mathcal{N}(\mathbf{0},\sigma^2\mathbf{I})$,
% \end{align}
where $\odot$ denotes elementwise multiplication and $\sigma^2 = 1$. With the utility estimator $U(\cdot;\theta_U)$ (STA-FNN or TRFI-FNN) frozen, $\theta_N$ is fitted over $n$ samples by minimizing a fidelity–noise objective,
% \begin{align}
$\mathcal{L}(\theta_N) = \frac{1}{n}\sum_{i=1}^{n} \big\|\mathbf{y}_i - U(\tilde{\mathbf{x}}_i;\theta_U)\big\|_2^2 - \lambda \sum_{k=1}^{2K}\log b_{i,k}$,
% \end{align}
with the trade-off coefficient $\lambda>0$. The MSE term suppresses noise on informative subcarriers (driving their $b_{i,k}$ downward), while the $\log$ penalty inflates $b_{i,k}$ for uninformative inputs; thresholding $\mathbf{b}_i$ therefore yields a faithful per-subcarrier relevance map that explains the estimator’s dependence structure.}

\paragraph{Model-Driven and Unfolded Inference for Channel Estimation}
To address beam squint and limited RF chains in hybrid mmWave massive MIMO, the authors of~\cite{10075639} unfold sparse Bayesian learning (SBL) into a DNN, where each layer updates angular-domain channel variance parameters via trainable modules. 
% This captures sparsity across angles and delays, improving estimation accuracy and reducing overhead. 
Although the unfolded DNN follows interpretable SBL steps, the neural layers obscure decisive angular-delay components. 
In \cite{11165341}, the authors propose deep unfolding-based radar-assisted channel estimation, where an MAML with CNN achieves high-precision direction-of-arrival (DOA).
Since the estimated DOA is injected as prior information into the unfolded sparse channel estimator, spurious features learned by the MAML-CNN under imperfect arrays can directly affect the angular support and the recovered channel.
\textit{To enhance transparency, one can apply Deep k-Nearest Neighbors (DkNN) as a post-hoc explainability tool: by comparing test inputs’ latent activations to nearest training exemplars in hidden layers, DkNN provides human-interpretable similarity-based explanations.}
% , revealing whether an estimated channel pattern aligns with prior support or deviates unexpectedly. 

In \cite{10091798}, a design is put forth for a practical RSMA receiver based on model-based DL (MBDL) to unite the simple structure of the conventional SIC receiver and the robustness and model agnosticism of DL. In \cite{9779569}, the authors first develop a new CS-based algorithm for sparse channel estimation, which requires no \textit{a-priori} knowledge of channel statistics. After the initial channel estimation, two DL networks, DnNet and DnLSTM, are utilized for denoising, so the final estimate combines a model-based CS frontend with data-driven refinement.

{\textit{For these two block-structured pipelines, SHAP \cite{NIPS2017_8a20a862} is an adequate choice to enhance explainability, as it operates directly on input descriptors and measurements without changing the model. In \cite{10091798}, treating channel and interference descriptors as SHAP features allows one to quantify their contributions to stream splitting and SIC ordering. In \cite{9779569}, SHAP with pilot/subcarrier occlusion on the measurement or denoised grids can highlight which time–frequency regions dominate the CS recovery and which regions are the most influential for DnNet/DnLSTM, separating the roles of the linear estimator and the nonlinear denoisers.
}

The authors of \cite{he2018deep} apply a learned denoising-based approximate message passing (LDAMP) network to estimate MIMO channels with limited RF chains. An analytical framework based on state evolution (SE) analysis of the asymptotic performance is provided.
\textit{DeepLIFT \cite{shrikumar2017learning} is suited to complement SE analysis; it propagates relevance scores through the fixed iterative graph and decomposes the final channel estimate into contributions from the residuals and denoiser outputs at each layer. Applying DeepLIFT \cite{he2018deep} makes it possible to see how much each LDAMP iteration contributes to the final CSI, and whether the learned update dynamics are consistent with the SE-predicted behavior.}

\paragraph{Structure-Aware Attention and Beamspace Learning}
% The authors of \cite{gao2021attention} employ attention networks in DL-based channel estimation for massive MIMO and automatically realize the ``divide and conquer" policy. The attention mechanism is integrated into the fully connected NN for the hybrid analog-digital structure. The gain of the scheme comes from the exploitation of the distribution characteristics of highly separable channels with narrow angular spread. Similarly, the authors of \cite{2019CS} first estimate the beamspace channel amplitude using an offline-trained NN, and then sort the estimated beamspace channel amplitude in descending order to select the indices of dominant entries, and finally reconstruct the channel according to the selected indices. It avoids a greedy manner since the scheme estimates dominant entries simultaneously instead of sequentially. In \cite{10570856}, the authors propose a UNet model with attention blocks, which selectively focus on the dominant part in order to capture the sparsity of the beamspace channel.

The authors of \cite{gao2021attention} employ attention networks in DL-based massive MIMO channel estimation and realize the ``divide and conquer'' policy. The attention mechanism is integrated into the fully connected NN for the hybrid analog–digital structure. 
% The gain comes from exploiting the distribution characteristics of highly separable channels with narrow angular spread. 
Similarly, the authors of \cite{2019CS} estimate the beamspace channel amplitudes using an offline-trained NN, sort the estimates in descending order to select dominant entries, and reconstruct the channel according to the selected indices.
% This avoids a greedy manner, since the scheme estimates dominant entries simultaneously instead of sequentially.

\textit{For these attention- and beamspace-driven estimators, SHAP \cite{NIPS2017_8a20a862} should be viewed as a post-hoc diagnostic rather than as an explanation already adopted in \cite{gao2021attention,2019CS}. 
By treating pilots, correlation vectors, or beamspace measurements as attribution features, SHAP can quantify which observations contribute to the predicted beamspace amplitudes and dominant-entry selection. 
Its explainability should be evaluated by attribution fidelity, stability under channel/noise perturbations, sparsity of the selected features, and consistency with the recovered angular support, while the original estimator performance is still measured by NMSE, spectral efficiency, or pilot overhead.}

In \cite{10570856}, the authors propose a deep-learning beamspace channel estimator with a switch-based selection network for mmWave massive MIMO. \textit{For such selection-driven estimators, explainability should focus on whether the learned switch or beamspace-entry selection is physically consistent with the dominant angular support, rather than on claiming that an XAI module directly improves estimation accuracy. 
If convolutional feature maps are available in the implementation, Grad-CAM \cite{selvaraju2017grad} can localize the beamspace regions that activate the estimator; otherwise, gradient- or perturbation-based saliency over candidate switches and beamspace entries is more directly aligned with the model output. 
The explanation quality can be measured by deletion/insertion tests, stability across channel and SNR realizations, and agreement with the selected support used for channel reconstruction, while performance should still be reported through NMSE, spectral efficiency, and pilot or RF-chain overhead.}

% {\color{red}\textit{For these attention- and beamspace-driven estimators, SHAP \cite{NIPS2017_8a20a862} is a natural choice to enhance explainability; it is model-agnostic and operates on the network inputs used to predict the beamspace coefficients. 
% 'this is more speculative, the reviewers may criticize...'}

% By treating the input features (e.g., pilot) as SHAP features and computing their contribution to the predicted beamspace amplitudes, one can identify which parts of the observation domain the attention-based network relies on when separating angular components in \cite{gao2021attention}, and which measurements drive the selection of large beamspace entries in~\cite{2019CS}.}

% In \cite{10570856}, the authors propose a UNet model with attention blocks, which selectively focus on the dominant part to capture beamspace sparsity. {\color{red}\textit{For this convolutional, attention-enhanced architecture, Grad-CAM \cite{selvaraju2017grad} is suitable: applied to intermediate feature maps, it produces beamspace-domain heatmaps that indicate which spatial or angular regions activate the attention blocks and dominate the reconstructed channel. }
% 'The review must explain how the explainability is measured and how the proposed approaches is better in terms of performance and explainability'
% }

% \begin{figure}[t]
% \centering
% \includegraphics[width=2.8in]{2019CS.png}
% \caption{System model of \cite{2019CS}}
% \end{figure}

\paragraph{Image-Inspired Super-Resolution and Denoising for CSI}

Considering the sparse mmWave channel matrix as a natural image, the authors of \cite{8815888} propose a practical and accurate channel estimation framework based on a fast and flexible denoising convolutional neural network (FFDNet). Unlike previous DL-based channel estimation methods, FFDNet is suitable for a wide range of signal-to-noise ratio (SNR) levels with a flexible noise level map as the input. 
In~\cite{10319696}, the channel response is treated as a low-resolution image and the resolution is increased during training so that the CSI contained in the image is progressively enriched. 

\textit{For the CNN-based SR/denoising designs \cite{8815888,10319696}, Grad-CAM \cite{selvaraju2017grad} can enhance explainability, as it operates on convolutional feature maps and produces heatmaps in the same image domain where the channel is represented. Applied to FFDNet in \cite{8815888} or the progressive-resolution network in \cite{10319696}, Grad-CAM can highlight the time–frequency patches that most influence the denoised or upsampled CSI.}

The study \cite{soltani2019deep} treats the time–frequency response of a fast fading channel as a 2D image and estimates the channel with pilots; a DL-based method using image super-resolution (SR) and image restoration (IR) is employed, wherein an SR network is cascaded with a denoising IR network and the pilot values are regarded as a low-resolution image. 
% For the cascaded SR–IR pipeline in \cite{soltani2019deep}, 
\textit{Occlusion-based sensitivity analysis \cite{MD2014Visualizing} can offer a complementary, input-focused perspective. By masking patches of the “low-resolution pilot image’’ and tracking the degradation in the reconstructed channel, one can quantify how pilot placement and local SNR patterns contribute to the SR stage and how much additional gain is provided by the IR denoiser.} 

In RIS-assisted MIMO-OFDM systems, the authors of~\cite{10025776} model CSI estimation as an image SR problem to recover and denoise the channel matrix with a designed SR CNN and a denoising CNN. The SR CNN extracts coarse features of the channel matrix. The denoising CNN recovers channel coefficients.
% For the RIS-assisted architecture in \cite{10025776}, 
\textit{Grad-CAM on the SR and denoising branches can locate which regions of the channel matrix drive the coarse reconstruction and which regions dominate the subsequent noise removal. In all three cases, the beamspace or time–frequency heatmaps remain in the same geometric domain as the CSI, providing a spatially grounded view of which structures the CNNs rely on when performing SR and denoising.}

\paragraph{Denoise-Then-LS Channel Estimation}
The authors of \cite{20215G} adopt a DNN for channel estimation of a frequency-selective 5G MIMO-OFDM system. It takes the LS estimate of the channel as input and decomposes the input into real and imaginary parts to improve the LS estimator with the DNN. 
% \begin{figure}[t]
% \centering
% \includegraphics[width=2.8in]{20215G.png}
% \caption{System model of \cite{20215G}}
% \end{figure}
In \cite{balevi2019deep}, the authors propose a DL-based channel estimation method for high-dimensional signals that does not require any training, e.g., MIMO-OFDM systems. The DNN 
% achieves advantages in various ways, compared with conventional estimators mainly, because it 
exploits correlations in the time-frequency grid to denoise the received signal for LS-type channel estimation. 
To obtain accurate CSI in low Earth orbit (LEO) satellite communication
systems, the authors of \cite{10200015} propose a denoising CNN to reduce channel estimation errors from the LS
estimator.

\textit{Model-agnostic SHAP  \cite{NIPS2017_8a20a862} can help elucidate the factors driving these LS-refined pipelines. Applying SHAP to the real and imaginary components of the LS estimate fed to the DNN in \cite{20215G} can reveal which subcarrier–time regions the refinement relies on when correcting LS errors. For \cite{balevi2019deep}, computing SHAP on the received-signal time–frequency grid before the denoise–then–LS step quantifies which samples the denoiser exploits. For \cite{10200015}, treating the LS output as SHAP features at the input of the denoising CNN can highlight which parts of the preliminary estimate are the most influential in reducing the residual error. 
% SHAP is suited to this setting, as its time–frequency perturbations and additive attributions align with the linear LS back end and provide physically interpretable importance measures without additional retraining. 
In addition, DeepLIFT \cite{shrikumar2017learning} can complement SHAP in these systems by propagating relevance scores through the DNN/CNN layers, making it explicit how intermediate representations transform LS inputs or received samples into refined channel estimates.}

\paragraph{End-to-End Autoencoder Communications}

In \cite{2017Power}, the authors employ a closed-box NN to handle channel estimation and signal detection in OFDM systems.
% This approach estimates the CSI with offline training to recover the signal directly in online deployment. where the NN is a closed box. 
In \cite{kang2020deep}, the authors estimate the MIMO channel at the transmitter with SNR feedback from the receiver. Under quasi-static block fading, a CNN constructs a deep autoencoder for joint channel estimation and pilot signal design; under time-varying fading, an RNN is additionally connected to a CNN to cope with the time-varying characteristic of the channel. 
% In both scenarios, GANs are used to address the scarcity of channel samples. 
In \cite{kang2018deep}, the authors employ a DL for MISO channel estimation at the transmitter side based on the harvested energy feedback from the receiver.
In the training phase, an autoencoder takes the CSI and the pilot symbols as the input, and the decoder outputs the estimated channel. 
In \cite{10120965}, a convolutional denoising autoencoder with an attention mechanism is designed for channel prediction in IRS-assisted millimeter-wave massive MIMO-OFDM systems. The attention module aggregates long-range dependencies in the effective channel and emphasizes subcarrier-interference patterns; the autoencoder performs denoising and dimensionality reduction.
% , so that spatial–temporal distribution features of the channel can be predicted more accurately. 
In \cite{10042061}, an attention-empowered residual autoencoder is developed, where residual blocks and attention modules at both encoder and decoder to enhance feature learning and cross-layer information fusion.
% , leading to improved reconstruction performance under various channel conditions.

\textit{To improve the transparency of these end-to-end and autoencoder schemes, one can derive low-complexity global surrogates via model reduction \cite{wen2016learning,2016DeepRED}. The learned mappings can be tested under controlled perturbations without retraining the primary networks. For \cite{2017Power}, a surrogate that maps time–frequency pilot structures and received samples to symbol decisions can be distilled and checked against the original closed-box.
% , turning the implicit decision behavior into auditable rules. 
For \cite{kang2020deep} and \cite{kang2018deep}, reduced surrogates can quantify how SNR or energy feedback and pilot design influence the reconstructed channel, and express the gains as conditions on feedback quality and pilot patterns. For the attention-based autoencoders \cite{10120965}, \cite{10042061}, the model-reduction pipeline can inspect the attention weights as importance scores over channel coefficients or latent features, revealing how the architectures weight spatial–temporal structures during channel prediction and symbol reconstruction.}

\begin{figure}[t]
\centering
\includegraphics[width=3.4in]{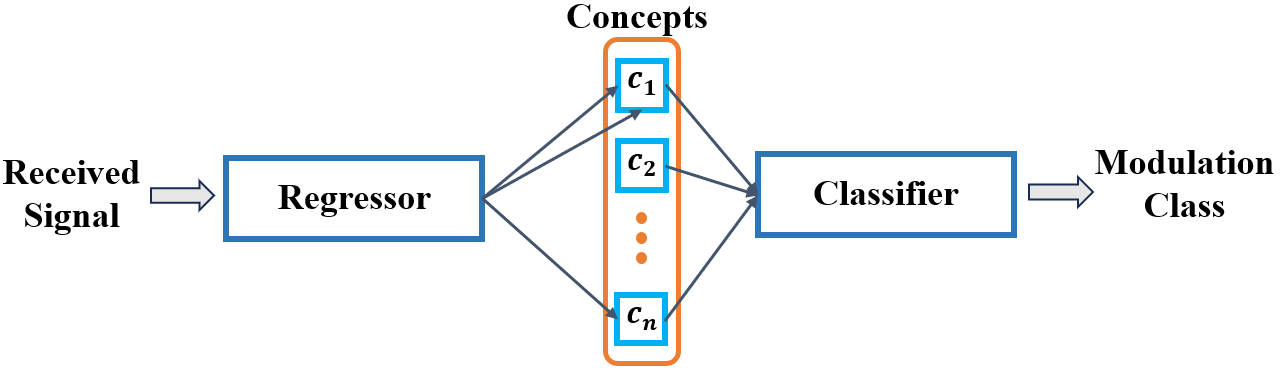}
\caption{\small The concept bottleneck model of \cite{9376108}. 
% Here, $c_1,\cdots,c_n$ are generated pre-defined, understandable concepts.
In this model, the Regressor first maps raw IQ signals into a set of pre-defined, human-understandable physical concepts ($c_1 \dots c_n$). Subsequently, the Classifier predicts the final modulation class based on these concepts. This structure forces the model to reason using explicit physical attributes, making the decision-making process transparent and verifiable.
}
\label{conceptbottleneck}
\end{figure}
 
\subsubsection{Modulation Classification}

Traditional automatic modulation classification (AMC) methods can be categorized into decision-theoretic methods and feature-based methods \cite{abdel2021survey}. 
The former is designed to evaluate the equality of different signal distributions to determine the modulation type~\cite{dobre2007survey}.
The latter first extracts the features of the received signal, such as spectral features and statistical features, and the extracted features are utilized to perform modulation classification with classical ML techniques, e.g., decision trees~\cite{zheng2019fusion}.

The use of DL allows for better classification and is more efficient compared with conventional methods, e.g.,~\cite{peng2018modulation,wang2020deep}. 
% Explainability approaches can verify the basis of model classification, which is helpful to combine models' experience with their personal knowledge.
As shown in Fig.~\ref{conceptbottleneck}, the authors of \cite{9376108} utilize Concept Bottleneck (CB) models \cite{pmlr-v119-koh20a} to provide AMC decision explainability for NNs. CB models are composed of two networks: a multi-head regression network, $\hat{c}=g(x)$ to acquire a series of pre-defined concepts $c$ using $x$ as input, and a classification network, $\hat{y}=f(\hat{c})$ to predict the target label $y$ with the explainable concepts $\hat{c}$ as input. 
% These anthropogenic concepts help attribute the preferences for network classification and provide similar performance to that of a single-network CNN. 
To address the difficulties of feature extraction caused by short observation times in AMC, LSTM blocks are deployed in \cite{9822385} and \cite{hong2017automatic} to capture long-range dependencies in the short-length data samples~\cite{yu2019review}. 
In \cite{10298111} and \cite{10380547}, the authors design CNN-LSTM to extract features from signal patterns, where the LSTM blocks are utilized for fusing the time characteristics of
signals.
% \textit{For better access to the explainability of using the LSTM blocks locally, SHAP \cite{NIPS2017_8a20a862} can be deployed to demonstrate the effect of each input time step on the modulation classification prediction. The SHAP library in Python can be adopted to build a visual interpreter and show the contribution scores of each feature to the output.}
\textit{For this class of AMC models that employ LSTM blocks locally, Integrated Gradients~\cite{pmlr-v70-sundararajan17a} and DeepLIFT~\cite{shrikumar2017learning} are more aligned with the recurrent decision path than perturbation-based SHAP. 
They can decompose the modulation logit with respect to I/Q samples, symbols, or short temporal windows, thereby indicating which transient amplitude, phase, or frequency variations are used by the LSTM when classifying short observations. 
The resulting explanations should be evaluated by temporal attribution fidelity, stability under SNR and channel perturbations, and consistency with known modulation-relevant segments, rather than being presented as a universally applicable improvement.}
% One can test whether decisions hinge on early or late segments under short observations without modifying training.

Transformer-based network structures can be utilized to extract more multiscale information to achieve more reliable AMC \cite{10038598}, where a hybrid ConvNeXt~\cite{liu2022convnet} and self-attention Transformer~\cite{yang2022lite} are designed. 
In \cite{10130733}, self-attention layers are integrated with a CNN to extract global features and generate the final classification vector.
The authors of \cite{10478085} exchange key and value in multiple heads of attention to incorporate linguistic features into visual representations.
% \textit{It is possible to deploy the glass-box transformers \cite{NEURIPS2023_1e118ba9} to replace the original transformer blocks, which is helpful for the transparent determination of the classification decision threshold. }
% For Transformer-based wireless AI models, r
\textit{Replacing standard blocks with glass-box Transformers \cite{NEURIPS2023_1e118ba9} can retain the self-attention inductive bias while rendering the token-to-logit computation decomposable into auditable components. This affords end-to-end traces of how multiscale features are weighted across heads and layers, yielding transparent criteria for class separation and decision thresholds without altering the training objective.}

DNNs with complementary outputs from different layers are considered in \cite{zheng2019fusion}. A nonlinear fusion of the layer outputs is proposed to improve overall AMC performance. 
% \textit{LIME \cite{ribeiro2016should} could be deployed to generate a local surrogate model for each layer for a deeper exploration of the predictive variances; Feature similarity between these layers can also be examined via Centered Kernel Alignment (CKA) \cite{ni2023learning} to judge the complementary degree between layers to improve explainability.}
\textit{For such AMC models that fuse complementary layer outputs, CKA \cite{ni2023learning} can be deployed. It operates directly on intermediate activations to quantify representational similarity and divergence across layers. Computing layer–by-layer (and, if needed, class-conditional), CKA can verify that the fused branches contribute non-redundant information and flags layers whose outputs are effectively overlapping. CKA is architecture- and loss-agnostic, requires no retraining, and aligns with the design goal of “complementarity”. }

% \begin{figure}[t]
% \centering
% \includegraphics[width=2.8in]{Supervised Learning/XAI and Wireless/Xiao_ref_figs/meng2018automatic.png}
% \caption{System model of \cite{meng2018automatic}}
% \end{figure}

Some CNNs with optimized input data have been considered for improving AMC training. A feature image-based AMC method is developed in \cite{8669002} using a CNN. \textit{By converting received signals to matrix graphics using higher-order moment information, researchers can use Grad-CAM \cite{selvaraju2017grad} or TensorBoard to realize feature visualization to improve the classification interpretability.}
A cascade of two CNNs is designed in \cite{wang2019data}, where the second CNN distinguishes the modulation modes not classified by the first CNN.
Some studies have also attempted to optimize the model structure. Based on the received modulated signal samples, the authors of \cite{10224342} utilize neural architecture search to search for a lightweight network. 
\textit{To enhance explainability of \cite{wang2019data}, DeepLIFT \cite{shrikumar2017learning} can reveal different neuronal contributions of the first CNN, to assess what meaningful inputs are to the second one.
For \cite{10224342}, applied to the final classifier logits with respect to the IQ input, DeepLIFT can highlight which time samples and I/Q components are responsible for each modulation decision.}

Based on deep belief networks, the authors in \cite{ma2016dbn} propose an AMC method for RFID signals based on DBN.
% which can achieve better performance than traditional BP ANNs under low SNR. 
The DBN has three hidden layers. It is relatively easy to examine their contributions by extracting features with transfer learning.
Another DBN classifier with low complexity is proposed in \cite{mendis2019deep} to satisfy the hardware requirements. Operations, such as activation function simplification, streamline the model to meet hardware constraints.

\textit{Despite the diversity of AMC models, the choice of explainability tools admits commonality. 
Perturbation-based attribution like SHAP \cite{NIPS2017_8a20a862} matches the sampling geometry of AMC, e.g., symbols, short windows, and time–frequency tiles, yielding contribution scores. DeepLIFT \cite{shrikumar2017learning} and Grad-CAM \cite{selvaraju2017grad} can reveal
which specific symbols carry the final decision. For fusion designs, CKA \cite{ni2023learning} can examine whether the fused layers provide complementary information.} 
% These methods provide model-agnostic and explainability across a series of model architectures without altering training.

\subsubsection{Anti-Jamming}
Deploying jamming at the PHY layer is increasingly concerning due to the vast availability of inexpensive USB dongle devices and SDRs \cite{9733393}. DNNs have been used for classification of jamming. 
% Explainable approaches can help the classification and prevent misclassification.

\begin{figure*}[bp]
\centering
\includegraphics[width=5.5in]{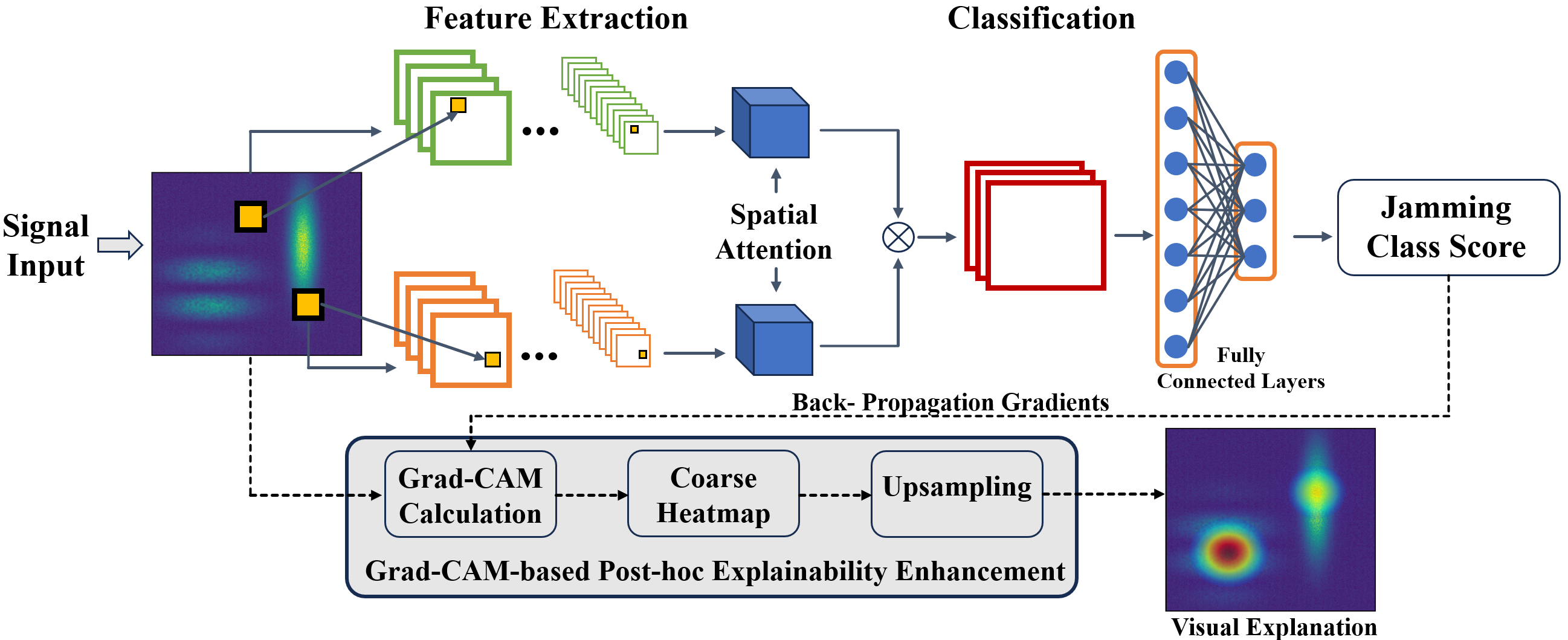}
\caption{\small 
The network of \cite{xiao2021active} with Grad-CAM. 
For the upper part, two CNNs extract the features of global time-frequency morphology and local fine-grained textures, respectively.
The lower part (in grey) details the explainability process: gradients of the predicted jamming class are backpropagated to compute channel-wise importance weights. The resulting heatmaps are upsampled and overlaid onto the original input to visualize which time–frequency patches contribute the most to the jamming decision, verifying that the learned spatial attention effectively highlights meaningful regions of the interference signal while suppressing background noise.
}
\label{xiao2021active with cam}
\end{figure*}

Based on CNN and its variants, some studies have conducted effective anti-jamming schemes.
As depicted in Fig.~\ref{xiao2021active with cam}, the authors of \cite{xiao2021active} design a bilinear CNN network with spatial attention to solve jamming recognition problems. By focusing on the most informative areas of input data, the time- and frequency-domain effective features are extracted.
In \cite{9079547}, a Siamese CNN (S-CNN), having four 1D-CNN branches that share parameters, extracts features to reuse training samples. With the $\ell_1$ or $\ell_2$ distance as the metric, the network drives the distance of similar samples in the feature space to decrease.
The authors of \cite{liu2019pattern} utilize a sliding window to control the input data according to the speed requirement of real-time jamming classification. Transfer learning helps improve the model's generalization ability under new environments.

\textit{For these CNN-based anti-jamming models, Grad-CAM~\cite{selvaraju2017grad} can produce saliency maps in the time–frequency or sample-time domain. Applied to the bilinear CNN with spatial attention in \cite{xiao2021active}, Grad-CAM can visualize time–frequency patches contributing the most to the jamming decision and verify whether the learned attention highlights physically meaningful regions. For the Siamese architecture~\cite{9079547} and the sliding-window scheme~\cite{liu2019pattern}, Grad-CAM on the last convolutional layers can reveal which segments of the input traces dominate the similarity metric or classification output.}
% , providing an intuitive check that the networks focus on informative jamming structures rather than spurious artifacts.

In \cite{10004699}, the authors utilize an RNN with multiple GRU units to predict the channels occupied by the jammers in upcoming time slots. More recently, in \cite{10955213}, to unleash the potential of movable antennas in anti-jamming communication, movable antenna arrays are deployed at the receiver side, where antennas are repositioned by training an MLP. 
\textit{For such sequence- and state-driven anti-jamming schemes, SHAP \cite{NIPS2017_8a20a862} provides a complementary, model-agnostic view. By treating future time slots or channel indices as features in \cite{10004699}, SHAP can quantify which parts of the recent interference history influence the predicted occupied channels most.}
% In both cases, the additive SHAP attributions align naturally with the discrete channel/position choices and yield time- and state-resolved importance scores without altering the original training or deployment procedures.

% Real-time requirements of anti-jamming systems impose higher demands on the transparency of modeling decisions. Researchers must design explainable networks that can specify different response mechanisms. For example, distinguishing between narrowband and wideband jamming, as well as identifying complex jamming patterns such as frequency hopping or pulsed jamming, requires a nuanced understanding of the signal environment. Explainable networks can provide insights into how these distinctions are made.

\subsubsection{Beam Selection} 
It is crucial to optimize beam pairs in massive MIMO to compensate for the high path loss in mmWaves. NNs can predict signal quality with only partial beam pair data, reducing beam selection overhead. 

CNNs have been applied to predict beams. 
In \cite{9027103}, a CNN learns the dual variables of the optimization problem. The features output from the model sense as expert knowledge. 
% This makes it more interpretable than other models whose output is the beamforming matrix. 
In~\cite{10118716}, the authors utilize a structured network composed of residual blocks, convolutional blocks, and an MLP to predict reference signal receiving power (RSRP) to reduce system overhead. In \cite{8642397}, a connected vehicle leverages its LIDAR data to suggest a set of beams selected via a deep CNN. 
The authors of \cite{10388426} propose Omni-CNN to expedite beam selection in vehicular networks. It constructs a convolutional backbone with modality-specific sparse subnetworks to ingest multi-modal data in a common architecture. Each modality operates on a disjoint subset of weights learned via gradient-based capacity search and ADMM pruning. 

% \textit{
% For each block, explainability techniques, e.g., \cite{MD2014Visualizing}, can be applied separately to imply the contribution from different model layers.
% GradCAM++ or LRP can help demonstrate the specific portion of input data that determines classification decisions.}
\textit{For these CNN-based beam selection and RSRP prediction, 
% two XAI techniques become widely suitable. 
Grad-CAM++ \cite{8354201} can exploit convolutional feature maps to locate the angle–delay (or LIDAR–spatial) regions and subcarrier groups that drive the predicted beam/RSRP.
% , yielding block-wise heatmaps across residual and convolutional stages. Second, 
Occlusion-based saliency \cite{MD2014Visualizing} can mask candidate sectors, subcarrier bands, or image patches to measure score drops, providing architecture-agnostic attributions aligned with beam codebook.
For Omni-CNN \cite{10388426}, SHAP \cite{NIPS2017_8a20a862} at the modality-aggregated feature level can disentangle the contributions of each sensor modality to the predicted beam quality.}
% , clarifying how sparse subnetworks cooperate under heterogeneous or missing sensor configurations.

\iffalse
\begin{figure}[t]
\centering
\includegraphics[width=3.3in]{new_figs/9736600.png}
\caption{{\color{cyan} The system model of \cite{9736600}. The first part tries to predict the missing data with $\Phi^{\text{miss}}(\cdot)$; with the learner $\Phi^{\text{group}}(\cdot)$, the second part tries to learn the best beam group that includes the best beam; for the third part, the data is classified into $B$ groups and each group is trained with an LSTM learner.}}
\end{figure}
\fi

Constrained by communication resources, the lack of synchronization signal blocks (SSBs) information could hinder accurate CSI prediction for NNs. The authors of \cite{9736600} propose a group-based LSTM network that exploits the spatial correlation among beams in the received signal strengths. Accurate beam selection can be obtained from sequentially processed, partially observed SSB measurements. 
In \cite{10149176}, an LSTM-aided selective beam tracking scheme is developed for mmWave communication systems.
% for multi-cell mmWave system. 
Selective beam tracking is a partially observed Markov decision process, and a data-driven LSTM is trained to choose which links to observe.
% based on previously measured links and their channel qualities.
% , thereby approximating the RL solution in a supervised, greedy manner with reduced complexity. 
In \cite{10511063}, LSTM-based predictive mmWave beam tracking is designed for vehicle-to-infrastructure communications. 
For co-located deployments, a sequence of historical sub-6~GHz CSI is fed into an LSTM classifier to predict the optimal beam index. For heterogeneous networks, an LSTM fusion model combines temporal sub-6~GHz CSI and 
% a limited number of 
mmWave wide-beam measurements through an attention-based fully connected module to jointly select the serving BS and beam. 

\textit{For this family of LSTM-based models~\cite{9736600,10149176,10511063}, model-agnostic SHAP~\cite{NIPS2017_8a20a862} provides an explainability tool because it operates directly on sequential inputs without modifying the training pipelines. Treating SSB groups, selected links, or historical sub-6~GHz CSI snapshots as features, SHAP assigns additive contribution scores to each time step and beam-related component for a given predicted beam index. In \cite{9736600}, such attributions can reveal which SSB groups are consistently responsible for accurate predictions under limited measurements; in \cite{10149176}, they can clarify how the LSTM aggregates past link qualities when choosing the next beams to probe under different mask constraints. }
% This unified, time-resolved view helps relate the learned beam policies back to spatial correlation patterns and resource constraints, improving the transparency and auditability of LSTM-based beam management.

% Constrained by limited communication resources, the lack of synchronization signal blocks (SSBs) information could prevent the accurate CSI prediction for NNs. The authors of \cite{9736600} proposed to explore the spatial correlation among beams in the received signal strengths by a group-based LSTM network to obtain an accurate solution from the sequentially processed inputs. 
% For mmWave communication systems,

% One key to explainability enhancement lies in the accuracy of the prediction of missing SSBs, as it affects the sequential assumptions of the input to the LSTM network data and hence the model explainability.

MLP and its variants are commonly adopted to perform beam predictions and classifications. In \cite{9013188}, a two-step MLP model captures the relationship between the receiving signal and the corresponding LoS path angle of arrival (AoA). An MLP model with dropout operations is designed in \cite{9121328} and used for fitting a mapping function that can predict the optimal mmWave beam directly from the sub-6 GHz channel. 
\textit{The interpretability of the MLP-based models lies in the effectiveness of the mapping function due to the Universal Approximation Theorem for NNs~\cite{yarotsky2017error}. }

% For the beam selection tasks, an NN with interpretability needs to specify which data are critical to influence the final classification, which also places severe demands on how the input data are processed. Proper selection and mapping of the input data can better match the expert knowledge and thus increase the confidence of the user.
\subsection{Reinforcement Learning}

% As the foundation of modern communication networks, functionality blocks in the physical layer are considered to be extremely difficult but important to design. This is because a large number of advanced theories are involved. 
As a substitute for complicated traditional functions, RL allows entities to learn and build knowledge about the communication and networking environment in the PHY layer. 
% In this section, we introduce RL methods used to implement in the physical layers. However, responsibility issues arise.

% \begin{itemize}
%     % \item The physical layer problems are at a bottom level, that is, most observations are at the symbol or waveform level. For issues like modulation or demodulation, the RL methods must be designed properly; otherwise, irresponsible decisions may be made.
%     \item For the RL system used in a physical network, the state variables can be hard to obtain precisely, due to the dynamically degraded wireless environment. For instance, perfect CSI is hard to acquire under a low SNR environment. For RL methods in the physical layer, additional robustness tests must be done.
%     \item A fast convergence rate of RL systems is highly desirable in the physical layer, since the wireless environment is highly dynamic. Unreliable decisions might be made during the slow convergence period.
% \end{itemize}

\subsubsection{Modulation and Coding}

Q-learning-based AMC in \cite{10826945} improves the coarse channel state partitioning in dynamic channel conditions and enhances system throughput.
A cognitive HetNet is considered in \cite{8703432} in which a user (PT) transmits uplink data to the BS (PR) on a certain spectrum band, and multiple STs adopt a sensing-based approach to access the same spectrum band. 
% Each ST may access the spectrum band with imperfect spectrum sensing and cause interference to the PR. 
Q-learning is employed for Modulation and Coding Scheme (MCS) selection, which defines MCS levels as the action space, current and previous SINR at the BS as the state space, and the number of transmitted data bits as the immediate reward.
The study \cite{8721074} expands to underwater communication, applying Q-learning to choose MCS and optimize the long-term expected utility of the underwater transmitter, e.g., BER, energy consumption, delay, and QoS.
% , without knowing the underwater channel model. 
The study \cite{9490648} further considers a scenario where the feedback delay in CSI severely degrades MCS performance, and addresses outdated CSI using Q-learning.

More recent works refine MCS and link adaptation via bandit and DRL formulations. 
In \cite{10026820}, link adaptation and channel selection are modeled as a piecewise-stationary multi-armed bandit, where each arm corresponds to an MCS–channel pair. A discounted structured and sleeping Thompson sampling scheme exploits monotone structure in the reward while discounting old observations and handling volatile (sleeping) arms.
In \cite{10024770}, DRL-based link adaptation is proposed for LTE/NR systems, where an agent observes link-quality indicators (e.g., control- and data-channel strengths) and selects the MCS to maximize throughput under BLER constraints.
% , avoiding reliance on static SNR–BLER lookup tables and improving robustness to model mismatch. 
In \cite{10841838}, PPO is employed for adaptive MIMO transmission in nonstationary environments, where the agent selects the MIMO transmission mode and modulation order based on instantaneous SNR and spatial-correlation features.
% , and is rewarded according to spectral efficiency subject to BER requirements, effectively tracking time-varying propagation conditions.

% From an explainability perspective, t
\textit{These RL- and bandit-driven MCS schemes share a common structure: low-dimensional channel or QoS indicators as states, discrete MCS (and sometimes mode) choices as actions, and scalar rewards aggregating throughput, reliability, and possibly energy. This structure is suited to a small set of RL-oriented XAI tools. 
Reward-decomposition methods, such as drQ~\cite{juozapaitis2019explainable}, can rewrite the scalar return (e.g., a composite score reflecting throughput, error rates, and energy cost} in Q-learning and DRL-based link adaptation~\cite{10826945,8703432,8721074,9490648,10024770,10841838} into labeled components (e.g., goodput gain, BLER penalty, delay cost), exposing the performance trade-offs that drive the selection of 
% high- or low-rate
MCS levels in different channel regimes. }

\textit{For algorithms with parametric policies or value networks, like the agent in \cite{10024770} and the PPO-based design in \cite{10841838}, DSP~\cite{landajuela2021discovering} can distill the learned behavior into compact symbolic expressions or decision rules that map SINR, interference, and correlation indicators to MCS or mode choices, revealing structure like implicit SNR thresholds and hysteresis. Complementarily, programmatically interpretable RL (PIRL)~\cite{verma2018programmatically} can fit simple program templates to the behavior of Q-learning and bandit-based schemes~\cite{8721074,9490648,10026820}, recovering human-readable if–then policies that approximate the learned strategies across operating scenarios. }
% Together, these techniques respect the discrete action spaces and handcrafted state features typical of MCS selection, and turn otherwise opaque RL controllers into policy descriptions that can be inspected, validated against link-adaptation guidelines, and stress-tested for robustness before deployment.

\iffalse
\begin{figure}[t]
\centering
\includegraphics[width=3.4in]{new_figs/10841838.png}
\caption{System model of \cite{10841838}}
\end{figure}
\fi

\subsubsection{Beam Selection}
A critical problem in mmWave massive MIMO is beam selection.
In \cite{10032173}, a DQN-based agent selects relay vehicles and beam directions in mmWave vehicular networks to maximize long-term rate and fairness.
While effective, the decision process lacks interpretability under dynamic mobility.
% To address this, 
Integrated Gradients~\cite{pmlr-v70-sundararajan17a} can be employed post-hoc to attribute beam–relay actions to environmental features, e.g., SINR or vehicle positions, revealing which state components drive specific beam choices and making the DQN policy more transparent.
In \cite{9766078}, a policy-gradient DRL framework performs blind beam alignment in mmWave vehicular networks using RF fingerprints and SINR inputs.
% The model outputs relay and beam configurations without explicit CSI, learning to align beams from interaction with the environment.
In \cite{9925080}, a double DQN (DDQN) under a federated learning (FL) framework solves this problem in ultra-dense mmWave networks, enabling adaptive beam management while sharing knowledge across distributed BSs.
In \cite{9042821}, a DDQN-based framework dynamically optimizes sector-specific MIMO broadcast beams in cellular networks, where the agent autonomously updates beam parameters based on user mobility patterns and evolving user distributions.
% Both single-sector and multi-sector deployments are considered, with a Q-network architecture whose complexity grows only linearly with the number of BSs.
Moreover, joint design of beam selection and precoding for a downlink mmWave MU-MIMO system with discrete lens arrays is investigated in \cite{9448095}, where a DRL-based NN and a deep-unfolding NN jointly optimize the beam selection and digital precoding matrices.
% , casting beam selection as an MDP solved via DDQN.

Several works have focused on learning beam tracking policies that exploit temporal structure and multi-cell coordination.
In \cite{10210616}, an LSTM-aided selective beam tracking method is proposed for multi-cell mmWave systems, where the network learns to track dominant beams across neighboring cells while probing only a subset of candidate beams.
In \cite{10041945} and \cite{10643601}, joint beam tracking and rate adaptation are modeled as an online RL problem.
The authors design a contextual multi-armed bandit framework in which each beam–rate configuration is an arm, and an asynchronous Thompson sampling strategy updates arm statistics whenever feedback is available.
% This allows the agent to adapt beam and rate decisions to time-varying link conditions while keeping the exploration cost manageable.

\textit{The learned policies and the DRL and bandit methods are generally difficult to interpret, which complicates validation under mobility, blockage, and heterogeneous deployment conditions.
PIRL \cite{verma2018programmatically} offers a suitable avenue to enhance the explainability of such beam selection agents.
By fitting a symbolic surrogate (e.g., a compact program or decision structure) to a trained DQN, policy-gradient, or bandit policy, PIRL can express beam and rate decisions as human-readable rules over measurable quantities, e.g., SINR, historical ACK/NACK patterns, or estimated user locations.}
% In this way, PIRL preserves the performance of the original RL controllers while exposing how state features and long-term objectives jointly shape beam choices, providing an interpretable layer that facilitates debugging and trustworthy deployment in real mmWave systems.

\textit{To improve transparency, DSP \cite{landajuela2021discovering} is appropriate by distilling a trained network into a compact, human-readable program over the same state variables (e.g., SINR summaries, user-distribution features, sector identifiers) and a discrete action set, including beam/sector indices. The resulting piecewise and threshold rules align with beam codebooks, indicating when specific state ranges trigger beam switches.
DSP can also apply to other DRL-based beam selection frameworks, e.g., the DQN and DDQN agents in \cite{10032173,9925080,9042821,9448095}, and the policy-gradient scheme \cite{9766078}, to extract comparable symbolic beam-selection rules from their trained policies.}

\iffalse
\begin{figure}[b]
\centering
\includegraphics[width=2.5in]{new_figs/9042821.png}
\caption{ {\color{cyan}The system model of \cite{9042821}, where RL is employed for the BSs to dynamically adjust MIMO broadcast beams.}}
\end{figure}
\fi
\subsubsection{Interference Alignment}
In \cite{10052069}, the authors propose to use DRL for interference management via joint beamforming and power control in multi-cell networks. 
% The goal is to maximize the SINR in dynamic millimeter-wave multi-cell networks.
The authors of \cite{he2017optimization} use a DQN to solve interference alignment problems under a finite-state Markov channel, where the complexity of the system is high. The optimal user selection policy can be derived using the network to analyze the collected CSI.

RL emerges as a viable solution to many MAC operations, but can encounter the following responsibility issues:
% when used for decision-making in the MAC layer:
On the one hand, 
% \begin{itemize}
% \item 
MAC layer performance metrics are diverse, encompassing QoS, throughput, and fairness among multiple users. 
% A comprehensive approach is essential for effective system design.
% , necessitating heightened attention to system-wide responsibility.
On the other hand, 
% \item 
early-stage decisions by online RL agents, before model convergence,
can have a significant impact,
particularly concerning the fairness of resource allocation and scheduling among multiple users.
% \end{itemize}

\subsection{Transfer Learning}

\subsubsection{Channel Estimation}
In~\cite{9175003}, the downlink channel prediction is cast as a transfer learning problem; a fully-connected NN is trained and then fine-tuned for new environments. 
In~\cite{9388873}, transfer learning is utilized for channel estimation of low-resolution
MIMO systems. 
% It exploits a pre-trained model for new task adaptation.
\textit{For both designs, LRP~\cite{Chefer_2021_CVPR} can visualize informative spectrogram regions and identify uplink CSI subbands contributing most to downlink prediction, revealing implicit learned mapping structures.}

\subsubsection{Modulation Classification}
% In \cite{meng2018automatic}, the authors resort to transfer learning to reuse learned prior higher-level information, which avoids the retraining process once the modulation set is changed.

Automatic modulation classification must stay accurate under diverse channels while adapting to changed modulation sets without costly retraining. The study \cite{meng2018automatic} addresses this via transfer learning: a CNN-based AMC first learns reusable high-level representations on a source dataset, and is then fine-tuned to new classes/environments, reducing data and adaptation time. 
Similarly, with a 2-D CNN and GRU, the work \cite{10102600} first trains audio signals to extract the spatial and temporal feature information, and obtains the transfer model to classify the modulated signals with a few training samples.

\textit{To enhance the explainability of these CNN-based transfer models, Grad-CAM \cite{selvaraju2017grad} can apply to the convolutional layers to highlight I/Q or time–frequency regions driving decisions, complemented with layer-wise relevance propagation or integrated gradients for per-sample attribution. SHAP \cite{NIPS2017_8a20a862} can help assess how input features jointly influence class logits.}

\subsubsection{Anti-Jamming}
The authors of~\cite{9250656} propose a deep transfer learning method using a CNN pre-trained on one ambient backscatter communication setting and fine-tuned on another. 
% This approach eliminates the need for explicit channel estimation and enables robust tag signal recovery under environmental shifts.
The authors of \cite{10873840} deploy transfer learning to fine-tune the designed signal separation model, which is composed of a transformer-based architecture.
% \textit{To enhance the explainability of transfer learning in~\cite{9250656} 
% % \cite{9868088} 
% for signal detection, one can embed prototype layers or attention mechanisms into the newly added modules, which allows each output to be linked to semantically meaningful features or spatial patterns.}
\textit{For these models, embedding prototype layers or attention in the fine-tuned head~\cite{Chefer_2021_CVPR,NEURIPS2019_adf7ee2d} is effective because prototypes anchor each decision to target-domain exemplars. 
% This exposes per-output similarity weights that indicate which learned patterns justify detection, while attention reweights time–frequency regions conditioned on the new environment to reveal the spectral patches that drive robustness. 
Both modules sit atop the pre-trained backbone with minimal changes, preserve transfer efficiency, and provide stable, auditable signals under domain shift without modifying the ambient feature extractor.}

The authors of \cite{10419906} design an RNN-based DRL model to avoid prior information from the environment, and use transfer learning to enable the DRL agent to learn fast in dynamic wireless networks.
\textit{To improve the explainability of this transfer step, CKA \cite{ni2023learning} can compare layer-wise representations before and after transfer, which quantifies how much of the source model’s temporal features are preserved or adapted, and thus provides a generic, training-free way to audit how transfer learning reshapes the anti-jamming policy.}

\subsubsection{Beam Selection}
The authors of \cite{9367008} utilize transfer learning to optimize the phase shifts on the IRS side, where the network is first trained offline using richly labeled source scenarios, and fine-tuned in a new environment with minimal labeled data.
For mmWave beam selection, the study \cite{10279177} employs transfer learning to adapt to new scenarios with a different number of beams.
\textit{SHAP can reveal which input features (e.g., propagation paths or channel coefficients) impact predicted phase configurations predominantly. By comparing relevance distributions before and after domain adaptation, one can observe whether the model bases its beamforming decision on environment-invariant physical features (e.g., geometry angles) or scenario-specific cues.}

\subsection{Meta-Learning}

Meta-learning can learn new tasks faster by observing how different ML methods perform on a wide range of learning tasks and learning from this experience or metadata.

\subsubsection{Channel Estimation}
In \cite{10308721}, a knowledge-driven meta-learning framework accelerates CSI feedback by leveraging spatial–frequency priors across tasks. While the method enables rapid adaptation, it is unclear which channel structures the model encodes. 
In \cite{9878313}, an MAML-based framework is developed for channel estimation in MIMO-OFDM systems, where a super-resolution CNN is trained across multiple static channel distributions to enable fast adaptation with limited pilots. Similarly, the study \cite{8761319} introduces RoemNet, a meta-learning estimator designed for OFDM systems with varying Doppler shifts and pilot configurations. 
% % Both methods demonstrate strong generalization and fast convergence under dynamic conditions. 
% However, their internal adaptation mechanisms remain opaque, especially regarding how pilot inputs influence predictions. 

% To address this, t-SNE \cite{t-SNE} visualization of latent features before and after fine-tuning can reveal whether the meta-learner captures physically consistent CSI embeddings, enhancing interpretability across deployment scenarios.
\textit{For these knowledge-driven meta-learning frameworks, t-SNE \cite{t-SNE} is suitable as a model-agnostic visualization of latent features. Applying t-SNE to embeddings before and after adaptation enables a training-free comparison across tasks.
% , indicating whether task-related structure becomes more organized and what the meta-learner retains from the spatial–frequency priors. Meanwhile,
On the other hand, gradient-based saliency maps can be applied post-hoc to visualize the impact of individual pilot symbols on the estimated channel, offering insights into the model's reliance on physically meaningful features.}

For RIS-assisted scenarios, the authors of \cite{10623434} propose a gradient-based meta-learning method with the gradients of the precoding matrix and phase shifting matrix as the input to enhance robustness.
\textit{For such a type of gradient-based meta-learning, SHAP \cite{NIPS2017_8a20a862} can treat gradient entries as features and provide per-component contribution scores. This helps clarify which precoder and phase-shift directions the meta-learner relies on when adapting to new RIS-assisted channels.}

\subsubsection{Modulation Classification}
In \cite{10049409},  meta-learning is used for a few-shot AMC with distribution bias. A multi-frequency octave ResNet (MFOR) is constructed to learn coarse (low-frequency) and fine (high-frequency) features, which can efficiently identify the modulation type of the signal while saving computational resources. 
To improve label mislabeling in AMC, meta-learning is utilized in~\cite{10681537} to correct untrusted samples with trusted few-shot labeled samples.
\textit{Similar to \cite{10308721,9878313,8761319,10623434,10445164}, across these few-shot, meta-learning AMC models, t-SNE \cite{t-SNE} is suitable:
% as a training-free, model-agnostic probe of episode-level embeddings: 
visualizing support and query features reveals whether adaptation yields clear class separation and tight clusters under limited data.}

In \cite{10423669}, supervised contrastive learning is combined with meta learning to amplify inter-class distinctions and reinforce intra-class compactness for few-shot AMC.
\textit{CKA 
% \cite{10423669}, 
\cite{ni2023learning} can quantify representation alignment of \cite{10423669} by measuring inter-class separation and intra-class compactness at the feature layers used by the supervised contrastive head.}
% , providing a complementary, numeric audit of the learned geometry.}

\subsubsection{Beam Selection}

In \cite{10847789}, an MAML-based adaptive beam alignment is proposed to enable swift adaptation to unknown scenarios.
In \cite{9257198}, a joint transfer and meta-learning framework enables fast downlink beamforming adaptation in wireless environments. By combining model initialization through meta-learning and rapid fine-tuning via transfer learning, the method achieves high spectral efficiency with limited data. The study \cite{9954418} applies meta-learning for beam prediction in dual-band systems, where a bilevel MLP network learns transferable representations from sub-6 GHz CSI to predict mmWave beam indices. Both approaches reduce adaptation time and training cost.
\textit{For these transfer- and meta-learning schemes, CKA \cite{ni2023learning} can compare representations pre- or post-adaptation in a model-agnostic, training-free manner. Computing layer-wise CKA between source and target tasks, researchers can evaluate beam-index partitions and assess whether adaptation preserves geometry tied to CSI structure and where features are reconfigured for new scenarios.}
% This provides a compact model understanding without modifying the pipeline.}
% {\color{cyan}
% Similar to \cite{10308721}, t-SNE \cite{t-SNE} is broadly applicable as a model-agnostic visualization of latent features. Applying it to source/target (or pre-/post-adaptation) embeddings enables a training-free check of whether the representation retains consistent geometry with respect to beam indices or CSI-driven clusters, clarifying how adaptation restructures the feature space across domains.
% }

\subsubsection{MIMO Detection}
In \cite{9234100}, a meta-learning-based MIMO detector combines an unfolded NN with an LSTM optimizer to adapt damping factors for iterative detection. This enables fast adaptation to varying channel conditions with minimal online data. However, the learned optimizer introduces interpretability concerns, as it implicitly governs update dynamics without a clear physical meaning. 
% \textit{Layer-wise relevance propagation (LRP)~\cite{} can be applied to the unfolded network to visualize which input signals or channel coefficients influence the damping decisions, helping reveal the model’s internal adaptation mechanism.}
\textit{LRP \cite{Chefer_2021_CVPR} can decompose the detector’s output through the unrolled computation, assigning relevance to inputs and intermediate variables at each step. This can reveal which received samples and channel coefficients drive the iterative updates and how the LSTM-adjusted damping influences them.}
% while remaining training-free and operating directly on the existing unfolded graph.}

\iffalse
\subsubsection{Channel Decoding and End-to-End Communication}
\begin{figure}[tbp]
\centering
\includegraphics[width=3.4in]{new_figs/9053252.png}
\caption{System model of \cite{9053252}}
\end{figure}
\fi
In \cite{9053252}, meta-learning enables fast adaptation of encoder–decoder communication systems over fading channels. 
% By learning a common initialization, the model can quickly adapt to new channel conditions with minimal updates, reducing the need for retraining. 
In \cite{9290055}, a meta-learned decoder supports channel-adaptive digital communication, where few-shot gradient updates allow adaptation to dynamic environments. 
The study \cite{10978447} introduces meta learning to help channel autoencoders (CAEs) enhance adaptability to varying channel conditions.
While these methods demonstrate adaptability and low training overhead, the adaptation mechanisms remain opaque.

\textit{SHAP \cite{NIPS2017_8a20a862} can be applied to the decoder modules in these frameworks to attribute symbol decisions to specific input components or channel descriptors. For meta-learned encoder–decoder systems, e.g., \cite{9053252,9290055,10978447}, SHAP 
% is broadly suitable because it is model-agnostic and operates via input perturbations, 
can yield feature-wise contribution scores for each decoded symbol. Comparing these attributions before and after meta-updates can reveal how the decoder’s reliance on particular observations or channel conditions changes across tasks.}
% offering a means to audit adaptation behavior.
% , thereby clarifying the learned adaptation behavior and supporting more trustworthy deployment of meta-learned communication systems.

Although meta-learning enables rapid adaptation across diverse wireless tasks, its internal mechanisms often remain opaque, especially during fast adaptation with limited data. This lack of interpretability raises concerns in dynamic or safety-critical deployments, where understanding the model’s reliance on domain-invariant or spurious features is essential. Enhancing explainability, e.g., via post-hoc attribution or latent representation analysis, not only improves transparency but also supports the responsible and trustworthy application of meta-learned models in practical wireless systems.

\begin{figure*}[bp]
\centering
\includegraphics[width=5in]{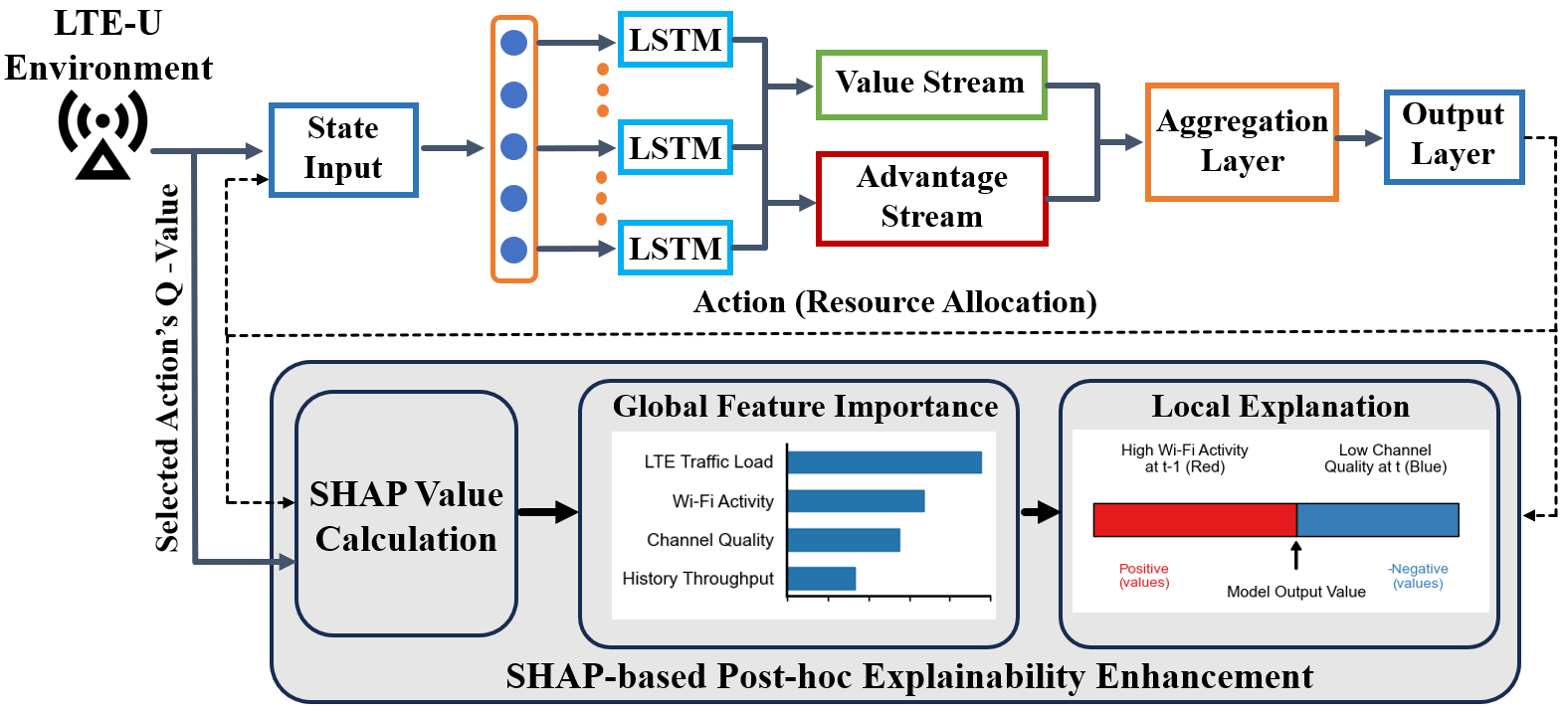}
\caption{ 
\small The network mode of \cite{8359094}, where SHAP is utilized to visualize the feature contribution analysis.
The plot ranks state features by their global importance, revealing how the model weighs both current and historical observations. 
The horizontal dispersion of the points indicates each feature’s impact on the Q-value output. 
By highlighting the significant negative contribution (red points with negative SHAP values), the visualization explicitly verifies that the LSTM layers effectively capture and utilize temporal dependencies to predict channel availability. This transforms the ``closed-box" DRL policy into an interpretable model, demonstrating alignment with the physical constraints of the LTE-U coexistence environment.
}
\label{8359094 with shap}
\end{figure*}

\subsection{Lessons Learned}

Although a variety of XAI techniques can be introduced into PHY learning tasks, their value in wireless systems does not mainly lie in producing generic feature-importance visualizations, but in revealing whether the model decision can be traced back to communication-relevant structures, such as pilots, subcarriers, antennas/beams, delay--Doppler components, and unfolded iterative steps. In this sense, explainability at the PHY should be judged not only by how understandable an explanation is to an ML practitioner, but also by whether it is aligned with radio semantics and can help communication engineers diagnose model behavior in terms of channel variation, interference conditions, resource usage, and receiver processing logic.

At the same time, no single XAI technique is universally suitable for all PHY tasks. Attribution-based methods are more natural for structured inputs such as CSI tensors, resource grids, and beam-domain features, whereas representation analysis and surrogate modeling are more suitable when one aims to understand what invariances, adaptation patterns, or decision rules have been learned. More importantly, explanations in wireless systems are not purely auxiliary outputs; they are coupled with communication utility. An explainability method that incurs excessive computation, latency, or signaling overhead may be difficult to deploy in real-time PHY pipelines, even if it is informative in an offline analysis setting. Therefore, future PHY-oriented XAI should jointly consider faithfulness, stability, and communication cost, and should be evaluated by whether it improves debugging, robustness assessment, and trustworthy deployment without undermining reliability, spectral efficiency, or latency.

\section{ Explainable AI supporting Wireless PHY}
The PHY decisions discussed above are rarely deployed in isolation; their realized utility depends on MAC/RAN control functions that allocate spectrum, power, time-frequency resources, and scheduling opportunities according to PHY-state information.
% Fairness, as a key performance indicator, refers to the equitable distribution of network resources among competing devices.
In this sense, XAI for resource allocation and scheduling supports PHY AI by exposing how PHY observations, such as SINR, CSI, queue state, and interference indicators, are translated into control actions that affect link reliability, latency, and user fairness.
Such explanations help clarify whether performance changes originate from PHY conditions, resource-control policies, or their interaction, thereby making the end-to-end PHY operation more diagnosable and accountable.
% % More specifically, a MAC-layer explanation ought to reveal which elements of the state---such as individual queue lengths, estimated channel qualities, slice-level weights, packet deadlines, battery levels, or interference indicators---drive a given scheduling or resource allocation decision, and how these drivers change across traffic regimes and load conditions. 
% % MAC decisions sit between the physical layer and the network layer.
% % Explainable MAC policies can help diagnose whether a perceived performance degradation stems from unfavorable physical conditions, from mismatched traffic priorities in the reward design, or from spurious correlations learned during training.
% MAC explainability contributes to fairness auditing, transparency in resource-tradeoff decisions, and the ability to expose when a policy is overfitting to particular traffic patterns or network compositions.

Feature- and state-attribution methods, e.g., SHAP~\cite{NIPS2017_8a20a862} and what-if/counterfactual explanations~\cite{bicalearning}, can be applied to supervised learning- and RL-based MAC controllers to quantify how perturbations in queues, SINR, or slice demands affect action selection. 
For supervised learning schemes, relevance-propagation tools, e.g., deep Taylor decomposition~\cite{2016Deep}, attribute rate or power decisions to input features, while multi-agent visualization tools (e.g., Dot-to-Dot~\cite{8968488}) reveal coordination across users and cells. 
For value-based and recurrent RL schemes, tree- and rule-based surrogates, e.g., CUSTARD and soft decision trees~\cite{topin2021iterative,coppens2019distilling}), temporal reward-processing~\cite{jenner2022preprocessing}, and symbolic policy extraction~\cite{landajuela2021discovering}, can transform neural policies into scheduling rules that expose thresholds on load, energy, and interference. 
% % Combined with fairness- and QoS-aware aggregation of explanations over time and user groups, 
% These XAI techniques enable systematic auditing of learning-based MAC systems for fairness, QoS, and regulatory compliance prior to deployment.
Table~\ref{xai mac layer} summarizes representative AI-enabled MAC/RAN control studies that support PHY operation, and lists practical XAI techniques for explaining how their decisions depend on PHY-state and traffic information.

\subsection{Supervised Learning for PHY Resource Control}
Supervised learning can approximate resource-control mappings that translate PHY-state information, e.g., channel gains, interference levels, and user demands, into power, subchannel, and user-association decisions.
In the supporting role considered here, explainability is needed to verify whether these learned mappings use PHY observations in a manner consistent with communication constraints and whether their outputs preserve throughput, energy efficiency, and fairness.
Achieving consistent evaluation across these indicators remains challenging.
\subsubsection{Resource Allocation}
The authors of \cite{9673099} approximate the optimal resource allocation strategy for arbitrary channel conditions using DNN models. 
The authors of \cite{8993757} utilize MLPs in place of traditional subgradient algorithms to realize power allocation for distributed antenna systems.
Recently, the sutdy in~\cite{10817307} treats optimization problems of varying size (e.g., different numbers of active users) as distinct tasks and proposes a multi-task learning framework with modular sharing. A single base DNN is shared across tasks, while a router activates task-specific subsets of input and output nodes so that one network can handle resource allocation problems of different dimensionalities.
% ; numerical results on average sum-capacity maximization show performance close to task-specific DNNs and clearly better than zero-padding baselines.

\textit{For DNN/MLP-based resource allocation~\cite{9673099,8993757,10817307}, knowledge distillation~\cite{2015Distilling,Haselhoff_2021_CVPR,LI2022108345} can enhance explainability. A high-capacity DNN (or the shared multi-task backbone with router in \cite{10817307}) can serve as a teacher; simpler student models (e.g., shallow networks or even linear/piecewise-linear surrogates) can be trained to mimic its power-allocation outputs over representative channel samples. Inspecting the distilled students can reveal which channel statistics or constraint summaries they effectively use. 
For \cite{10817307}, task-specific students can reveal how dimensionality (i.e., the number of users) alters the mapping from channel state to resource allocation.
In particular, for each dimensionality,
the subnetwork selected by the router in the multi-task DNN is distilled into an independent student model.
By using feature attribution or representation similarity methods,
and comparing the intermediate feature representations and final allocation outputs of these students,
we can obtain distilled artifacts that characterize how the decision logic evolves as users are added or removed, such as changes in power prioritization, sensitivity to channel gains, or constraint-dominated behaviors.}
Since the students operate on the same inputs and outputs as the original models, they provide auditable, low-complexity descriptions of the learned allocation policies.
% without modifying the primary inference paths.
The method can improve the stability of communication,
and incurs additional computational cost only during training due to the teacher’s forward pass.
For PHY-supporting resource control, these distilled descriptions help check whether channel and constraint information is converted into power-allocation decisions that are consistent with link reliability and resource-efficiency requirements.

In~\cite{9123600}, the authors apply MLP to NOMA mmWave networks for user association, subchannel allocation, and power control, with an integration of the Lagrange dual decomposition method.
Sun \emph{et al.}~\cite{8227766} propose an MLP-based resource allocation scheme that achieves throughput close to traditional iterative algorithms, e.g., WMMSE, while reducing computation time.
% For these MLP-based resource allocation models~\cite{9123600,8227766}, 
\textit{To produce explainability, deep Taylor decomposition~\cite{Chefer_2021_CVPR} can decompose each predicted allocation (e.g., user power levels or NOMA resource shares) into relevance scores assigned to the original input features, e.g., channel gains, interference levels, QoS weights, or dual variables from the Lagrangian formulation. Deep Taylor decomposition can clarify how user association and subchannel/channel-state descriptors drive the final power-control decisions in \cite{9123600}, and  how the learned MLP of \cite{8227766} internalizes the behavior of the WMMSE iterations by exposing which channel and noise terms dominate each resource update. Such relevance scores further indicate whether the allocation policy reacts to PHY conditions in a way that supports reliable transmission rather than merely optimizing an opaque surrogate objective.}
% Since the relevance scores are additive in the input space and respect the network’s layer-wise computations, DTD yields physically meaningful attributions that can be checked against classical KKT-based reasoning and resource-allocation heuristics, improving trust in these data-driven MAC-layer controllers.}

\begin{table*}[t]
\caption{Summary of XAI applications for MAC/RAN control supporting PHY operation.}
\centering
\label{xai mac layer}
\setlength{\tabcolsep}{3pt}
\renewcommand{\arraystretch}{1.05}
\scriptsize
\begin{tabularx}{\textwidth}{|p{3.8cm}|Y|p{6.75cm}|}
\hline
\textbf{Use Case} & \textbf{AI Algorithm and Ref.} & \textbf{Recommended XAI Techniques} \\ \hline

Resource Allocation (Supervised Learning)
& DNN-based resource allocation~\cite{9673099};
MLP power allocation for DAS~\cite{8993757};
multi-task learning with router for varying dimensions~\cite{10817307}
& Knowledge distillation~\cite{2015Distilling,Haselhoff_2021_CVPR,LI2022108345} to obtain low-complexity student models;
task-specific students to expose dimensionality effects~\cite{10817307} \\ \hline

Resource Allocation (NOMA / Iterative Approximation)
& Semi-supervised MLP with Lagrange dual decomposition~\cite{9123600};
MLP approximating WMMSE updates~\cite{8227766}
& Deep Taylor decomposition~\cite{Chefer_2021_CVPR} to attribute power/user association decisions to channel gains, interference, and dual variables \\ \hline

Resource Allocation (DRL-based)
& DRL for scheduling~\cite{9785600};
network slicing~\cite{9676621};
channel allocation~\cite{9047860};
RAN slicing and caching~\cite{9878155};
MEC orchestration~\cite{10091508};
multi-agent DRL for edge computing~\cite{10465255}
& What-if / counterfactual explanations~\cite{bicalearning} to probe sensitivity to queues, traffic load, and slice demand \\ \hline

Resource Allocation (Value-based / Recurrent RL)
& DQN/DDQN for latency, energy, and DSA~\cite{9678008,8868117,8532121,8809381};
multi-agent value-based RL for O-RAN slicing~\cite{10702574};
LSTM-based RL for LTE-U~\cite{8359094}
& SHAP~\cite{NIPS2017_8a20a862} on state features to attribute Q-values and actions;
temporal contribution analysis for LSTM-based policies~\cite{8359094} \\ \hline

Scheduling (Utility--Fairness Trade-off)
& CS-DLMA for fair spectrum sharing~\cite{9079169}
& Dot-to-Dot~\cite{8968488} to decompose Q-functions into critical state--action trajectories and fairness-driving patterns \\ \hline

Scheduling (Distributed / Partial Observability)
& DRQN for distributed eMBB scheduling~\cite{9127161};
DRQN for dynamic spectrum access~\cite{9089307}
& Reward Processing~\cite{jenner2022preprocessing} to decompose temporal rewards and trace influential past observations \\ \hline

Dynamic Spectrum Access (Policy-gradient / DQN)
& RF-powered ambient backscatter access~\cite{8703118};
spectrum sensing and aggregation with ACK feedback~\cite{9044839}
& DSP~\cite{landajuela2021discovering} to distill neural policies into symbolic decision rules over energy, occupancy, and ACK history \\ \hline

\end{tabularx}
\end{table*}

\subsection{RL for PHY Resource Control}
RL-based controllers support PHY operation by adapting scheduling, slicing, access, and computation-resource decisions to time-varying channel, traffic, and interference states.
Their explanations should connect learned state-action policies to the PHY quantities that trigger control changes and to the resulting latency, reliability, and fairness outcomes.
\subsubsection{Resource Allocation}
DRL has been utilized in various IoT networks for scheduling~\cite{9785600}, network slicing~\cite{9676621},
channel allocation~\cite{9047860}, and radio and cache resource allocation in 5G RAN slicing~\cite{9878155}, where DRL can resolve NP-hard problems and yield a solution at a faster speed.
In \cite{10091508}, DRL jointly orchestrates computation offloading and communication resources in MEC-enabled networks, learning a long-term scheduling policy under time-varying traffic and channel conditions.
In \cite{10465255}, a multi-agent DRL framework is designed for distributed edge computing systems, where individual agents at edge nodes coordinate task admission and local computing resources to improve the global service performance.
\textit{To enhance the interpretability of such DRL-empowered resource allocation algorithms, ``what-if'' explanations~\cite{bicalearning} help probe policy sensitivity: By perturbing selected state components (e.g., traffic load, queue length, or slice demand) and observing changes in the learned action, one can obtain a quantitative view of how different QoS and congestion indicators drive the agent’s allocation decisions, without altering the underlying policies.}
In a PHY-supporting setting, these perturbation-based explanations help diagnose whether changes in traffic or channel-related states lead to reasonable scheduling, slicing, or offloading adjustments before they affect latency and link reliability.

DQN has been employed 
% to optimize wireless communication systems 
for latency minimization in maritime UAV communication networks~\cite{9678008},
energy usage and inter-cell interference minimization under SINR constraints~\cite{8868117},
dynamic spectrum access (DSA)~\cite{8532121},
and constrained MDP formulations~\cite{8809381}.
DDQN is incorporated in~\cite{8532121} to improve the estimated Q-value in the presence of bad states regardless of the action taken.
In \cite{10702574}, a multi-agent value-based RL scheme is proposed for resource allocation in 6G O-RAN slicing, where agents learn communication policies and resource allocation decisions that balance slice-level utilities and RAN constraints.
LSTM cells are integrated into RL in~\cite{8359094} to proactively allocate LTE resources over unlicensed spectrum, where the BSs can predict a sequence of interdependent actions over a long-term time horizon, as illustrated in Fig.~\ref{8359094 with shap}.
% achieving long-term fairness among different underlying technologies. 

\textit{For these value-based and recurrent RL schemes~\cite{9678008,8868117,8532121,8809381,8359094,10702574}, SHAP~\cite{NIPS2017_8a20a862} offers a way to assess the contribution of state features (e.g., user traffic history, SINR, interference levels, and slice demand indicators) to the Q-values or policy outputs.
By treating each state component as a feature and computing SHAP values for the chosen actions, one can visualize which traffic or channel patterns most strongly influence resource allocation decisions, and how the temporal information encoded by LSTM in~\cite{8359094} or multi-agent observations in~\cite{10702574} shape the learned policies. This connects SINR, interference, and slice-demand features to latency, energy, and fairness outcomes, making the PHY-facing consequences of the resource-control policy explicit.}
% This yields post-hoc yet quantitatively grounded explanations for when and why specific resource allocation strategies are selected.

% \begin{figure}[t]
% \centering
% \includegraphics[width=2.8in]{Zhu/uchallita.png}
% \caption{System model of \cite{8359094}}
% \end{figure}

% \begin{figure}[t]
% \centering
% \includegraphics[width=2.8in]{Zhu/naparstek.png}
% \caption{System model of \cite{8532121}}
% \end{figure}

% \begin{figure}[t]
% \centering
% \includegraphics[width=2.8in]{Supervised Learning/XAI and Wireless/Xiao_ref_figs/9047860.png}
% \caption{System model of \cite{9047860}}
% \end{figure}

% Previous Version
% \begin{table*}[h!]
% \caption{Explainable RL in MAC Layer}
% \centering
% \begin{tabular}{ ||c c c c c||}
% \hline
% Scenario & Reference & Algorithm & Suitable XRL Method & Remark \\
% \hline
% \multirow{3}{4em}{Resource Allocation} & \cite{8868117} & DQN & ... & ... \\
% & \cite{8809381} & DRL & ... & ... \\ 
% & \cite{8359094} & RL-LSTM & ... & ...\\ 
% & \cite{8532121} & DQN & ... & ... \\
% & \cite{8793119} & RL & ... & ... \\
% \hline
% &\cite{9127161} & DRQN & ... & ... \\
% &\cite{9089307} & DRQN & ... & ... \\
% &\cite{9079169} & DQN & ... & ... \\
% &\cite{8703118} & Policy Gradient Descent & ... & ... \\
% &\cite{9044839} & DQN & ... & ... \\
% \hline
% \end{tabular}
% \end{table*}

\subsubsection{Scheduling}

\begin{figure*}[htbp]
\centering
\includegraphics[width=4.5in]{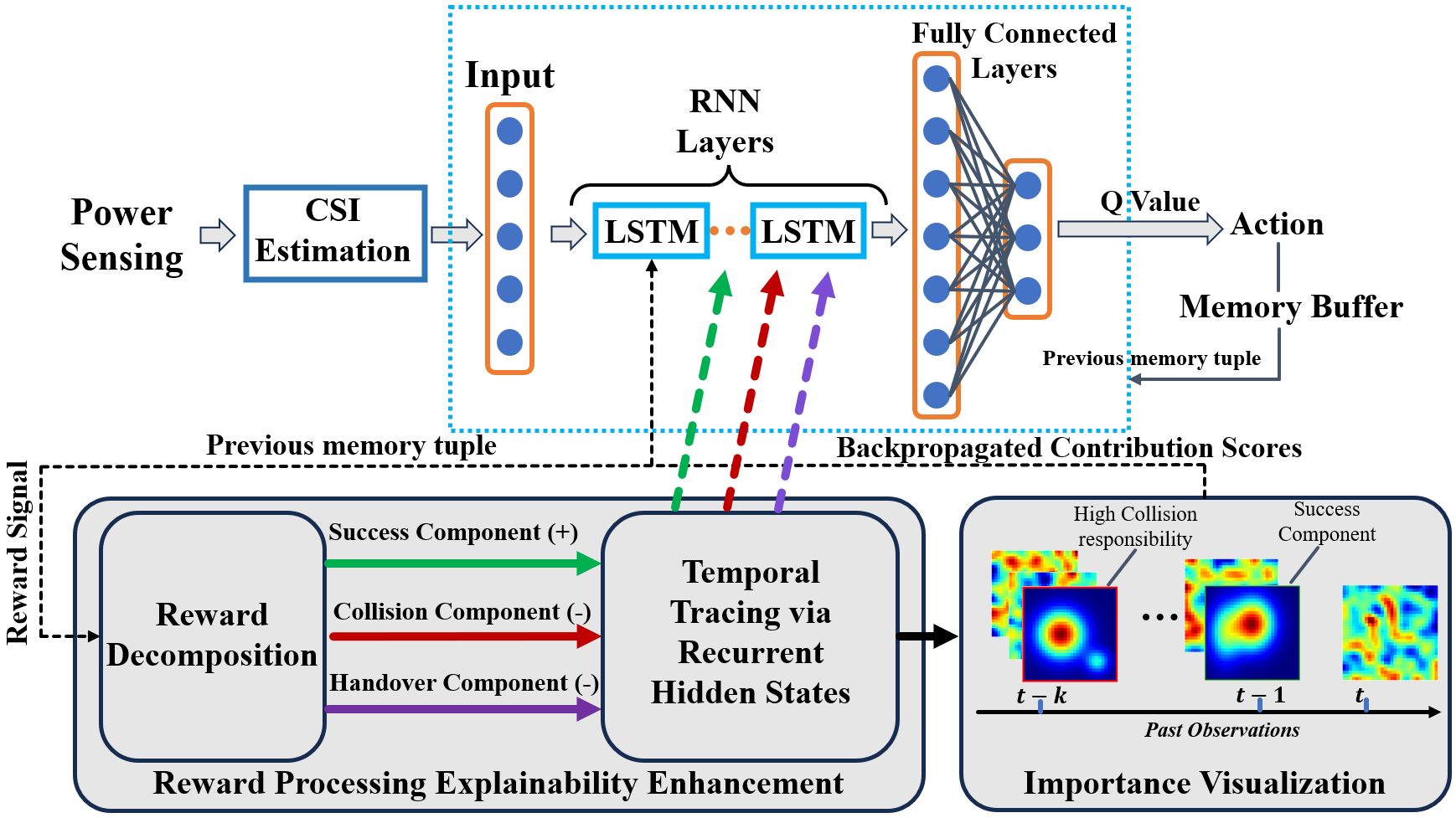}
\caption{\small System model of \cite{9089307}. The upper part depicts the DRQN structure, where the inputs are processed by LSTM to make channel access decisions under partial observability. 
The lower part (in grey) details the Reward Processing explainability mechanism, which decomposes the reward signal into semantically meaningful components such as ``Success" and traces their propagation back through the recurrent hidden states. By backpropagating these contribution scores, the framework generates an importance visualization highlighting which specific past observations are most responsible for the current action, thereby elucidating the temporal logic of the decision-making process.}
\label{9089307 with rp}
\end{figure*}

Due to non-convex and combinatorial characteristics, it is generally challenging to design scheduling and spectrum-sharing policies for heterogeneous wireless networks. In \cite{9079169}, a new class of DRL-based CSMA protocols, termed carrier-sense DRL multiple access (CS-DLMA), enables efficient and equitable spectrum sharing among co-located heterogeneous wireless networks. By adopting $\alpha$-fairness as the objective, CS-DLMA targets sum-rate maximization, proportional fairness, or max–min fairness when coexisting with TDMA, ALOHA, and Wi-Fi.
% , and relies on a DQN-based formulation with a non-uniform time-step structure.

\textit{For this DQN-based scheduling scheme that jointly considers utility and fairness, Dot-to-Dot~\cite{8968488} can serve as a post-hoc diagnostic tool. By decomposing learned Q-functions into a few critical state–action trajectories and visualizing how these trajectories contribute to returns, Dot-to-Dot can reveal which traffic-load patterns, contention outcomes, and neighbor configurations drive the decisions of CS-DLMA \cite{9079169}.} 
% This exposes how the agents coordinate to approach a fair equilibrium and clarifies the distinct roles of the dueling heads and double-Q updates, while leaving the original training pipelines unchanged.

% \begin{figure}[htbp]
% \centering
% \includegraphics[width=1.8in]{Zhu/nzhao.png}
% \caption{System model of a DRL-based scheduling scheme}
% \end{figure}

Designing low-complexity policies with local observations, yet able to adapt user association to the global network dynamics, is challenging. In \cite{9127161}, a distributed algorithm based on a deep recurrent Q-network (DRQN) maximizes the network sum-rate for eMBB services, where each UE acts as an independent agent to operate in a fully distributed manner. 
% UEs learn from experience, using only local and partial observations, to map environment states into connection requests to surrounding BSs and thereby maximize the network sum-rate for eMBB services. The framework naturally incorporates environment dynamics during learning, such that user association self-reorganizes when mmWave channels or traffic patterns change. 
As depicted in Fig.~\ref{9089307 with rp}, DRQN is also used in \cite{9089307} for dynamic spectrum access under independent channels and heterogeneous primary users. 
% A node has no knowledge of other nodes and predicts future channel states to learn a channel access strategy with low collision probability and high utilization.

\textit{For these DRQN-based schedulers, Reward Processing~\cite{jenner2022preprocessing} is particularly suitable to acquire interpretability. By decomposing the temporal reward signal into components aligned with key performance indicators (e.g., successful transmissions, collisions, or handovers) and tracing how these components propagate through the recurrent hidden states over time, Reward Processing can clarify which past observations are most responsible for current actions in \cite{9127161} and \cite{9089307}. }
% This helps distinguish genuinely long-term strategies from myopic reactions to recent rewards and explains how the learned DRQN policies internalize delayed feedback in highly dynamic wireless environments.

In \cite{8703118}, a low-complexity dynamic spectrum access framework is proposed for an RF-powered ambient backscatter system, where the secondary transmitter harvests energy from ambient signals and reflects them with modulated data. 
% Under the dynamics of ambient signals, an MDP formulation is first used to derive an optimal policy that maximizes system throughput, assuming full knowledge of environment parameters. To overcome the impracticality of complete prior knowledge, a low-complexity online RL algorithm based on policy gradient is designed, allowing the secondary transmitter to learn an effective access strategy directly from its interaction with the environment. 
In \cite{9044839}, dynamic spectrum sensing and aggregation are studied in a wireless network with correlated channels whose occupancies follow an unknown two-state Markov model. The secondary users receive binary ACK feedback after each transmission and sequentially choose sensing and access to maximize successful transmissions without affecting the primary users. 
% This problem is cast as a POMDP and tackled via a DQN to handle both unknown system dynamics and the computational burden of the high-dimensional belief space.

\textit{For these policy-gradient- and DQN-based dynamic access schemes, DSP~\cite{landajuela2021discovering} can offer a complementary, policy-level view of explainability. By distilling the learned neural policies in \cite{8703118} and \cite{9044839} into compact symbolic expressions over observable variables (e.g., estimated channel occupancy, harvested energy level, recent ACK history), DSP produces human-readable rules to approximate the agent’s decisions. These symbolic rules can reveal how channel availability, collision feedback, and energy states trigger access or harvesting actions, thereby clarifying how dynamic access policies support PHY-layer transmission opportunities.} 

\subsection{Lessons Learned}
The main role of XAI in this section is to make the resource-control loop that supports PHY AI more transparent.
Resource allocation, scheduling, and slicing policies translate PHY observations, e.g., channel state, interference, and queue information, into actions that determine whether the gains of PHY learning can be realized under latency, reliability, and fairness constraints.
Explanations therefore should not only describe a MAC/RAN model in isolation, but also reveal how PHY-state variations propagate through allocation decisions and affect link-level performance and user-level QoS.

Existing attribution, surrogate, distillation, and counterfactual tools provide useful starting points for this purpose.
They can expose which channel or traffic features dominate power allocation, which state-action patterns drive scheduling, and whether a learned policy is sensitive to particular channel realizations or traffic regimes.
Supporting PHY operation imposes additional requirements: explanations must remain lightweight enough for real-time control, stable across temporal and multi-user dynamics, and interpretable at the protocol level rather than only at the feature or neuron level.
Future XAI methods for MAC/RAN control should couple explanation fidelity with communication utility, so that explanations help diagnose whether performance degradation stems from PHY impairments, resource-control policies, or their interaction.

\section{Standardizations and Industrial Projects}

\subsection{Existing XAI Standards for 6G}
Several international standards organizations have initiated efforts to formalize the concept of XAI through technical, legal, and ethical frameworks. These initiatives aim to provide rigorous definitions, design guidelines, and implementation protocols that ensure AI-driven decisions are transparent, auditable, and aligned with human-centric values.
Notable examples include IEEE's foundational standard for XAI (P2976)~\cite{ieee_p2976_xai_2025}, ETSI's security and trust principles incorporating explainability, and regulatory alignment under the EU AI Act via CEN/CENELEC~\cite{cencenelec_ai_jtc21_2021,eu_ai_act_article40_2024} and ITU~\cite{itu_who_ai4h_whitepaper}. Collectively, these standards efforts provide a critical foundation for embedding explainability in next-generation wireless systems, enabling AI deployments that are not only functional but also transparent and accountable, as summarized in Tab.~\ref{tab:xai-standards}.

\begin{table*}[!t]
\centering
\caption{Existing XAI standards and regulatory frameworks applicable to 6G systems.}
\label{tab:xai-standards}
\renewcommand{\arraystretch}{1.05}
\begin{tabular}{ |p{3.2cm}<{\centering}| p{1.8cm}<{\centering} | p{4.4cm}<{\centering} | p{6.8cm}<{\centering} |}
\hline
\textbf{Standard / Framework} & \textbf{Authority} & \textbf{Scope} & \textbf{Relevance to 6G} \\
\hline
IEEE P2976~\cite{ieee_p2976_xai_2025} & IEEE & Standard for XAI system design
& Defines explainability types, use-case classes, fidelity, and usability; foundational for 6G RIC agents and xApps. \\ \hline

ETSI SAI/AI cybersecurity guidance~\cite{etsi104223} & ETSI
& AI cybersecurity and trustworthy AI lifecycle
& Embeds explainability as a requirement for secure, auditable telecom AI systems such as those in O-RAN. \\ \hline

ISO/IEC TR 24028~\cite{iso24028} & ISO/IEC JTC 1/SC 42
& Trustworthiness in AI
& Addresses transparency and explainability within the broader context of AI reliability and lifecycle assurance. \\ \hline

EU AI Act~\cite{eu_ai_act_article40_2024} & European Commission
& Legal framework for transparency and risk-based AI classification
& Applies to high-risk AI systems, including autonomous 6G functions, requiring human-interpretable outputs. \\ \hline

ITU AI guidance~\cite{ituTrustAI2023} & ITU
& AI for autonomous and health systems
& Develops explainability profiles for mission-critical AI use cases, with emerging relevance to telecom and 6G. \\
\hline
\end{tabular}
\end{table*}

\subsection{Ongoing Projects of XAI for 6G}
Recent projects have implemented concrete XAI techniques to enhance the interpretability and trustworthiness of learning-based 6G systems. These initiatives cover network slicing, anomaly detection, and graph-based control, with tools like SHAP, surrogate models, and XRL. We summarize four representative efforts that exemplify the integration of XAI into next-generation wireless systems.
% , with insights into architecture design.
 
The Turbo Explainable Federated Learning (TEFL) framework targets the challenge of deploying trustworthy federated DRL in 6G RAN slicing~\cite{tefl2022}.
TEFL introduces post-hoc explanation techniques, namely, SHAP and IG, to interpret local model decisions under non-IID data distributions.
By quantifying feature importance and providing client-side explainability,
TEFL enables service-level agreement (SLA)-aware orchestration that is both adaptive and transparent.

The SliceOps framework presents an MLOps-centric solution for managing the full AI lifecycle of 6G network slicing agents, emphasizing explainability across deployment, monitoring,
and retraining stages~\cite{sliceops2023}.
It employs XRL to maintain interpretable slice decision-making, including surrogate models, such as decision trees and saliency maps, for tracking agent behavior.
SliceOps also features confidence and fidelity metrics for monitoring explanation quality in real time. 
By integrating explainability into continuous deployment workflows, SliceOps supports regulatory compliance and end-to-end transparency in AI-driven slice orchestration.

The XAInomaly system addresses real-time anomaly detection in O-RAN-based 6G networks by embedding lightweight, XAI models into time-sensitive control loops~\cite{xainomaly2024}.
It leverages a customized variant of SHAP tailored for low-latency environments,
along with counterfactual analysis to contextualize detected anomalies.
The approach supports URLLC, offering explainable forecasts for deviations in RAN behaviors and enabling interpretable feedback for system operators. 

The EXPLORA framework integrates explainability into graph-based DRL
for resource management in Open RAN~\cite{explora2023}.
It models the network environment as a dynamic graph and applies attribution methods over node and edge representations to interpret policy behaviors.
Post-hoc explanation tools, including node saliency heatmaps~\cite{Pope_2019_CVPR}, are used to visualize and understand multi-hop resource allocation decisions.
This allows stakeholders to validate the reasoning of autonomous agents in spectrum assignment or routing, contributing to both trust and auditability.

\section{Challenges and Future Directions}

A range of challenges are identified to further advance the area of responsible AI technologies for future 
% communication and network systems. 
wireless PHY. 

\subsubsection{Explainability-Performance Tradeoff}
% {\color{red}Although the introduction of NNs in communication systems provides better performance guarantees for a variety of tasks, their corresponding theoretical limits of the explainability-utility tradeoff have not been well explored. 
% The exploration of relevant theoretical limits will be the key to whether NN-driven tasks can be generalized in the new generation of mobile communication networks.
% 'The directions are very general and abstract. For example, how does explainability affect the communication performance (not only the model execution performance)?'
% }

Although the introduction of NNs in communication systems provides better performance guarantees for a variety of physical tasks, the limits of the explainability-utility tradeoff have not been well explored.
Here, ``utility" represents the task metrics of communication, including spectral efficiency, BER, and latency.

In general, explainability can affect communication utility through the following aspects.
First, adding additional explainability blocks to DNNs, e.g., disentangled factors~\cite{2017Interpretable}, could reduce the model capacity and limit the adaptability of the learned strategy to the environment.
% First, adding intrinsic explainability blocks to DNNs, e.g., enforcing disentangled latent factors~\cite{}, can reduce model capacity and limit how well the learned strategy adapts to environmental changes. 
For instance, in a DNN-based link adaptation module, forcing the internal features to be separated into a few fixed factors can make the model miss the combined effect of fading and interference, yielding biased MCS decisions, which causes a conservative rate, both degrading effective throughput and latency.
Utilizing XAI explanations can also induce protocol and air-interface overhead. For instance, generating counterfactual explanations with~\cite{sharma2019certifai} typically relies on an iterative search over input perturbations, which may require multiple model queries and cause extra on-device/edge computation and energy consumption, and longer decision latency to construct and validate the counterfactual. 
% Moreover, if explanations are audited or consumed by an edge controller, the counterfactual itself (e.g., perturbed feature vectors, time--frequency masks, confidence deltas, or selected exemplars) must be reported through the control plane or backhaul, occupying uplink resources and increasing signaling load; in extreme cases, validating a counterfactual against communication objectives (e.g., BLER/outage under a candidate MCS) may further trigger extra probing/decoding trials or auxiliary measurements, consuming time--frequency resources that would otherwise carry payload data.
Meanwhile, in some cases, explainability can also help improve communication performance. For instance, explanation signals generated by attribution maps~\cite{10.1145/3704413.3764466} can expose when a receiver is relying on unreasonable cues 
% (such as transient interference patterns or dataset-specific artifacts) 
rather than physically meaningful features, thereby enabling timely debugging and model correction before the incorrect outputs cause system performance degradation.

It indicates that the tradeoff is not merely between computation and accuracy, which could be non-monotone at the system level.
The exploration of relevant theoretical limits will be the key to whether NN-driven tasks can be generalized in the new generation of mobile communication networks.

\subsubsection{Explainability Enhancement of Data Processing}
The diversity and complexity of input data in each layer significantly impact the utility and explainability of NN models.
A promising future research direction is to investigate how time-frequency conversion, dimension conversion, and feature extraction normalization of network input data affect model explainability.
One can analyze the effects of various signal processing methods on the feature distribution of input data, the model's internal structure, and decision-making, with techniques, e.g., feature visualization and decision path visualization.
Feature engineering grounded in physical knowledge can also enhance the physical explainability of the input data, aiding in model training and interpretation.
An enhanced data processing framework, based on stochastic optimization tools, is anticipated to cover key input data types in the PHY layer to improve input explainability.

\subsubsection{Customized Explainability for Communication Systems}% {\color{red}Most explainability techniques have been applied to fields like computer vision and natural language processing. NNs used in communication require new explainability theories and algorithms due to their training time and input complexity. By leveraging information theory and utility-explainability trade-off, researchers may model an objective function that simultaneously characterizes the explainability of both the input data and the network. This facilitates analyzing the explainability performance limits of NNs. Using stochastic optimization theory, a customized explainability enhancement algorithm for these NNs can be designed, resulting in a comprehensive analysis and algorithm design.

% 'customized in which ways? it is good to explain'
% }

Most explainability techniques were originally developed for computer vision and natural language processing.
However, NNs in communication systems call for customized explainability theories and algorithms since their inputs are highly structured (e.g., I/Q samples and CSI tensors).
Meanwhile, the learned strategies are executed in real-time, and the corresponding decisions immediately affect link reliability and resource usage. 
Therefore, it is essential to make specialized improvements based on these characteristics.

First, explanations should manage to align with communication factors, rather than generic pixel-/token-level saliency. 
In particular, explanations are expected to localize the decision reasons on key factors such as subcarriers, OFDM symbols, and pilot signals. 
Such explanations based on communication structures are more direct for diagnosing model decision-making and fault modes.
Explainability should be defined and jointly optimized with communication utility as well. 
Unlike conventional XAI settings, wireless systems are evaluated with utilities like spectral efficiency, and experiences explainability-performance tradeoff, as discussed in Section VII-A.
To this point, information theory provides techniques to formalize this goal, e.g., by quantifying how much task-relevant information is retained in the input representation and how much of the decision can be attributed to physically meaningful factors, which further enables the study of explainability performance limits.
Meanwhile, customized explainability should be lightweight and based on practical protocols. 
For computational-constrained tasks, producing explanations must introduce little computation overhead, which requires the explanations to be directly generated by protocol actions, such as conservative fallback and safe-mode switching.

\subsubsection{Interpretability in PHY Layer}
NNs for channel estimation, modulation classification, and beamforming remain opaque to domain engineers, hindering trusted deployment. A promising direction is to embed physics via deep‐unfolding and invariance constraints so that layer operations align with recognized signal-processing steps, while attribution tools quantify which spectral, temporal, or angular features drive decisions. Fairness should be formalized by multi-objective training (e.g., rate–fairness tradeoffs) and certifiable constraints for power/beam allocation to avoid systematic disadvantage to cell-edge or low-tier devices. Privacy can be enforced through on-device or federated training on raw CSI and/or using explanation sanitization to prevent leakage of location or mobility cues. Together, these yield interpretable, fair, and privacy-preserving PHY models.

\subsubsection{Cross-Layer Explainability and Governance}
More recently, end-to-end QoS emerges from coupled decisions across the physical, MAC, and network layers. Siloed explanations are insufficient. A cross-layer agenda should build causal attribution across the stack, aligning per-layer rationales into a coherent path from signal conditions to scheduling and routing outcomes. Fairness must be specified globally (e.g., QoE parity) and decomposed into layer-level constraints, with explanations that quantify each layer’s contribution to disparities and propose remedies. Privacy-by-design requires minimal cross-layer data sharing, federated coordination, and tiered explanations that adapt granularity to authorization levels. Standardized interfaces, metrics, and audits for multi-layer explanations will enable transparent, fair, and privacy-preserving AI governance in future wireless systems.

% \subsection{\color{red}LLM, agentic AI, etc.}
% Large language models (LLMs) are increasingly deployed across a wide range of communication network scenarios and applications to enhance productivity~\cite{lee2024llm,zhang2025communication,shao2024wirelessllm}.
% For the next-generation wireless network, there exist two potential research directions: AI4Net and Net4AI~\cite{9770094,11103489}. Specifically, AI4Net involves leveraging the strong inference ability of LLMs and agentic AI to efficiently resolve a series tasks of different communication layers. Net4AI entails optimising communication transmission and resource allocation, specifically tailored to the model characteristics of LLMs, thereby enhancing network training efficiency.

% The high complexity and inherent hallucinatory nature of LLMs persistently challenge researchers and diminish users' confidence and trustworthiness~\cite{wu2024usable,bilal2025llms}.

\subsubsection{XAI for LLM and Agentic AI}  
Large language models (LLMs) are increasingly deployed across communication network scenarios and applications~\cite{lee2024llm,zhang2025communication,shao2024wirelessllm}.  
Beyond providing natural-language reasoning or prompt-based assistance, the paradigm of \emph{agentic AI} has recently emerged~\cite{tong2025wirelessagent}, which not only interprets a user’s intent via an LLM but also formulates sub-goals, plans multi-step actions, invokes external tools or interfaces, and executes those actions with minimal human supervision~\cite{tong2025wirelessagent}.

Agentic AI can sense CSI, plan suitable modulation/coding and power configurations, and adjust transmit parameters; plan scheduling and resource allocation decisions; and automatically execute routing decisions~\cite{lu2025agenticgraphneuralnetworks}.
For the next-generation wireless networks, potential research directions involve AI4Net and Net4AI~\cite{9770094,11103489}. AI4Net leverages the strong inference ability of LLMs and agentic AI to efficiently resolve a series tasks of different communication layers. Net4AI optimizes communication transmission and resource allocation, specifically tailored to the model characteristics of LLMs/agentic AI, enhancing network training efficiency.

The high complexity and inherent hallucinatory nature of LLMs/agentic AI persistently diminish users’ confidence and trustworthiness~\cite{wu2024usable,bilal2025llms}. A key question is whether the existing XAI techniques developed for DNNs can be readily deployed for LLM/agentic AI systems. 
Some preliminary works have applied methods, such as feature‐attribution (e.g. SHAP/LIME on token embeddings), chain-of-thought tracking, and natural-language rationale generation by the LLM itself, thereby providing transparency into which inputs influenced the agent’s decision and how the reasoning proceeded~\cite{tong2025wirelessagent,cambria2024xaimeetsllmssurvey}.
However, there has yet to be a mature body of work that adapts XAI to network agentic systems with stringent latency in the wireless network domain. 

New XAI methods are required that can (i) provide real-time, lightweight explanations compatible with live network control loops, (ii) correlate high-dimensional wireless input features with agent actions in a human-interpretable way, and (iii) validate that the generated rationales truly reflect the agent’s internal decision-making rather than post-hoc plausible narratives. 
Addressing these gaps is critical to realize responsible AI  since transparency, causality, confidence, privacy, and human trust are all required for deployment of LLM/agentic AI systems in wireless communications.

% For XAI techniques applied to LLMs, their responsibility lies not only in unravelling their closed-box nature, but also in further enhancing the applicability of LLMs in practical tasks. For example, given that the input prompts employed by LLM during inference are flexible and multimodal, XAI techniques can be employed to acquire more transparent input content to enhance the transparency of their outputs, which can reveal the decision-making processes of LLMs, thereby improving model explainability.

\section{Conclusion}

This paper provided a responsible-AI–oriented survey of explainable learning for wireless PHY layer. We formalized key responsibility goals (trustworthiness, causality, privacy, fairness, transferability, informativeness, and confidence) and developed a taxonomy that connects explanation forms, validation criteria, and deployment constraints. We reviewed representative neural models and explainability practices, and summarized how explanations support reliability, robustness, and risk-aware control under non-stationary channels. Open issues include radio-native explanation semantics and validation metrics, real-time and lightweight explanations for edge/RAN control, cross-layer explanation consistency, and faithful explanations under channel dynamics and distribution shift. We further outlined challenges for LLM and agentic AI in wireless PHY, where current XAI tools remain insufficient to ensure faithful, causal, and privacy-preserving rationales.

\bibliographystyle{IEEEtran}
\bibliography{changeref2}

\end{document}